\documentclass[aps,prd,longbibliography,preprint,amsmath,amssymb,floatfix]{revtex4-1}
\usepackage{graphicx}

\usepackage{bm}
\usepackage{xcolor}
\PassOptionsToPackage{hyphens}{url}\usepackage{hyperref}
\usepackage{url}
\usepackage{bookmark}
\usepackage{lineno}

\newif\ifcolorenabled
\colorenabledtrue   

\ifcolorenabled
\else

\renewcommand{\color}[1]{}
\renewcommand{\definecolor}[3]{}
\fi

\newcommand{\be}{\begin{equation}}
\newcommand{\ee}{\end{equation}}
\newcommand{\eq}[1]{Eq.~(\ref{#1})}
\newcommand{\fig}[1]{Fig.~\ref{#1}}
\def\bea{\begin{eqnarray}}
\def\eea{\end{eqnarray}}

\def\bra{\langle}
\def\ket{\rangle}
\def\vq{{\bf q}}

\def\vk{{\bf k}}

\def\qp{{\bf q}_{\parallel}}

\def\tj{$t$-$J$ }
\def\tjv{$t$-$J$-$V$ }

\def\ii{\mathrm{i}}
\def\lN{large-{\it N} }
\def\xop{$\hat{X}$}

\begin{document}

\title{Strong-coupling theory of bilayer plasmon excitations} 

\author{
Hiroyuki Yamase$^{1}$, Luciano Zinni$^{2}$, Mat\'{\i}as Bejas$^{3}$, and Andr\'es Greco$^{3}$
}
\affiliation{
{$^1$}Research Center of Materials Nanoarchitectonics (MANA), National Institute for Materials Science (NIMS), Tsukuba 305-0047, Japan \\
{$^2$}Facultad de Ciencias Exactas, Ingenier\'{\i}a y Agrimensura (UNR-CONICET), Avenida Pellegrini 250, 2000 Rosario, Argentina \\
{$^3$}Facultad de Ciencias Exactas, Ingenier\'{\i}a y Agrimensura and Instituto de F\'{\i}sica Rosario (UNR-CONICET), Avenida Pellegrini 250, 2000 Rosario, Argentina
}

\date{\today}

\begin{abstract} 
Recently plasmon excitations in bilayer lattice systems were studied extensively in the weak-coupling regime. Unlike single-layer systems, these bilayers exhibit two distinct modes, $\omega_{\pm}$, which show characteristic dependences upon the momentum and hopping integrals along the $z$ direction. To apply them to cuprates, strong correlation effects should be considered, but a comprehensive analysis has not yet been investigated. In this work, we present a strong-coupling theory to analyze  the charge dynamics of a bilayer system, utilizing the $t$-$J$-$V$ model,  which includes the long-range Coulomb interaction, $V$, on a lattice. Although our theoretical framework is fundamentally different from the weak-coupling approach, we find that resulting plasmon excitations are similar to those of a weak-coupling theory. A key distinction is that our strong-coupling framework reveals a  noticeable suppression of particle-hole excitations, which allows the plasmon modes to remain well-defined over a wider region of momentum. We suggest that the experimentally reported plasmon excitations in Y-based cuprates can be described by the $\omega_{-}$ mode, although we call for more systematic experiments to verify this. 
\end{abstract}


\maketitle

\section{introduction}
The parent compounds of cuprate superconductors are widely known to be antiferromagnetic Mott insulators. Upon carrier doping, the charge degrees of freedom become active and the system transitions into a metallic state, which suppresses the antiferromagnetic order. A high-temperature superconducting state emerges at a carrier doping level of approximately 5 \% for hole-doped and 10--15 \% for electron-doped materials, reaching a maximal $T_{c}$ around 16 \% doping \cite{keimer15}.  

It is well established that electrons within the CuO$_{2}$ layers play a central role in high-temperature superconductivity. For this reason, the  two-dimensional $t$-$J$ and Hubbard models are considered the  minimal theoretical framework \cite{anderson87}. While the importance of spin-spin interaction is frequently emphasized, the charge degrees of freedom should be equally crucial to understanding the cuprate physics. The full charge dynamics in momentum-energy space has recently been revealed comprehensively through the advent of the resonant inelastic x-ray scattering (RIXS) technique \cite{ament11,degroot24}.  

In the charge excitation spectrum, two distinct features have been observed. The first feature is low-energy excitations around in-plane momentum $\qp=(0.6\pi, 0)$  in hole-doped cuprates  \cite{ghiringhelli12,chang12,achkar12,blackburn13,blanco-canosa14,comin14,tabis14,da-silva-neto14,hashimoto14,peng16,chaix17,arpaia19,yu20,wslee21,lu22,arpaia23,misc-LSCO} and around $(0.5\pi, 0)$ in electron-doped cuprates \cite{da-silva-neto15,da-silva-neto16,da-silva-neto18}. The second feature is a distinct V-shaped dispersion centered at $\qp=(0,0)$. While its origin was initially debated \cite{ishii05,ishii14,wslee14,greco16,ishii17, dellea17}, it is now consistently understood as an acousticlike plasmon, which is a characteristic feature of layered systems \cite{greco16,hepting18}. A particular important feature of the plasmon is the rapid decrease of its  energy with increasing momentum transfer perpendicular to the layers, $q_{z}$  \cite{hepting18,greco19}. Additionally, a gap at $\qp=(0,0)$ has been observed \cite{hepting22}, which is proportional to the interlayer hopping $t_{z}$ \cite{greco16}. This dependence provides a valuable way to extract the value of $t_{z}$, a parameter that is difficult to determine through other experimental techniques. 

The present paper focuses on the plasmon excitations in cuprate superconductors. While most experimental studies have been limited to single- and infinite-layer systems, where the unit cell contains one CuO$_{2}$ plane, a theoretical description requires two additional factors beyond the standard $t$-$J$ and Hubbard models: the long-range Coulomb interaction (LRC)---which in continuum space has a $1/r$ dependence---and the interlayer hopping, both of which are essential to correctly capture the observed $q_{z}$ dependence of the plasmon energy \cite{greco16,greco19,greco20}. 

A crucial theoretical question arises regarding how plasmon excitations change with an increasing number of layers per unit cell. Pioneering studies by Fetter \cite{fetter74} and Griffin and Pindor \cite{griffin89} for a layered electron gas model showed the existence of two modes, one of which has very low energy.  Concurrently, the superconducting onset temperature $T_{c}$ increases substantially when the number of CuO$_{2}$ planes is increased to two or three, up to 140 K; for four or more layers, $T_{c}$ plateaus around 110 K \cite{iyo07}.  

Experimental studies of low-energy plasmon modes for multilayer cuprates remain limited, with a single report in  bilayer Y-based compounds \cite{bejas24}—which has been extensively analyzed within the weak-coupling random phase approximation (RPA) \citep{bejas24,yamase25,sellati25}—and a subsequent observation in trilayer Bi-based cuprates  \cite{nakata25}.

The aim of the present work is to formulate a strong-coupling theory for the bilayer system. This work complements the very recent weak-coupling analysis in Ref.~\cite{yamase25}, allowing us to systematically clarify the similarities and differences between the two theoretical approaches. This point is important because some argue that weak-coupling approaches are inadequate for cuprates, given their strongly correlated nature. In fact, alternative frameworks have been proposed, including holographic descriptions of charge dynamics \cite{mitrano18,romero-bermudez19,vandeneede24}. Furthermore, the formalism developed here acquires renewed relevance following the recent discovery of high-$T_c$ superconductivity in bilayer nickelates \cite{sun23,hou23}.

The remainder of this paper is organized as follows. In Sec.~\ref{sec:formalism}, we formulate a large-$N$ theory of the layered $t$-$J$-$V$ model. The LRC  is treated on a lattice, respecting the bilayer structure \cite{yamase25}, rather than a continuum form used in a layered electron gas model \cite{fetter74,griffin89}. In Sec.~\ref{sec:results}, we present our results for charge excitation spectra, which may be compared to those obtained in the weak-coupling theory \cite{yamase25}. We also investigate the dependence of these excitation on the LRC, $V$. Discussions and conclusions are given in Sec.~\ref{sec:discussions} and \ref{sec.conclusions}, respectively. In Appendix \ref{app:self-energy}, the complete formalism of our strong coupling theory is presented. In Appendix \ref{app:zero-sound-data}, we discuss whether the collective modes obtained within the \tj model without LRC can account for the experimental data.

\section{Formalism}\label{sec:formalism}
\begin{figure}[h!]
	\centering
	\includegraphics[]{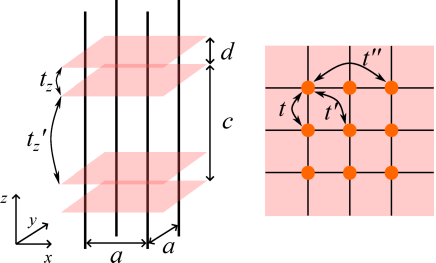}
	\caption{Schematic of the bilayer lattice and corresponding hopping integrals.  Each layer forms  a square lattice with lattice constant $a$; $d$ is the intrabilayer distance and $c$ is the lattice constant along the $z$ direction.
	}
	\label{fig:lattice}
\end{figure}

We begin with our theoretical analysis by defining the bilayer \tjv model on a square lattice. The Hamiltonian is given by:  
\begin{align}
	H=&-\sum_{i,j,\sigma,\alpha,\beta} t_{ij}^{\alpha\beta}\tilde{c}^\dagger_{i\sigma,\alpha}\tilde{c}_{j\sigma,\beta}
	-\mu\sum_{i,\alpha}n_{i,\alpha}     +J\sum_{\bra i,j \ket,\alpha}\left(\vec{S}_{i,\alpha}\cdot\vec{S}_{j,\alpha}-\frac{1}{4}n_{i,\alpha}n_{j,\alpha}\right) \nonumber\\    
	&+\frac{J_\perp}{2}\sum_{i,\alpha \ne \beta}\left(\vec{S}_{i,\alpha}\cdot\vec{S}_{i,\beta}-\frac{1}{4}n_{i,\alpha}n_{i,\beta}\right)
	+\frac{1}{2}\sum_{i \ne j,\alpha,\beta}V_{ij}^{\alpha\beta}n_{i,\alpha} n_{j,\beta} \; ,
	\label{eq:model}
\end{align}
where $i$ and $j$ run over the three-dimensional lattice sites, $\alpha,\beta=1,2$ denote the plane within a unit cell, and $\tilde{c}^\dagger_{i\sigma,\alpha}$ ($\tilde{c}_{i\sigma,\alpha}$) is the creation (annihilation) operator of an electron with spin $\sigma(=\uparrow,\downarrow)$ at site $i$ in layer $\alpha$. The hopping integrals $t_{ij}^{\alpha\beta}$ extend up to third-nearest neighbors on the square lattice of each layer,  denoted as $t$, $t'$ and $t''$, respectively. Along the $z$ direction, the hopping within a bilayer  is denoted as $t_z$ (intrabilayer hopping), and the hopping between bilayers is given by $t_{z}^{'}$ (interbilayer hopping)---see Fig.~\ref{fig:lattice}.  
$J$ is the strength of the in-plane spin exchange interaction between nearest-neighbor site $\bra i, j \ket$; $J_\perp$ is out-of-plane spin exchange and considered only within the intrabilayer; $V_{ij}^{\alpha\beta}$ is the three-dimensional LRC. $n_{i,\alpha}=\sum_\sigma \tilde{c}^\dagger_{i\sigma,\alpha} \tilde{c}_{i\sigma,\alpha}$ is the electron density operator at site $i$ in layer $\alpha$, $\vec{S}_{i,\alpha}$ is the spin operator, and $\mu$ is the chemical potential. In the $t$-$J$-$V$ model, all operators are defined in the Fock space without double occupancy, which yields the local constraint:  
\be
\sum_{\sigma} \tilde{c}^{\dagger}_{i, \sigma, \alpha} \tilde{c}_{i, \sigma, \alpha} \leq 1
\label{constraint}
\ee
for any site $i$ and layer $\alpha$.

Here we employ a large-$N$ technique in a path integral formalism in terms of the Hubbard operators \cite{foussats02}. In this scheme, the number of spin component is extended from 2 to $N$ and physical quantities are computed by the power of $1/N$ systematically.

Setting $t_z = t_{z}^{'} = J_\perp = V_{ij}^{\alpha \beta} = 0 \, (\alpha \ne \beta)$ in Eq.~\eqref{eq:model} reduces the system to two decoupled \tjv models. 
Such a model was successfully applied to analyzing the charge dynamics in cuprates \cite{greco16,greco19,greco20,nag20,yamase21c,hepting22,hepting23,nag24}. We extend this formalism to the bilayer lattice shown in Fig.~\ref{fig:lattice}, which was originally introduced to analyze bilayer nickelates \cite{bejas25}. Our interest lies in bilayer plasmon excitations and we now summarize the key steps of this formalism. The complete formalism is given in Appendix~A.

At leading order in the bilayer system, the electron Green's function is obtained as $2 \times 2$ matrix: 
\begin{eqnarray}\label{eq:G0}
	G^{(0)}_{\alpha \beta}(\vk,\ii\nu_n)= \left(
	\begin{array}{cc}
		\ii\nu_n -\varepsilon^{\parallel}_{\vk} & -\varepsilon^{\perp}_{\vk}\mathrm{e}^{\ii k_z \frac{d}{c}}\\
		-\varepsilon^{\perp *}_{\vk}\mathrm{e}^{-\ii k_z \frac{d}{c}} & \ii\nu_n-\varepsilon^{\parallel}_{\vk}\
	\end{array}
	\right)^{-1} \;, 
\end{eqnarray}
where $\nu_{n}$ is a fermionic Matsubara frequency, and the electron dispersions are obtained as 
\begin{align}{\label{eq:ek}}
	\varepsilon_{\vk}^{\parallel} &= -2 \left( t \frac{\delta}{2}+\chi \right) \left(\cos k_{x}+\cos k_{y}\right)- 4t' \frac{\delta}{2} \cos k_{x} \cos k_{y} - 2t'' \frac{\delta}{2} \left(\cos 2k_{x} + \cos 2k_{y}\right) - \mu  \,, \\
	\varepsilon^{\perp}_{\vk} &=-\left[t_z \frac{\delta}{2}\left(\cos k_{x}-\cos k_{y}\right)^2 + \chi '\right]-t_{z}^{'} \frac{\delta}{2}\left(\cos k_{x}-\cos k_{y}\right)^2 \mathrm{e}^{-\ii k_z } \, .
\end{align}
Here $k_{x}$, $k_{y}$ and $k_{z}$ are in units of the inverse of the lattice constant $a$, $a$, and $c$, respectively. For a given doping $\delta$, the chemical potential $\mu$ and the values $\chi$ and $\chi'$ are determined self-consistently by solving: 
\begin{equation}{\label {chi}}
	\chi= \frac{J}{8N_s} \sum_{\vk} (\cos k_x + \cos k_y) \left[ n_{F}( \varepsilon_{\vk}^{1}) + n_{F}( \varepsilon_{\vk}^{2})\right] \;, 
\end{equation}
\begin{equation}{\label {chi'}}
	\chi'= -\frac{J_\perp}{4N_s} \sum_{\vk} \frac{\varepsilon^{\perp}_{\vk}\mathrm{e}^{\ii k_z \frac{d}{c}}}{ | \varepsilon^{\perp}_{\vk} |} \left[ n_{F}( \varepsilon_{\vk}^{1}) - n_{F}( \varepsilon_{\vk}^{2}) \right] \;,
\end{equation}
and
\begin{equation}
	1 - \delta = \frac{1}{N_s} \sum_{\vk} \left[n_{F}( \varepsilon_{\vk}^{1}) + n_{F}( \varepsilon_{\vk}^{2})\right] \;,
\end{equation}
where $N_s$ is the number of lattice sites; $n_{F}$ is the Fermi distribution function; from the determinant of Eq.~\eqref{eq:G0}, the bonding and antibonding bands ($\alpha=1,2$, respectively) can be obtained as
\begin{equation}
	\varepsilon_{\vk}^\alpha = \varepsilon_{\vk}^{\parallel}-(-1)^{\alpha} |\varepsilon^{\perp}_{\vk}| \; .
\end{equation}

Charge fluctuations are described by the $14 \times 14$ matrix of the boson propagator at the order of the $1/N$ in the bilayer system. But the on-site charge fluctuations including plasmons that we are interested in are described by a $ 4\times 4$ reduced matrix, 
\begin{align}
	D^{-1}_{ab}({\bf q}, \ii\omega_n) &=\left[ D^{(0)}_{ab}({\bf q}, \ii\omega_n) \right]^{-1} - \Pi_{ab}({\bf q}, \ii\omega_n) \, ,
	\label{eq:Dyson}
\end{align}
where $a$ and $b$ run from 1 to  4; $\vq$ is a three-dimensional vector; $\omega_{n}$ is a bosonic Matsubara frequency.  $D^{(0)}_{ab}({\bf q}, \ii\omega_n)$ is the bare bosonic propagator and is obtained as 
\begin{align}
	\left[D^{(0)}_{ab}(\vq,\ii\omega_{n})\right]^{-1} &= N \left(
	\begin{array}{cccc}
		\frac{\delta^2}{2} \left[\frac{V(\vq)}{2}- J(\vq)\right] & \frac{\delta}{2} & \frac{\delta^2}{2} \left[\frac{V'(\vq)}{2}- J'(\vq)\right] & 0 \\
		\frac{\delta}{2} & 0 & 0 & 0 \\
		\frac{\delta^2}{2} \left[\frac{V'^*(\vq)}{2}- J'^{*}(\vq)\right] & 0 & \frac{\delta^2}{2} \left[\frac{V(\vq)}{2}- J(\vq)\right] & \frac{\delta}{2} \\
		0 & 0 &  \frac{\delta}{2} & 0
	\end{array}
	\right) \; ,
\end{align}
where $J({\bf q}) = (J/2) (\cos q_x + \cos q_y)$ ($q_x$ and $q_y$ are in units of the inverse of the lattice constant $a$) and $J'(\vq) = (J_\perp/4) \mathrm{e}^{-\ii q_z \frac{d}{c}}$ ($q_z$ is in units of the inverse of the lattice constant $c$). $V(\vq)$ and $V'(\vq)$ are the intralayer and interlayer Fourier components of the LRC in the bilayer lattice, respectively, which are given by
\begin{align}
	V(\vq)=&\frac{V_c}{\mathrm{det}\tilde{V}}\left[\alpha\left(2-\cos q_x-\cos q_y\right)-\frac{1}{2}h_3-\frac{1}{2}h_1\cos q_z \right] \; , \label{eq:Vq}\\
	V'(\vq)=&\frac{1}{2}\frac{V_c}{\mathrm{det}\tilde{V}}\left\{ h_2\cos \left(q_z\frac{d}{c}\right) +h_4\cos \left[q_z\left(1-\frac{d}{c}\right)\right]-\ii h_2 \sin \left(q_z \frac{d}{c}\right) \right. \nonumber \\
	&\left.  + \ii h_4 \sin \left[q_z\left(1-\frac{d}{c}\right)\right] \right\} \; , \label{eq:Vqp}\\
	\mathrm{det}\tilde{V}=&\left[\alpha\left(2-\cos q_x-\cos q_y\right)\right]^2-\alpha\left(2-\cos q_x-\cos q_y\right)\left(h_1\cos q_z  +h_3\right)\nonumber \\
	&+\frac{6c^2}{(c-d)(2c-d)}\left(1-\cos q_z \right) \; . \label{eq:detV}
\end{align}
Here $V_c=\frac{e^2c}{2a^2\varepsilon_\perp}$, $\alpha=\frac{c^2\varepsilon_\parallel}{a^2\varepsilon_\perp}$ ($\varepsilon_\parallel$ and $\varepsilon_\perp$ are the parallel and perpendicular dielectric constants, respectively; $e$ is the electric charge of electron), and $h_1=\frac{2c(c-2d)}{(2c-d)(c-d)}$, $h_2=\frac{2c}{c-d}$, $h_3=-\frac{4c}{c-d}$, and  $h_4=\frac{2c(c+d)}{(2c-d)(c-d)}$. 
Equations~\eqref{eq:Vq} and \eqref{eq:Vqp} were first derived in Ref. \cite{yamase25} and they represent the only known analytical expressions for the LRC in the bilayer structure. The self-energy  components $\Pi_{ab}({\bf q}, \ii\omega_n)$ in \eq{eq:Dyson} are calculated in Appendix~A.

In the \lN formalism for bilayer systems, the charge-charge correlation function $\chi(\mathrm{r}_i-\mathrm{r}_j,\tau)=\langle T_\tau n_i(\tau)n_j(0)\rangle$ is related in $\vq$-$\omega$ space to the dressed bosonic propagator $D_{ab}({\bf q}, \ii\omega_n)$ as
\begin{equation}\label{eq:chi_lN}
	\chi\left(\vq,\ii\omega_n\right)=\frac{N}{2}\left(\frac{\delta}{2}\right)^2
	\left[D_{11}\left(\vq,\ii\omega_n\right)+
	D_{33}\left(\vq,\ii\omega_n\right)+
	D_{13}\left(\vq,\ii\omega_n\right)+
	D_{31}\left(\vq,\ii\omega_n\right)\right] \; . 
\end{equation}
The elements $D_{11}$ and $D_{33}$ ($D_{13}$ and $D_{31}$) give the intralayer (interlayer) contributions to $\chi\left(\vq,\ii\omega_n\right)$.

After performing the analytical continuation $\ii\omega_n \rightarrow \omega + \ii\Gamma$ and adopting the physical value $N=2$, we obtain the imaginary part of the charge-charge correlation function, Im$\chi(\mathbf{q},\omega)$, which can be directly compared with RIXS measurements. The parameter $\Gamma$ may account  for both experimental resolution and spectral broadening from electron correlations \cite{prelovsek99}.

\section{Results} \label{sec:results}

We present our results organized into three subsection. 
Focusing on cuprates and without losing generality, we adopt the following parameter set:  $t'=-0.3t$, $t''=0.15t$, $J=0.3t$, $J_{\perp}=0$ \cite{misc-perp}. The lattice parameters are set to $a=3.88$ \AA, $c=11.68$ \AA , $d=3.36$ \AA  \cite{bejas24}. The dielectric constants are chosen as $\epsilon_{\parallel} = 6 \epsilon_{0}$, $\epsilon_{\perp} = 2.25 \epsilon_{0}$. These values yield $V_{c}=38.75 t$ and $\alpha=24$. We take hole doping rate $\delta=0.21$, corresponding to the electron density $n=0.79$. The broadening parameter $\Gamma$ is set to $0.01t$ except for Fig.~\ref{Vc-dep-dispersion}. We take a temperature of $T=0.04t$ to avoid bond-charge instabilities \cite{bejas12,bejas14}. Our analysis first focuses on the case of $t_{z}^{'}=0$ before examining its effect. Unless stated otherwise, $t$ is the unit of energy.

\subsection{Charge excitation spectra with  \boldmath{$t_{z}^{'}=0$}} 
\subsubsection{Overall spectrum} 

\begin{figure}[t]
\centering
\includegraphics[]{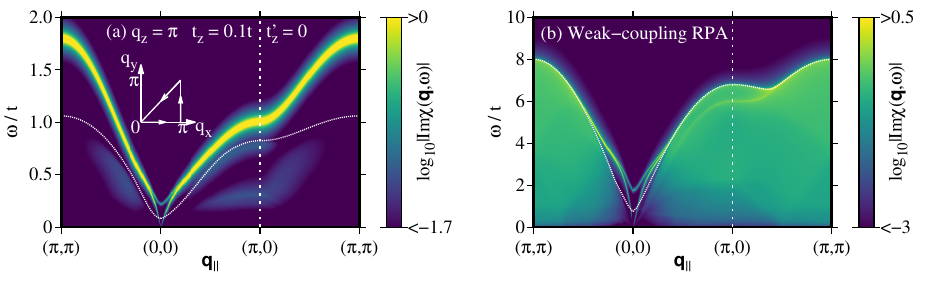}
\caption{(a) Intensity map of charge excitation spectrum $\log_{10} | {\rm Im}\chi(\vq, \omega)|$ in the plane of in-plane momentum $\qp$ and energy transfer $\omega$. The white dotted curve represents the upper boundary of the particle-hole continuum excitations. The strong intensity corresponds to plasmon modes. (b) The corresponding map in a weak-coupling analysis \cite{yamase25} for the same parameters as (a) except that $V_{c}=130$ is taken. 
}
\label{overall}
\end{figure}

Figure~\ref{overall}(a) presents the charge excitations spectrum across a broad range of in-plane momentum $\qp$ and energy transfer $\omega$. The white dotted curve shows the upper boundary of the particle-hole continuum. While weak signals are visible below this boundary, the most significant spectral weight is found above it, corresponding to plasmon excitations. These features are qualitatively similar to those observed in a single-layer system (see Fig.~1 in Ref.~\cite{greco16}). 
A crucial distinction emerges at low energy ($\omega <0.3t$) and small in-plane momentum, specifically around $\qp=(0,0)$, where two distinct plasmon modes are present for a fixed $q_{z}=\pi$. Upon closer inspection, the lower branch is found to be a gapless plasmon mode that extends into the continuum, a feature characteristic of bilayer systems \cite{yamase25}. Following the nomenclature of Griffin and Pindor \cite{griffin89}, we refer to the higher-energy and lower-energy  branches as $\omega_{+}$ and $\omega_{-}$ modes, respectively.

The corresponding weak-coupling RPA result \cite{yamase25} is shown in \fig{overall}(b) for exactly the same parameters as \fig{overall}(a) except for a value of $V_{c}$ so that two plasmon branches are well visible. A comparison between \fig{overall}(a) and (b) reveals that strong electronic correlations suppress the particle-hole continuum substantially, allowing the plasmon branches to remain well defined along the entire Brillouin-zone path, whereas in the weak-coupling scheme they are confined to the vicinity of $\mathbf{q}_{\parallel}=(0,0)$ and merge into the continuum away from $\qp=(0,0)$. 

The remainder of this section will focus on the detailed properties of $\omega_{+}$ and $\omega_{-}$ modes. 

\subsubsection{Plasmon excitations}

\begin{figure}[!ht]
\centering
\includegraphics[width=8cm]{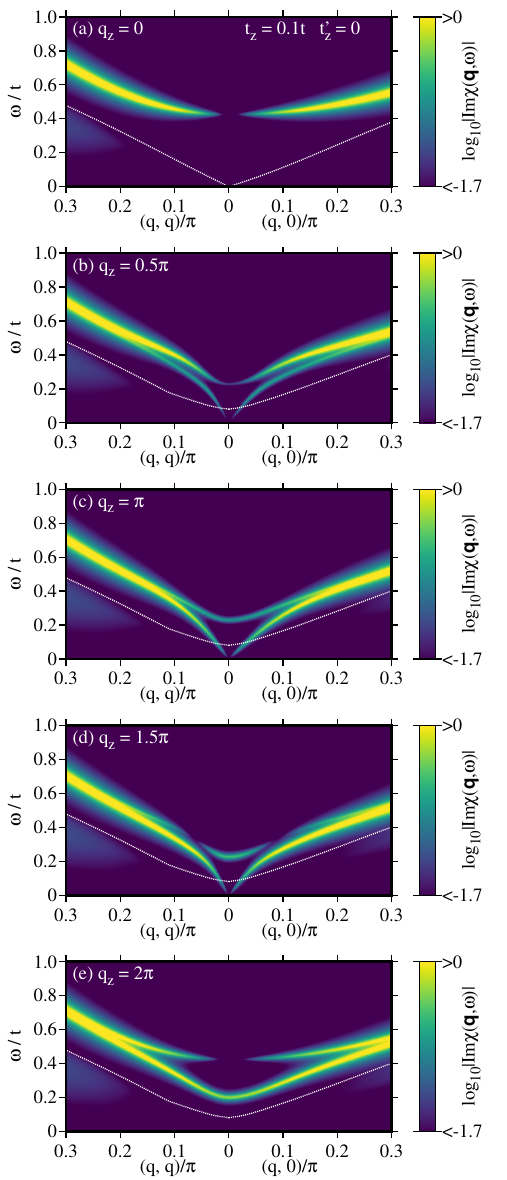}
\caption{Intensity maps of $\log_{10} | {\rm Im}\chi(\vq, \omega)|$ for a sequence of $q_{z}$ around a region of $\vq_{\parallel}=(0,0)$ for $t_{z}=0.1t$: (a) $q_{z}=0$, (b) $q_{z}=0.5\pi$, (c) $q_{z}=\pi$, (d) $q_{z}=1.5\pi$, and (e) $q_{z}=2\pi$. The white dotted curve denotes the upper boundary of the particle-hole continuum. It goes to zero at $\vq_{\parallel} =(0,0)$ and $q_{z}=0$ in  (a). 
}
\label{tz0.1}
\end{figure}

To investigate the plasmon excitations in more detail, we analyze the $\vq$-$\omega$ map around $\qp=(0,0)$ for a sequence of $q_{z}$ values as shown in Figs.~\ref{tz0.1}(a)--(e). At $q_{z}=0$ [\fig{tz0.1}(a)], only the $\omega_{+}$ mode is present. The charge conservation  makes the spectral weight vanish at $\qp =(0,0)$ because charge fluctuations between the two layers are in-phase. This confirms that the $\omega_{+}$ mode corresponds to the well-known optical plasmon. By symmetry, the $\omega_{-}$ mode corresponds to out-of-phase fluctuations and is not present for $q_{z}=0$, a feature that has the following physical interpretation: When applying infinitesimally small external electric field with $q_{z}=0$, this external field is strictly uniform along the out-of-plane direction and thus cannot couple to the out-of-phase charge oscillation, i.e., $\omega_{-}$ mode.

Although the spectral weight at $\vq_{\parallel}=(0,0)$ vanishes for $q_{z}=0$, a finite intrabilayer hopping $t_{z}$ results in a finite energy for the upper boundary of the continuum at $\qp=(0,0)$ for $q_{z} \ne 0$. This allows for the presence of two plasmon modes, the $\omega_{+}$ and $\omega_{-}$ modes, as seen in  Figs.~\ref{tz0.1}(b)--(d). These modes should not be confused with the even and odd modes, since the LRC couples fluctuations across all layers. Notably, the $\omega_{-}$ mode exhibits a gapless dispersion, which is particularly evident as it enters the continuum around $\vq_{\parallel}=(0,0)$. However, a significant change occurs at $q_{z} = 2n\pi$ where $n\ne 0$ [\fig{tz0.1}(e)], as the $\omega_{-}$ mode becomes gapped. A comparison between Figs.~\ref{tz0.1}(a) and (e) reveals that the plasmon excitations at $q_{z}=0$ are not generic. Instead, the behavior at $q_{z} = 2n\pi$ with $n\ne 0$ is more representative: both $\omega_{\pm}$ modes are present and gapped, but the intensity of the $\omega_{+}$ mode vanishes at $\vq_{\parallel}=(0,0)$ while it does not for the $\omega_{-}$ mode.

\begin{figure}[t]
\centering
\includegraphics[width=15cm]{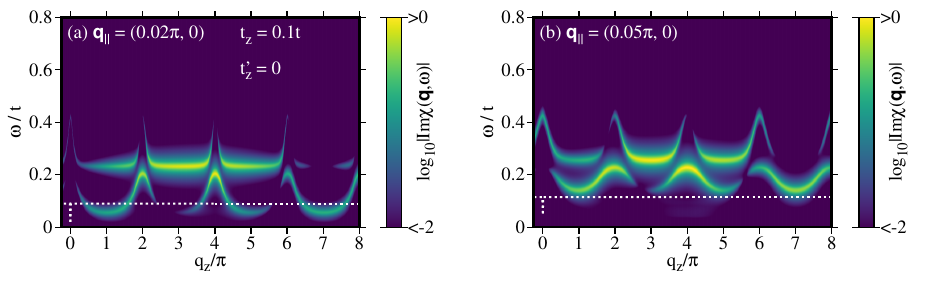}
\caption{$q_{z}$ dependence of $\omega_{+}$ mode (higher energy) and $\omega_{-}$ mode (lower energy) at (a) $\vq_{\parallel}=(0.02\pi, 0)$ and (b) $\vq_{\parallel}=(0.05\pi, 0)$ for $t_{z}=0.1t$. The white dotted line is the upper boundary of the continuum spectrum and exhibits a sharp drop at $q_{z}=0$ because of the vanishing of the $\omega_{-}$ mode there. 
}
\label{tz0.1-qz}
\end{figure}

We next examine the $q_{z}$ dependence of the $\omega_{\pm}$ modes more closely. Figure~\ref{tz0.1-qz}(a) shows results at a fixed in-plane momentum of $\vq_{\parallel}=(0.02\pi, 0)$ for $t_{z}=0.1t$. The white dotted line indicates the upper boundary of the particle-hole continuum. The $\omega_{+}$ mode is always present above the continuum. Its energy sharply peaks at $q_{z} = 2n\pi$ (corresponding to  the optical plasmon) and quickly decreases as $q_{z}$ moves away from these values, remaining largely $q_{z}$ independent until the next peak. The $\omega_{-}$ mode, which is absent at $q_{z}=0$, gains intensity immediately as $q_{z}$ increases. It shows a dispersive feature between $\omega = 0.06$--$0.21 t$ with a peak at $q_{z} = 2n\pi$ where $n \ne 0$. This dispersion is particularly interesting since it occurs by crossing the boundary of the continuum.

Figure~\ref{tz0.1-qz}(b) presents the same plot but for slightly larger in-plane momentum $\vq_{\parallel}=(0.05\pi, 0)$. The $\omega_{+}$ mode exhibits a $q_{z}$ dependence similar to that in \fig{tz0.1-qz}(a). The $\omega_{-}$ mode, however, shows a clear cosinelike dispersion along the $q_{z}$ direction  and is situated entirely above the continuum. 

In both Figs.~\ref{tz0.1-qz}(a) and (b), the intensity of the $\omega_{\pm}$ modes displays a characteristic $q_{z}$ dependence, with ``nodes''  where the spectral weight almost vanishes at specific $q_{z}$ values. For example, the $\omega_{+}$ mode loses intensity around $q_{z}=7\pi$. These positions are parameter dependent, especially on the ratio of $d/c$, though the vanishing intensity of the $\omega_{-}$ mode at $q_{z}=0$ is a robust feature.
It is also important to note that the intensity of both $\omega_{\pm}$ modes does not have $2\pi$ periodicity along the $q_{z}$ direction.  

\begin{figure}[th]
\centering
\includegraphics[width=8cm]{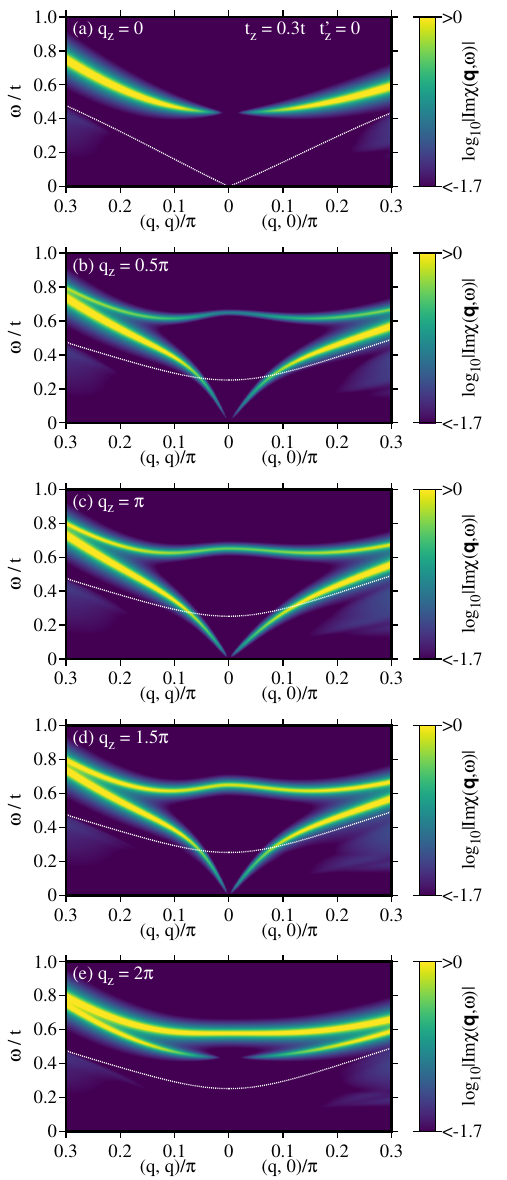}
\caption{Intensity maps of $\log_{10} | {\rm Im}\chi(\vq, \omega)|$  for a sequence of $q_{z}$ around a region of $\vq_{\parallel}=(0,0)$ for $t_{z}=0.3t$: (a) $q_{z}=0$, (b) $q_{z}=0.5\pi$, (c) $q_{z}=\pi$, (d) $q_{z}=1.5\pi$, and (e) $q_{z}=2\pi$. The white dotted curve denotes the upper boundary of the particle-hole continuum. It becomes zero at $\vq_{\parallel}=(0,0)$ and $q_{z}=0$ in (a). 
}
\label{tz0.3}
\end{figure}

For completeness, we also investigate the case of a large intrabilayer hopping $t_{z}=0.3t$. The spectral maps for various $q_{z}$ are shown in \fig{tz0.3}. Similar to the $t_{z}=0.1t$ case, only a single mode is present at $q_{z}=0$ [\fig{tz0.3}(a)], and its vanishing spectral weight at  $\qp=(0,0)$ identifies it as the $\omega_{+}$ mode. For $q_{z} \ne 0$ [Figs.~\ref{tz0.3}(b)--(d)], two modes appear. A remarkable feature in this large $t_{z}$ case is that the upper mode forms an upward-convex shape around $\vq_{\parallel}=(0,0)$ whereas the other mode is gapless and sharply defined even inside the continuum.  At $q_{z}=2\pi$ [\fig{tz0.3}(e)], two gapped modes are present. However, a key difference from the $t_{z}=0.1t$ case is that the lower-energy mode in \fig{tz0.3}(e) is the one with zero spectral weight at $\qp=(0,0)$, identifying it as the $\omega_{+}$ mode. Given the continuity of the modes with varying $q_{z}$, this suggests a reversal in the energy hierarchy of the $\omega_{\pm}$ modes for a large $t_{z}$. Specifically the $\omega_{+}$ mode becomes gapless for $q_{z} \ne 2 n \pi$, while $\omega_{-}$ mode is always gapped except for its vanishing at $q_{z}=0$. This inversion can be traced back to the competing energy scales set by the LRC---through $V_{c}$ and $\alpha$---and by the intrabilayer hopping $t_{z}$.  When $t_{z}$ becomes sufficiently large, the energy scale of the $\omega_{-}$ mode can exceed the optical plasmon energy at $q_{z}=2n\pi$, producing the observed reversal between the $\omega_{+}$ and $\omega_{-}$ modes. This situation can be realized for a system with a smaller intrabilayer distance. Designing materials in this way would be highly interesting.

\begin{figure}[t]
\centering
\includegraphics[width=15cm]{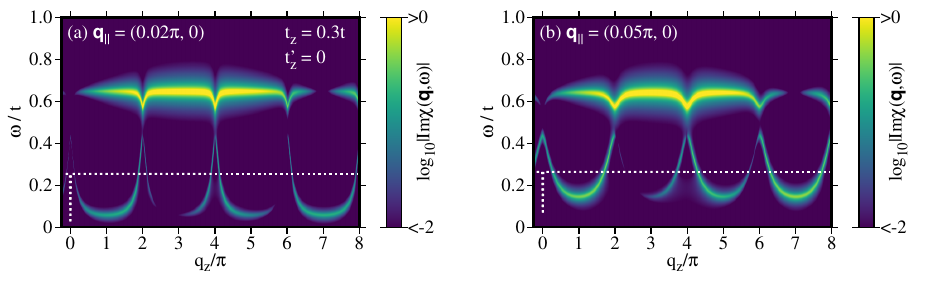}
\caption{$q_{z}$ dependence of $\omega_{+}$ mode (lower energy) and $\omega_{-}$ mode (higher energy) at (a) $\vq_{\parallel}=(0.02\pi, 0)$ and (b) $\vq_{\parallel}=(0.05\pi, 0)$ for $t_{z}=0.3t$. The corresponding results of $t_{z}=0.1t$ are shown in \fig{tz0.1-qz}. The white dotted line is the upper boundary of the continuum spectrum and exhibits a sharp drop at $q_{z}=0$ because of the vanishing of the $\omega_{-}$ mode there. 
}
\label{tz0.3-qz}
\end{figure}

As we observed in \fig{tz0.3}, the $\omega_{-}$ mode has higher energy than the $\omega_{+}$ mode for large $t_{z}$. This is confirmed by the results in Figs.~\ref{tz0.3-qz}(a) and (b), where the higher-energy mode shows zero spectral weight at $q_{z}=0$, characteristic of the $\omega_{-}$ mode. The $q_{z}$ dependence of the $\omega_{-}$ mode for large $t_{z}$ is markedly different from the $t_{z}=0.1t$ case, showing a dip rather than a peak at $q_{z} = 2 n \pi$. In contrast, the low-energy $\omega_{+}$ mode exhibits a significant dispersive feature, particularly at $\qp=(0.02\pi, 0)$, where its energy spans a wide range between $\omega = 0.06$--$0.44t$. This mode is sharply defined even when it lies within the continuum, a consequence of the very low spectral weight of the continuum itself near $\qp=(0,0)$.  At $\vq_{\parallel} = (0.05\pi, 0)$ shown in \fig{tz0.3-qz}(b), the $\omega_{+}$ mode is less dispersive, but shows a similar overall feature. 

\subsection{Charge excitation spectra with  \boldmath{$t_{z}^{'} \ne 0$}}
The previous analysis has focused on the case of zero interbilayer hopping ($t_{z}^{'}=0$), though the LRC is included. Here, we extend our study to investigate the effect of a finite $t_{z}^{'}$. We begin by assuming a moderate value of  $t_{z}^{'}=t_{z}/2=0.05t$ and later present results for a larger $t_{z}^{'}$ to capture the behavior observed for a large $t_{z}$ in the previous subsection. 

\begin{figure}[t]
\centering
\includegraphics[width=8cm]{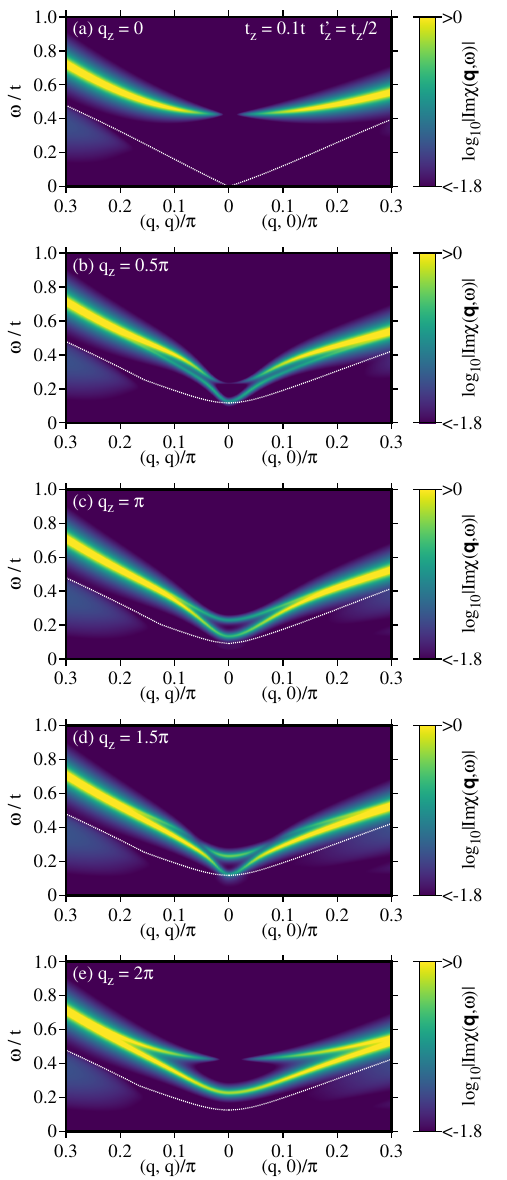}
\caption{Intensity maps of $\log_{10} | {\rm Im}\chi(\vq, \omega)|$ for a sequence of $q_{z}$ around a region of $\vq_{\parallel}=(0,0)$ for $t_{z}=0.1t$ and $t_{z}^{'}=t_{z}/2$: (a) $q_{z}=0$, (b) $q_{z}=0.5\pi$, (c) $q_{z}=\pi$, (d) $q_{z}=1.5\pi$, and (e) $q_{z}=2\pi$. The white dotted curve denotes the upper boundary of the particle-hole continuum. There is no continuum spectrum at $\vq_{\parallel}=(0,0)$ and $q_{z}=0$. 
}
\label{tz0.1tzp0.05}
\end{figure}

We present results in a manner consistent with  Figs.~\ref{tz0.1}(a)--(e), as shown in Figs.~\ref{tz0.1tzp0.05}(a)--(e). At  $q_{z}=0$ [\fig{tz0.1tzp0.05}(a)], the spectrum is nearly identical to the $t_{z}^{'}=0$ case in \fig{tz0.1}(a), with only the $\omega_{+}$ mode present. This indicates that the effect of $t_{z}^{'}$ is negligible at $q_{z}=0$. A similar minor effect is seen at  $q_{z}=2\pi$ [\fig{tz0.1tzp0.05}(e)], where the $\omega_{-}$ mode exists and is shifted to a slightly higher energy due to the inclusion of $t_{z}^{'}$. The most significant change induced by a finite $t_{z}^{'}$ is observed at $q_{z} \ne 2n\pi$, where the $\omega_{-}$ mode, which was gapless for $t_{z}^{'}=0$, now becomes gapped at $\qp=(0,0)$. Consequently,  for any $q_{z} \ne 0$, we have two gapped plasmon modes in the presence of $t_{z}^{'}$.

\begin{figure}[t]
\centering
\includegraphics[width=15cm]{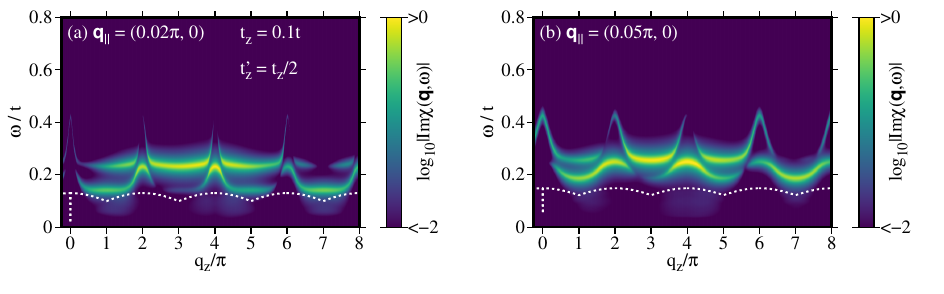}
\caption{$q_{z}$ dependence of $\omega_{+}$ mode (higher energy) and $\omega_{-}$ mode (lower energy) at (a) $\vq_{\parallel}=(0.02\pi, 0)$ and (b) $\vq_{\parallel}=(0.05\pi, 0)$ for $t_{z}=0.1t$ and $t_{z}^{'}=t_{z}/2$.  The white dotted curve denotes the upper boundary of the particle-hole continuum. There is a sharp drop at $q_{z}=0$ because of the disappearance of the $\omega_{-}$ mode there. 
}
\label{tz0.1tzp0.05-qz}
\end{figure}

The $q_{z}$ dependence of the $\omega_{\pm}$ modes is presented in   Figs.~\ref{tz0.1tzp0.05-qz}(a) and (b) for $\vq_{\parallel}=(0.02\pi, 0)$ and $(0.05\pi, 0)$, respectively. A comparison with Figs.~\ref{tz0.1-qz}(a) and (b) shows that the overall behavior is preserved, with the primary difference being a shift of the $\omega_{-}$ mode to higher energy due to the finite $t_{z}^{'}$. While the spectral intensity does not exhibit $2\pi$ periodicity, the location of the ``nodes'', where the intensity is strongly suppressed, remain largely unaffected by the inclusion of $t_{z}^{'}$. This suggests that the effect of $t_{z}^{'}$ is relatively weak on the intensity dependence of the plasmon modes. 

\begin{figure}[t]
\centering
\includegraphics[width=8cm]{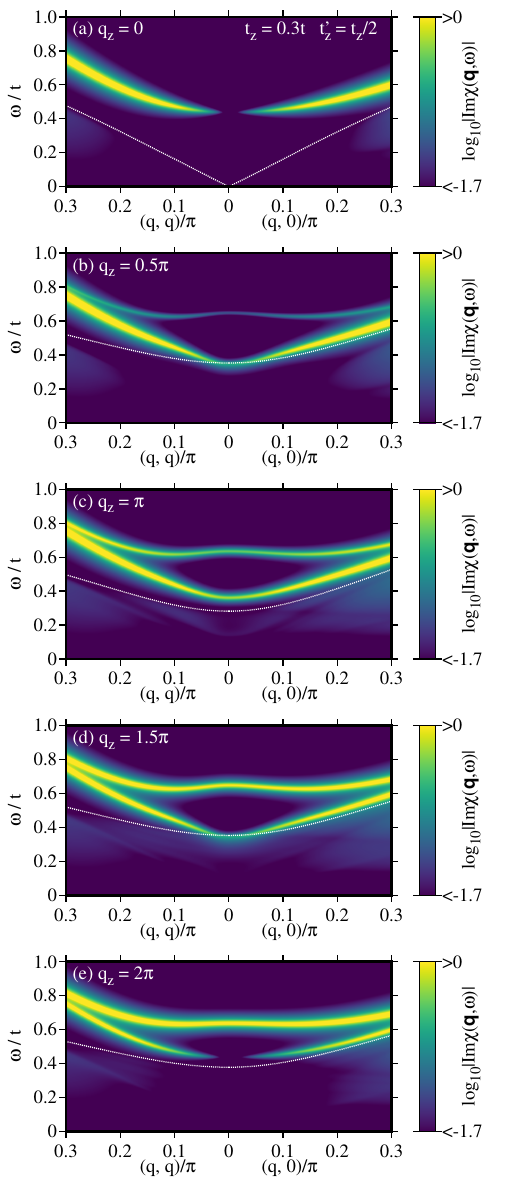}
\caption{Intensity maps of $\log_{10} | {\rm Im}\chi(\vq, \omega)|$ in the presence of a large $t_{z}=0.3t$ and $t_{z}^{'}=t_{z}/2$. (a)--(e) $\vq_{\parallel}$ dependence for a sequence of $q_{z}$ around a region of $\vq_{\parallel}=(0,0)$: (a) $q_{z}=0$, (b) $q_{z}=0.5\pi$, (c) $q_{z}=\pi$, (d) $q_{z}=1.5\pi$, and (e) $q_{z}=2\pi$. The white dotted curve denotes the upper boundary of the particle-hole continuum---there is no continuum spectrum at $\vq_{\parallel}=(0,0)$ and $q_{z}=0$ in (a). 
}
\label{tz0.3tzp0.15}
\end{figure}

For completeness, we also study the case of a large hopping $t_{z}=0.3t$ with $t_{z}^{'}=0.15t$. As we have previously established, the energy hierarchy of the $\omega_{\pm}$ modes is interchanged for a large $t_{z}$, and this behavior persists in the presence of $t_{z}^{'}$. 

$\vq$-$\omega$ maps for this case are shown in Figs.~\ref{tz0.3tzp0.15}(a)--(e). At $q_{z}=0$  [\fig{tz0.3tzp0.15}(a)], only the $\omega_{+}$ mode is present, with its spectral weight vanishing at $\vq_{\parallel}=(0,0)$ as a consequence of charge conservation. For $q_{z}\ne 0$, the particle-hole continuum gains spectral weight even at $\qp=(0,0)$, and the $\omega_{+}$ mode is realized close to this upper boundary around $\qp=(0,0)$. In contrast, the $\omega_{-}$ mode has a higher energy and is located above the continuum. As seen in \fig{tz0.3tzp0.15}(b), the $\omega_{-}$ mode has a relatively low-spectral weight near $q_{z}=0$ and forms an upward-convex shape centered at $\qp=(0,0)$ and $\omega \approx 0.65t$. This mode gains more spectral weight as $q_{z}$ increases as shown in  Figs.~\ref{tz0.3tzp0.15}(c)--(e), and its dispersion shows a small dependence on $q_{z}$. The low-energy $\omega_{+}$ mode enters slightly the continuum around $\vq=(0,0)$ in Figs.~\ref{tz0.3tzp0.15}(b) and (d). Its presence, despite being within the continuum, is due to the  low-spectral weight of the continuum around $\qp=(0,0)$. At $q_{z}=2 \pi$ [\fig{tz0.3tzp0.15}(e)], the $\omega_{+}$ mode is pushed up slightly above the continuum and its spectral intensity at $\qp=(0,0)$ vanishes.

\begin{figure}[t]
\centering
\includegraphics[width=15cm]{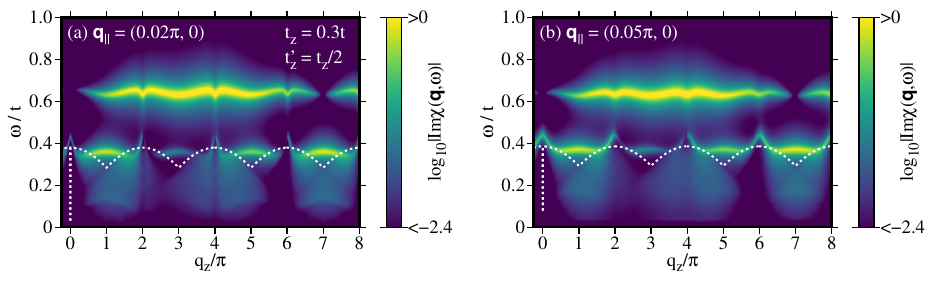}
\caption{$q_{z}$ dependence of $\omega_{+}$ mode (lower energy) and $\omega_{-}$ mode (higher energy)  at (a) $\vq_{\parallel}=(0.02\pi, 0)$ and (b) $\vq_{\parallel}=(0.05\pi, 0)$ for $t_{z}=0.3t$ and $t_{z}^{'}=t_{z}/2$. The white dotted curve denotes the upper boundary of the particle-hole continuum. There is a sharp drop at $q_{z}=0$ because of the disappearance of the $\omega_{-}$ mode there. 
}
\label{tz0.3tzp0.15-qz}
\end{figure}

Finally, we examine the $q_{z}$ dependence of the spectral intensity at fixed in-plane momenta in Figs.~\ref{tz0.3tzp0.15-qz}(a) and (b) for  $\vq_{\parallel} = (0.02\pi, 0)$ and $(0.05\pi, 0)$, respectively. The $\omega_{+}$ mode is well defined when it is located above the continuum, and is somewhat blurred within the continuum, but a sharp peak is discernible at $q_{z}=2n\pi$,  particularly in \fig{tz0.3tzp0.15-qz}(b). The $\omega_{-}$ mode, which vanishes at $q_{z}=0$, exhibits a less dispersive feature along the $q_{z}$ direction, with a small dip at  $q_{z}=2 n \pi$ $(n \ne 0)$. The strong suppression of the $\omega_{-}$ mode around $q_{z}=7\pi$ is a feature also observed in the previous subsection.

\subsection{\boldmath{$V_{c}$} dependence of dispersive modes} \label{sec:zs}

\begin{figure}[t]
\centering
\includegraphics[width=7cm]{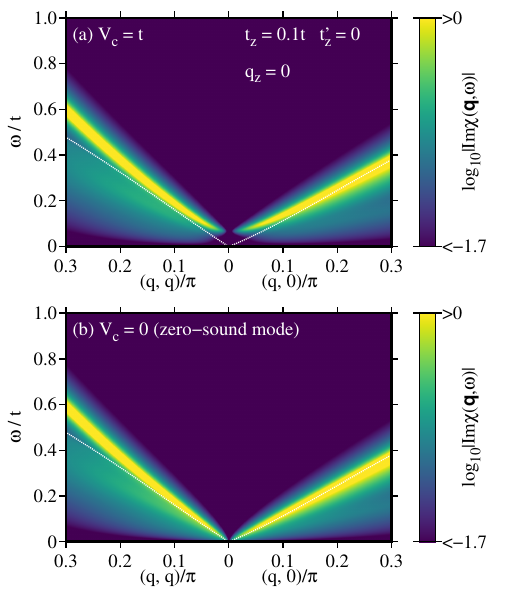}
\caption{Forming the zero-sound mode in the limit of $V_{c}\rightarrow 0$: (a) $V_{c}=t$ and (b) $V_{c}=0$. It is the $\omega_{+}$ mode that changes into the zero-sound mode. 
}
\label{zero-sound}
\end{figure}

\begin{figure}[ht]
\centering
\includegraphics[width=7cm]{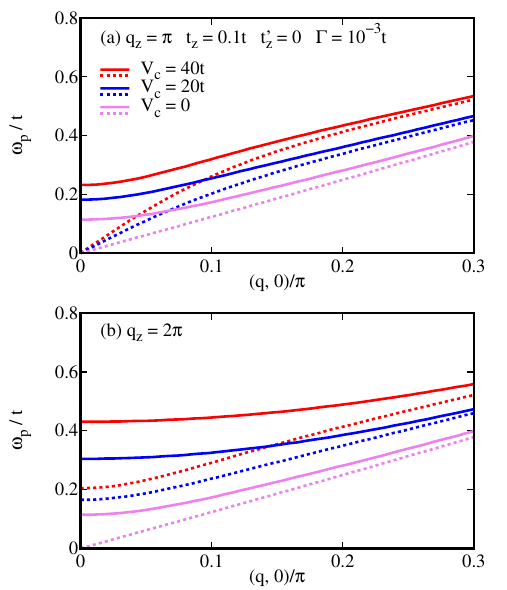}
\caption{Dispersive modes along the (0,0)-($0.3\pi$, 0) direction for a sequence of values of $V_{c}$. 
We here employed a smaller broadening $\Gamma=10^{-3}t$ to trace the dispersion precisely. 
}
\label{Vc-dep-dispersion}
\end{figure}

The $t$-$J$-$V$ model contains a short-range interaction of the $J$-term as well as from the local constraint, and thus we can study the interplay with the long-range interaction $V$. 

In the case of $q_{z}=0$, we have only the $\omega_{+}$ mode with a gap [\fig{tz0.1}(a)]. This mode eventually becomes a gapless mode in the limit of $V_{c}\rightarrow 0$, forming the zero-sound mode, as shown in \fig{zero-sound}, although the mode along the $q_{x}$ direction is realized very close to the upper boundary of the continuum in \fig{zero-sound}(b).  

At $q_{z}=\pi$, both $\omega_{+}$ and $\omega_{-}$ modes are present, with the $\omega_{-}$ mode being gapless as shown in \fig{Vc-dep-dispersion}(a).  As the interaction strength $V_{c}$ is reduced,  the energy of the $\omega_{+}$ mode decreases, but it remains a gapped mode in the limit of $V_{c} \rightarrow 0$. In contrast, the $\omega_{-}$ mode retains its gapless character, though the velocity is reduced as $V_{c}$ decreases. While it might appear that plasmon modes persist in the  $V_{c}=0$ limit, they are no longer plasmons. In particular, the $\omega_{-}$ mode is realized inside the continuum as a peak of the continuum. 
On the other hand, if we include $t_{z}^{'}$, both $\omega_{\pm}$ modes are gapped even 
in the limit of $V_{c} \rightarrow 0$.  

\begin{figure}[t]
\centering
\includegraphics[width=8cm]{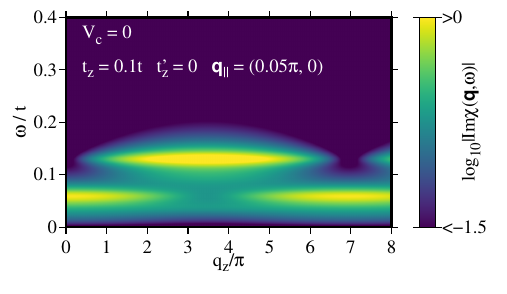}
\caption{$q_{z}$ dependence of spectral weight for $V_{c}=0$. While the zero-sound mode is well defined at $q_{z}=0$, it evolves smoothly into a lower-energy branch at $q_{z} \ne 0$. The higher-energy mode is absent at $q_{z}=0$, but is realized at a finite $q_{z}$. 
}
\label{Vc-qz}
\end{figure}

A unique feature is observed at $q_{z}=2 n \pi$. As previously shown in  Fig.~\ref{tz0.1}(e) both $\omega_{\pm}$ modes are gapped when $n \ne 0$.  As $V_{c}$ decreases [\fig{Vc-dep-dispersion}(b)], this gap is reduced. In contrast to the case of $q_{z}=0$ [\fig{zero-sound}(b)], the $\omega_{+}$ mode retains a gap in the limit of $V_{c}\rightarrow 0$. In this limit, it is the $\omega_{-}$ mode that changes into a gapless mode, namely evolves into the zero-sound-wavelike mode. This feature is the same even if there is  a finite $t_{z}'$.

Corresponding spectra showing the $q_{z}$ dependence at $\qp=(0.05\pi, 0)$ are presented in \fig{Vc-qz} in the limit of $V_{c}=0$, where there are two branches.
The analysis of this figure provides three important insights. First, while the lower-energy branch at $q_{z}=0$ is the zero-sound mode, it smoothly connects to a branch at a finite $q_{z}$. Second,  because the $V$-term is the primary source of three-dimensional coupling, the system effectively reduces to being two-dimensional in the $V_{c}=0$ limit, resulting in a very weak dependence on $q_{z}$. Third, interestingly, the spectral intensity still exhibits a strong $q_{z}$ dependence due to the kinetic hopping term along the $z$ direction. Therefore, depending on the value of $q_{z}$, the modes may be detected as only one branch, although in principle two branches exist. 

\section{Discussions}\label{sec:discussions}
Despite the fundamental difference between our large-$N$ theoretical framework and the RPA \cite{yamase25}, we have found that our results for charge dynamics are strikingly similar in many respects, with a few notable differences. 

We have used a parameter set closely matching that of Ref.~\cite{yamase25}, with the exception of the Coulomb potential $V_{c}$ and the anisotropic parameter $\alpha$, and  temperature. A key difference lies in our use of a significantly smaller value of $V_{c}$ (in units of $t$). This disparity reflects the strong correlation effects inherent in the $t$-$J$ model, which lead to a notable band narrowing effect. This strong correlation is a predominant factor, enabling the plasmon modes to exist across a wide range of momentum space as shown in \fig{overall}(a). This sharply contrasts with the RPA results [\fig{overall}(b)], where plasmon modes are typically confined to the vicinity of $\qp=(0,0)$ and becomes heavily damped into  the particle-hole continuum as they move further away.

\begin{figure}[b]
  \centering
  \includegraphics[width=8cm]{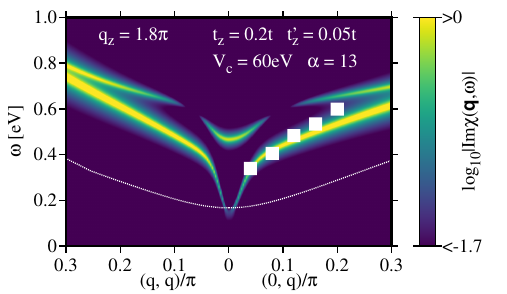}
\caption{Comparison with the plasmon energy (solid squares) reported in Y-based cuprate superconductors in Ref.~\cite{bejas24}. The experimental data are superimposed on the intensity map of $\log_{10} | {\rm Im}\chi(\vq, \omega)|$ computed at $q_{z}=1.8\pi$ by using $t_{z}=0.2t$, $t_{z}^{'}=0.05t$, $V_{c}=60$ eV, and $\alpha=13$ (corresponding to $\epsilon_{\parallel} = 1.68 \epsilon_{0}$ and $\epsilon_{\perp} = 1.16 \epsilon_{0}$), keeping the other parameters unchanged; $t/2$ is assumed to be 400 meV. The white dotted curve denotes the upper boundary of the particle-hole continuum. 
}
\label{fig:YBCO}
\end{figure}

So far there is only one experimental RIXS report \cite{bejas24} in bilayer cuprates where a single $q_{y}$-scan was presented. Similar to weak-coupling calculations \cite{yamase25,sellati25}, the present theory---a more adequate theory for cuprates---also suggests that the observed mode is the $\omega_{-}$ mode as shown in \fig{fig:YBCO}. Because of a finite $t_{z}^{'}$, the $\omega_{-}$ mode will be gapped at $\qp=(0,0)$. To reinforce this conclusion, we call for more comprehensive data such as $q_{x}$-,  $q_{y}$-, and  $q_{z}$-scans. 

In the bilayer \tj model two collective modes (zero-sound-wavelike modes) remain due to electron correlations. In Appendix \ref{app:zero-sound-data} we investigate whether these modes alone could account for the experimental dispersion. Our analysis indicates that this scenario is less consistent with the data, especially regarding the slope of the measured dispersion, and that the plasmon interpretation offers a more coherent description of the observed features.

\section{Conclusions}\label{sec.conclusions}
In this work, we have constructed a strong-coupling theory of bilayer plasmons by employing a large-$N$ formalism for the $t$-$J$-$V$ model. Our computational approach of charge excitation spectra was conducted in a matter that allows for a direct comparison with a recent RPA study \cite{yamase25}. Despite the fundamental differences in the theoretical framework, we have found a striking similarity in the plasmon dispersion and intensity maps. This agreement, however, is not without crucial distinctions. In our theory,  the strong correlation effects inherent in the $t$-$J$ model lead to a significant band narrowing, which in turn allows plasmon modes to remain well-defined across the entire Brillouin zone. This stands in sharp contrast to weak-coupling RPA calculations, where plasmons are typically heavily damped away from the zone center. 

A unique contribution of our model is the insight gained by systematically varying the Coulomb interaction $V_{c}$. We have shown that as $V_{c}$ is decreased, the plasmon modes change into two distinct modes. When $q_{z}=2n \pi$ $(n \ne 0)$, one of these modes remains gapped while the other becomes a gapless mode. When $q_{z}=0$, only the gapless mode---zero-sound mode---is  present. We have also found that while the spectral intensity of these modes shows a strong dependence on $q_{z}$, the mode energy itself is remarkably independent of $q_{z}$. 

The ability of our strong-coupling theory to reproduce plasmon modes  provides a comprehensive framework for interpreting experimental data. Our results may offer an explanation for the RIXS data recently obtained in Y-based cuprate superconductors. The ultimate test of the present strong-coupling theory will be the acquisition of more comprehensive RIXS data in bilayer cuprates. If our model provides a coherent explanation for these future results, it would offer compelling evidence that a strong-coupling approach is necessary for describing charge dynamics in these correlated systems.

\acknowledgments
The authors thank  M. Hepting and B. Keimer for valuable discussions about their data, and L. Benfatto,  F. Gonzalez, W. Metzner, and I. Pomponio for illuminating discussions.  H.Y. was supported by World Premier International  Research Center Initiative (WPI), MEXT, Japan. M.B. and A.G. are indebted to warm hospitality of Max-Planck-Institute for Solid State Research. M.B. also thanks MANA Short-Term Invitation Program and warm hospitality in NIMS. A part of the results presented in this work was obtained by using the facilities of the CCT-Rosario Computational Center, member of the High Performance Computing National System (SNCAD, MincyT- Argentina). 

\appendix

\section{Large-\boldmath{$N$} formalism of the bilayer \boldmath{$t$-$J$-$V$} model}\label{app:self-energy}

A major challenge in handling the $t$-$J$ model arises from the non-double-occupancy constraint. To rigorously enforce the local constraint [\eq{constraint}], we express the Hamiltonian in terms of the Hubbard operators \xop  \cite{hubbard63}. The constraint is then implicitly described by the algebra of these operators: $\tilde{c}^\dagger_{i\sigma,\alpha}=\hat{X}_{i\alpha}^{\sigma 0}$, $\tilde{c}_{i\sigma,\alpha}=\hat{X}_{i\alpha}^{0 \sigma }$, $S_{i,\alpha}^+=\hat{X}_{i\alpha}^{\uparrow\downarrow}$, $S_{i,\alpha}^-=\hat{X}_{i\alpha}^{\downarrow\uparrow}$, $n_{i,\alpha}=\hat{X}_{i\alpha}^{\uparrow\uparrow}+\hat{X}_{i\alpha}^{\downarrow\downarrow}$, and $\hat{X}_{i\alpha}^{00}$ describes the number of doped holes. The $z$ component of the spin operator is $S_{i,\alpha}^z=\frac{1}{2}(\hat{X}_{i\alpha}^{\uparrow\uparrow}-\hat{X}_{i\alpha}^{\downarrow\downarrow}$). The operators $\hat{X}_{i\alpha}^{\sigma 0}$ and $\hat{X}_{i\alpha}^{0\sigma}$ are called fermionlike, whereas the operators $\hat{X}_{i\alpha}^{\sigma\sigma'}$ and $\hat{X}_{i\alpha}^{00}$ are bosonlike.

The Hamiltonian in Eq.~\eqref{eq:model} can be expressed in terms of the Hubbard operators as: 
\begin{align}
	H=&-\sum_{i,j,\sigma,\alpha,\beta} t_{ij}^{\alpha\beta}X_{i\alpha}^{\sigma 0}X_{j\beta}^{0\sigma}
	-\mu\sum_{i,\sigma,\alpha}X_{i\alpha}^{\sigma\sigma}     +\frac{J}{2}\sum_{\bra i,j \ket,\sigma,\sigma',\alpha}\left(X_{i\alpha}^{\sigma\sigma'}X_{j\alpha}^{\sigma'\sigma}
	-X_{i\alpha}^{\sigma\sigma}X_{j\alpha}^{\sigma'\sigma'}\right) \nonumber\\       
	&+\frac{J_{\perp}}{4}\sum_{i,\sigma,\sigma',\alpha \ne \beta}\left(X_{i\alpha}^{\sigma\sigma'}X_{i\beta}^{\sigma'\sigma}
	-X_{i\alpha}^{\sigma\sigma}X_{i\beta}^{\sigma'\sigma'}\right)
	+\frac{1}{2}\sum_{i \ne j,\sigma,\sigma',\alpha,\beta}V_{ij}^{\alpha\beta}X_{i\alpha}^{\sigma\sigma}X_{j\beta}^{\sigma'\sigma'} \; .
	\label{eq:model-X}
\end{align}
The formalism starts with the construction of a first-order classical Lagrangian using the Faddeev-Jackiw and Dirac methods \cite{faddeev88,sundermeyer82,govaerts90}. In this representation, the fermionlike (bosonlike) Hubbard operators are associated with Grassmann (usual bosonic) variables. Next, a \lN expansion is applied to the spin projection, extending it from $\sigma=\uparrow,\downarrow$ to $p=1,...,N$ and rescaling the amplitude as $t_{ij}^{\alpha\beta}/N$, $J/N$, $J_\perp/N$, and $V_{ij}^{\alpha\beta}/N$ to ensure a finite theory in the limit of $N \rightarrow \infty$. Using the condition 
\begin{equation}
	X_{i\alpha}^{pp'}=\frac{X_{i\alpha}^{p 0}X_{i\alpha}^{0p'}}{X_{i\alpha}^{00}} \; 
\end{equation}
for $J$- and $J_{\perp}$-terms, we write the fermionlike fields as 
\begin{align}
	f^\dagger_{ip,\alpha}&=\frac{1}{\sqrt{N \delta /2}}X_{i\alpha}^{p0} \;,\\
	f_{ip,\alpha}&=\frac{1}{\sqrt{N \delta /2}}X_{i\alpha}^{0p} \;,
\end{align}
where $\delta$ is the hole doping away from half-filling. 
$X_{i\alpha}^{\sigma \sigma}$ in $J$-, $J_{\perp}$- and $V_{ij}^{\alpha \beta}$-terms are,  
on the other hand, treated by utilizing the local constraint $X_{i \alpha}^{00} + \sum_{p}X_{i \alpha}^{pp}=N/2$ and this constraint is imposed by introducing the Lagrange multiplier $\lambda_{i \alpha}$. 

The fields \( X_{i\alpha}^{00} \) and \( \lambda_{i\alpha} \) are expressed in terms of their static mean-field components and dynamic fluctuations: 
\begin{align}
	X_{i\alpha}^{00} &= N\frac{\delta}{2}\left(1 + \delta R_{i\alpha}\right),\\
	\lambda_{i\alpha} &= \lambda_0 + \delta \lambda_{i\alpha} \;,
\end{align}
where $\delta R_{i \alpha}$ denotes the fluctuation of the hole density at site $i$ in layer $\alpha$; 
$\delta \lambda_{i \alpha}$ is the fluctuation of the Lagrange multiplier to enforce the constraint against double occupancy.

The resulting effective Lagrangian includes two distinct four-fermion interaction terms, one from the in-plane exchange interaction $J$, and the other from the out-of-plane interaction \( J_\perp \). To decouple these terms, we introduce  Hubbard-Stratonovich fields $\Delta_{ij,\alpha}$ and $\Delta_i'$: 
\begin{align}
	\Delta_{ij,\alpha} &= \frac{J}{2} \sum_p \frac{f^\dagger_{jp,\alpha} f_{ip,\alpha}}{\sqrt{(1 + \delta R_{i\alpha})(1 + \delta R_{j\alpha})}} \;,\\
	\Delta_i' &= \frac{J_\perp}{4} \sum_p \frac{f^\dagger_{ip,1} f_{ip,2}}{\sqrt{(1 + \delta R_{i1})(1 + \delta R_{i2})}} \;.
\end{align}
The fields $\Delta_{ij,\alpha}$ and $\Delta_i'$ describe bond-charge fluctuations in the intralayer and intrabilayer, respectively. Since $i$ and $j$ are nearest-neighbor sites on the square lattice, we may write $\Delta_{i j, \alpha}= \Delta_{i \alpha}^{\eta}$ where $\eta=x$ or $y$. 
We  parametrize those fields as: 
\begin{align}
	\Delta_{i\alpha}^\eta &= \chi \left(1 + r_{i\alpha}^\eta + \ii A_{i\alpha}^\eta\right) \;,\\
	\Delta_i' &= \chi' \left(1 + r_{\perp,i} + \ii A_{\perp,i}\right) \;,
\end{align}
where $r_{i\alpha}^\eta$ and $A_{i\alpha}^\eta$ ($r_{\perp,i}$ and $A_{\perp,i}$) represent the real and imaginary parts of the in-plane (out-of-plane) bond-field fluctuations, respectively, and $\chi$ ($\chi'$) is the corresponding static mean-field value.

Finally, the terms involving $1/\sqrt{1 + \delta R_{i\alpha}}$ are expanded perturbatively in powers of $\delta R_{i\alpha}$. This expansion systematically organizes the interactions in powers of $1/N$, thus controlling the hierarchy of contributions in the \lN formalism. The effective theory of Eq.~\eqref{eq:model} is then described in terms of fermions, bosons, and their interactions.

In the large-$N$ formalism for bilayer systems, we introduce a $14$-component bosonic field as a basis: 
\begin{equation}
	\delta X^{a} = (\delta R_1,\;\delta{\lambda_1},\;
	\delta R_2,\;\delta{\lambda_2},\;
	r_1^{x},\;r_1^{y},\; A_1^{x},\;A_1^{y},\;
	r_2^{x},\;r_2^{y},\; A_2^{x},\;A_2^{y},\;
	r_\perp,\; A_\perp)\, ,
	\label{eq:boson-field}
\end{equation}
where the site index is omitted for clarity.

Following Refs.~\cite{foussats04,yamase21a,bejas25}, the Feynman rules applied to the effective theory give a $14 \times 14$ bare bosonic propagator $D^{(0)}_{ab}({\bf q},\mathrm{i}\omega_{n})$ that is $O(1/N)$: 
\begin{align}
	\left[D^{(0)}_{ab}(\vq,\ii\omega_{n})\right]^{-1} &= 
	N \left(
	\begin{array}{ccccc}
		D^{(0)}_A & D^{(0)}_B & 0     & 0     & 0 \\
		D^{(0) *}_B & D^{(0)}_A & 0     & 0     & 0 \\
		0 &      0 & D^{(0)}_C & 0     & 0 \\
		0 &      0 & 0     & D^{(0)}_C & 0 \\
		0 &      0 & 0     & 0     & D^{(0)}_D
	\end{array}
	\right) \; ,
	\label{eq:D0}
\end{align}
\noindent where $\omega_n$ is a bosonic Matsubara frequency and the matrices $D^{(0)}_{A\text{-}D}$ are: 
\begin{align}
	D^{(0)}_A &= \left(
	\begin{array}{cc}
		\frac{\delta^2}{2} \left[\frac{V(\vq)}{2}-J(\vq)\right] & \frac{\delta}{2} \\
		\frac{\delta}{2} & 0
	\end{array}
	\right)  \; ,
\end{align}
\begin{align}
	D^{(0)}_B &= \left(
	\begin{array}{cc}
		\frac{\delta^2}{2} \left[\frac{V'(\vq)}{2}- J'(\vq)\right] & 0 \\
		0 & 0
	\end{array}
	\right)  \; ,
\end{align}
\begin{align}
	D^{(0)}_C &= \left(
	\begin{array}{cccc}
		\frac{4\chi^2}{J} & 0 & 0 & 0 \\
		0 & \frac{4\chi^2}{J} & 0 & 0 \\
		0 & 0 & \frac{4\chi^2}{J} & 0 \\
		0 & 0 & 0 & \frac{4\chi^2}{J}
	\end{array}
	\right)  \; ,
\end{align}
\begin{align}
	D^{(0)}_D &= \left(
	\begin{array}{cc}
		\frac{4\chi'^2}{J_{\perp}} & 0 \\
		0 & \frac{4\chi'^2}{J_{\perp}}
	\end{array}
	\right) \; .
\end{align}

The \lN formalism yields the bare bosonic propagator as a $14\times14$ matrix [see Eq.~\eqref{eq:D0}]. However, because we focus on on-site charge excitations, including plasmons \cite{bejas17}, we restrict our analysis to the corresponding $4\times4$ submatrix of the bare bosonic propagator given by
\begin{align}
	\left[D^{(0)}_{ab}(\vq,\ii\omega_{n})\right]^{-1} &= N \left(
	\begin{array}{cccc}
		\frac{\delta^2}{2} \left[\frac{V(\vq)}{2}- J(\vq)\right] & \frac{\delta}{2} & \frac{\delta^2}{2} \left[\frac{V'(\vq)}{2}- J'(\vq)\right] & 0 \\
		\frac{\delta}{2} & 0 & 0 & 0 \\
		\frac{\delta^2}{2} \left[\frac{V'^*(\vq)}{2}- J'^{*}(\vq)\right] & 0 & \frac{\delta^2}{2} \left[\frac{V(\vq)}{2}- J(\vq)\right] & \frac{\delta}{2} \\
		0 & 0 &  \frac{\delta}{2} & 0
	\end{array}
	\right) \; ,
\end{align}
where the matrix elements are limited to $a,b=\delta R_{a}, \delta \lambda_{a}$. 

\begin{figure}[t]
\centering
\includegraphics[width=8.1cm]{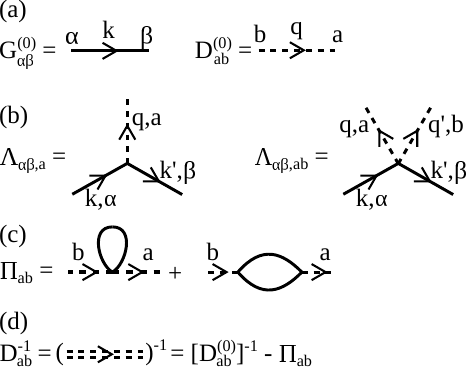}
\caption{(a) Bare fermionic $G^{(0)}_{\alpha \beta}$ and bosonic $D^{(0)}_{ab}$ propagators, solid and dashed lines, respectively. (b) $\Lambda_{\alpha\beta,a}$ and $\Lambda_{\alpha\beta,ab}$ are the three- and four-legs vertices, respectively.  (c) $\Pi_{ab}$ is the bosonic self-energy. (d) Double dashed line is the dressed bosonic propagator $D_{ab}$. 
}
\label{fig:dyson}
\end{figure}

Interactions between bosons and fermions are governed by three- and four-leg vertices. The three-leg vertex $\Lambda_{\alpha\beta,a}$ represents the interaction of a fermion from layer $\alpha$ that ends at  layer $\beta$ after interacting with a boson $\delta X^a$. We can write this vertex as
\begin{equation}\label{eq:threeleg0}
 \Lambda_{\alpha\beta,a}(k,k',q) =\tilde{\Lambda}_{\alpha\beta,a}(k,k',q)\mathrm{e}^{\ii (-k_z d_\alpha+k_z'd_\beta+q_zd_a)}\delta(k-k'-q) \;
\end{equation}
where $k\equiv \vk, \ii\nu_n$, $q\equiv\vq,\ii\omega_n$ and $d_a$ is equal to $d_1$ or $d_2$ depending on the plane of the boson $\delta X^a$, with $d_1=0$ and $d_2=d/c$. After executing $\delta(k-k'-q)$, the nonzero components of $\tilde{\Lambda}_{\alpha\beta,a}(k,k',q)$ in the charge sector become 

\begin{equation}\label{eq:threeleg1}
 \tilde{\Lambda}_{\alpha\alpha,a}(k,q) =- \left[
 \ii\nu_n-\frac{\ii\omega_n}{2} + \mu + 2\chi \sum_{\eta=x,y} \cos\left(k_\eta-\frac{q_\eta}{2}\right) \cos \frac{q_\eta}{2},\; 1\;
 \right]
\end{equation}
\noindent for each component $a = \delta R_\alpha$, $\delta \lambda_\alpha$. For $\alpha \ne \beta$ 
\begin{align}\label{eq:threeleg2}
 \tilde{\Lambda}_{\alpha\beta,a}(k,q) =& - \left(  \frac{\chi'}{2}, \frac{\chi'}{2} \right) 
\end{align}
\noindent for each component $a = \delta R_1$, $\delta R_2$. 

The four-leg vertex $\Lambda_{\alpha\beta,ab}$ represents a fermion from layer $\alpha$ that ends in layer $\beta$ after interacting with the bosons $\delta X^a$ and $\delta X^b$. We write this vertex as
\begin{equation}\label{eq:fourleg0}
 \Lambda_{\alpha\beta,ab}(k,k',q,q') =\tilde{\Lambda}_{\alpha\beta,ab}(k,k',q,q')\mathrm{e}^{\ii (-k_z d_\alpha+k_z'd_\beta+q_zd_a+q_z'd_b)}\delta(k-k'-q-q') \; , 
\end{equation}
where, after executing $\delta(k-k'-q-q')$, the nonzero components of $\tilde{\Lambda}_{\alpha\beta,ab}(k,k',q,q')$ in the charge sector become
\begin{align}\label{eq:fourleg}
 \tilde{\Lambda}_{\alpha\alpha,ab}(k,q,q') =& \left[
 \begin{array}{cc}
  F_{k,q,q'} & 1/2 \\
  1/2      & 0
 \end{array}
 \right]
\end{align}
\noindent for each component $a,b = \delta R_\alpha$, $\delta \lambda_\alpha$, where
\begin{align}
 F_{k,q,q'} &= \ii\nu_n-\frac{\ii\omega_n+\ii\omega'_n}{2} + \mu + \chi \sum_{\eta=x,y}
 \cos\left(k_\eta - \frac{q_\eta+q'_\eta}{2} \right)\nonumber \\
 &\times \left[
 \cos\left(\frac{q_\eta+q'_\eta}{2} \right) +
  \cos \frac{q_\eta}{2}  \cos \frac{q'_\eta}{2}
 \right] \; .
\end{align}
For $\alpha \ne \beta$ 
\begin{align}\label{eq:lam12ab}
 \tilde{\Lambda}_{\alpha\beta,ab} = \frac{\chi'}{8}\left[
 \begin{array}{cc}
  3 & 1  \\
  1 & 3  \\
 \end{array}
 \right]
\end{align}
\noindent for each component $a,b = \delta R_1$, $\delta R_2$. Note that the vertices are $O(1)$.

Using the propagators [\eq{eq:G0} and Fig.~\ref{fig:dyson}(a)] and vertices [Fig.~\ref{fig:dyson}(b)] the bosonic self-energy $\Pi_{ab}({\bf q}, \ii\omega_n)$ [Fig.~\ref{fig:dyson}(c)] is computed, considering both Hartree and bubble diagrams. The Dyson equation [Fig.~\ref{fig:dyson}(d)] yields the dressed bosonic propagator $D_{ab}({\bf q}, \ii\omega_n)$ as 
\begin{align}
 D^{-1}_{ab}({\bf q}, \ii\omega_n) &=\left[ D^{(0)}_{ab}({\bf q}, \ii\omega_n) \right]^{-1} - \Pi_{ab}({\bf q}, \ii\omega_n) \, .
 \label{eq:Dyson2}
\end{align}

Focusing on the $4\times4$ charge sector of the self-energy, the analytical expressions of each component are:   
 \begin{align}\label{eq:self-energy}
 \Pi_{11}(\vq, \ii\omega_n) =& -\frac{N}{16N_s}\sum_{\alpha,\beta=1}^{2} \sum_{\vk}  \left[n_F\left(\varepsilon_{\vk-\vq}^{\alpha}\right) - n_F\left(\varepsilon_{\vk}^{\beta}\right)\right]\left(\tilde{\varepsilon}_{\vk}^{\beta} - \tilde{\varepsilon}_{\vk-\vq}^{\alpha} \right) + \left(\tilde{\varepsilon}_{\vk}^{\beta} + \tilde{\varepsilon}_{\vk-\vq}^{\alpha}\right)^2 g^{\alpha\beta} \;,  \\
 \Pi_{12}(\vq, \ii\omega_n) =& -\frac{N}{8N_s} \sum_{\alpha,\beta=1}^{2} \sum_{\vk} \left(\tilde{\varepsilon}_{\vk}^{\beta} + \tilde{\varepsilon}_{\vk-\vq}^{\alpha}\right)  g^{\alpha\beta} \;, \\ 
 \Pi_{13}(\vq, \ii\omega_n) =& -\mathrm{e}^{-\ii q_z \frac{d}{c}}\frac{N}{16N_s}\sum_{\alpha,\beta=1}^{2} \left(-1\right)^{\alpha+\beta} \sum_{\vk}\frac{\varepsilon_{\vk-\vq}^{\perp*} \varepsilon_{\vk}^\perp}{|\varepsilon_{\vk-\vq}^\perp||\varepsilon_{\vk}^\perp|}  \left\{\left[n_F\left(\varepsilon_{\vk-\vq}^{\alpha}\right) - n_F\left(\varepsilon_{\vk}^{\beta}\right)\right]\left(\tilde{\varepsilon}_{\vk}^{\beta} - \tilde{\varepsilon}_{\vk-\vq}^{\alpha} \right)\right. \nonumber \\
 &+ \left.\left(\tilde{\varepsilon}_{\vk}^{\beta} + \tilde{\varepsilon}_{\vk-\vq}^{\alpha}\right)^2 g^{\alpha\beta}\right\} \;, \\
  \Pi_{31}(\vq, \ii\omega_n) =& -\mathrm{e}^{\ii q_z \frac{d}{c}}\frac{N}{16N_s}\sum_{\alpha,\beta=1}^{2} \left(-1\right)^{\alpha+\beta} \sum_{\vk}\frac{\varepsilon_{\vk-\vq}^{\perp} \varepsilon_{\vk}^{\perp*}}{|\varepsilon_{\vk-\vq}^\perp||\varepsilon_{\vk}^\perp|}  \left\{\left[n_F\left(\varepsilon_{\vk-\vq}^{\alpha}\right) - n_F\left(\varepsilon_{\vk}^{\beta}\right)\right]\left(\tilde{\varepsilon}_{\vk}^{\beta} - \tilde{\varepsilon}_{\vk-\vq}^{\alpha} \right)\right. \nonumber \\
 &+ \left.\left(\tilde{\varepsilon}_{\vk}^{\beta} + \tilde{\varepsilon}_{\vk-\vq}^{\alpha}\right)^2 g^{\alpha\beta}\right\} \;, \\ 
 \Pi_{14}(\vq, \ii\omega_n) =& -\mathrm{e}^{-\ii q_z \frac{d}{c}}\frac{N}{8N_s}\sum_{\alpha,\beta=1}^{2}\left(-1\right)^{\alpha+\beta} \sum_{\vk}\frac{\varepsilon_{\vk-\vq}^{\perp*} \varepsilon_{\vk}^\perp}{|\varepsilon_{\vk-\vq}^\perp||\varepsilon_{\vk}^\perp|} \left(\tilde{\varepsilon}_{\vk}^{\beta} + \tilde{\varepsilon}_{\vk-\vq}^{\alpha}\right)  g^{\alpha\beta} \\
  \Pi_{41}(\vq, \ii\omega_n) =& -\mathrm{e}^{\ii q_z \frac{d}{c}}\frac{N}{8N_s}\sum_{\alpha,\beta=1}^{2}\left(-1\right)^{\alpha+\beta} \sum_{\vk}\frac{\varepsilon_{\vk-\vq}^{\perp} \varepsilon_{\vk}^{\perp*}}{|\varepsilon_{\vk-\vq}^\perp||\varepsilon_{\vk}^\perp|} \left(\tilde{\varepsilon}_{\vk}^{\beta} + \tilde{\varepsilon}_{\vk-\vq}^{\alpha}\right)  g^{\alpha\beta} \\
 \Pi_{22}(\vq, \ii\omega_n) =& -\frac{N}{4N_s}\sum_{\alpha,\beta=1}^{2}\sum_{\vk}  g^{\alpha\beta} \\
 \Pi_{24}(\vq, \ii\omega_n) =& -\mathrm{e}^{-\ii q_z \frac{d}{c}}\frac{N}{4N_s}\sum_{\alpha,\beta=1}^{2} \left(-1\right)^{\alpha+\beta} \sum_{\vk} \frac{\varepsilon_{\vk-\vq}^{\perp*} \varepsilon_{\vk}^\perp}{|\varepsilon_{\vk-\vq}^\perp||\varepsilon_{\vk}^\perp|}     g^{\alpha\beta} \;, \\
  \Pi_{42}(\vq, \ii\omega_n) =& -\mathrm{e}^{\ii q_z \frac{d}{c}}\frac{N}{4N_s}\sum_{\alpha,\beta=1}^{2} \left(-1\right)^{\alpha+\beta} \sum_{\vk} \frac{\varepsilon_{\vk-\vq}^{\perp} \varepsilon_{\vk}^{\perp*}}{|\varepsilon_{\vk-\vq}^\perp||\varepsilon_{\vk}^\perp|}     g^{\alpha\beta} \;, \\
 \Pi_{21}(\vq, \ii\omega_n)=& \Pi_{34}(\vq, \ii\omega_n)= \Pi_{43}(\vq, \ii\omega_n)=\Pi_{12}(\vq, \ii\omega_n) \;, \\
 \Pi_{23}(\vq, \ii\omega_n)=&\Pi_{14}(\vq, \ii\omega_n) \;, \\
 \Pi_{32}(\vq, \ii\omega_n)=&\Pi_{41}(\vq, \ii\omega_n) \;, \\
 \Pi_{33}(\vq, \ii\omega_n)=&\Pi_{11}(\vq, \ii\omega_n)\;, \\
 \Pi_{44}(\vq, \ii\omega_n)=&\Pi_{22}(\vq, \ii\omega_n) \;, 
 \end{align}
\noindent where $\tilde{\varepsilon}_{\vk}^\alpha$ is equal to $\varepsilon_{\vk}^\alpha$ with $\mu=\chi=\chi'=0$, and 
\begin{equation}
    g^{\alpha\beta}= \frac{n_F\left(\varepsilon_{\vk-\vq}^{\alpha}\right) - n_F\left(\varepsilon_{\vk}^{\beta}\right)}{\ii\omega_n + \varepsilon_{\vk-\vq}^{\alpha} - \varepsilon_{\vk}^{\beta}} \; .
\end{equation}

\section{Analysis of experimental data with \boldmath{$V_c=0$}}\label{app:zero-sound-data}
\begin{figure}[b]
	\centering
	\includegraphics[width=15cm]{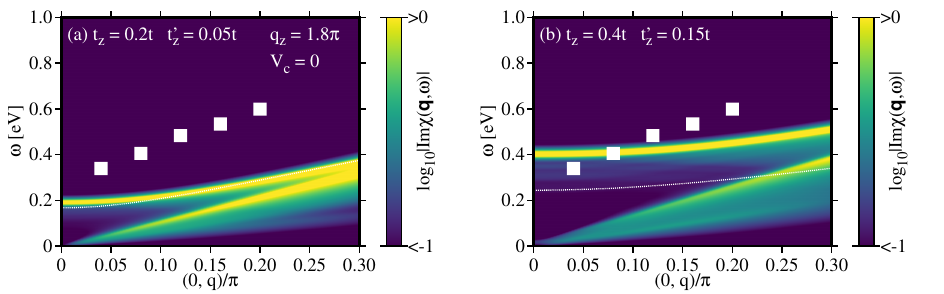}
	\caption{Comparison with the plasmon energy (solid squares) reported in Y-based cuprate superconductors in Ref.~\cite{bejas24}. The experimental data are superimposed on the intensity map of $\log_{10} | {\rm Im}\chi(\vq, \omega)|$ computed at $q_{z}=1.8\pi$ with $V_{c}=0$ by using (a) $t_{z}=0.2t$, $t_{z}^{'}=0.05t$ (same as in Fig.~\ref{fig:YBCO}), and (b) $t_{z}=0.4t$, $t_{z}^{'}=0.15t$, while keeping the other parameters unchanged; $t/2$ is assumed to be 400 meV. The white dotted curve denotes the upper boundary of the particle-hole continuum.  
	}
	\label{fig:fit-YBCO-zs}
\end{figure}
In Sec.~\ref{sec:zs} we showed that two collective modes persist even when the long-range Coulomb interaction is switched off ($V_{c} = 0$). In this section we examine whether the experimental dispersion in Fig.~\ref{fig:YBCO} could be interpreted within this scenario.

A first qualitative consideration already points to a limitation: the lower-energy zero-sound-wavelike mode necessarily decreases linearly to zero as $\qp \to (0,0)$, whereas the experimental data show no indication of such behavior. This implies that, if a zero-sound-based interpretation were to be considered, the relevant branch would need to be the higher-energy mode.

Figure~\ref{fig:fit-YBCO-zs}(a) shows the intensity map for $V_c = 0$ using the same band parameters that successfully describe the plasmon dispersion in Fig.~\ref{fig:YBCO}. The resulting zero-sound-wavelike branches lie at substantially lower energies than the experimental data, indicating that these parameters cannot account for the measured spectrum. To counteract this effect, we consider larger values of $t_z$ and $t_{z}^{'}$ and display the corresponding results in Fig.~\ref{fig:fit-YBCO-zs}(b). While the overall energy scale then becomes comparable to the measured one, the slope of the computed dispersion remains inconsistent with the observed trend. Moreover, the required values of $t_z$ and $t_{z}^{'}$ appear too large for bilayer cuprates, suggesting that a description based solely on zero-sound-wavelike modes is unlikely.

These analyses lead us to conclude that the experimental features cannot be satisfactorily described by the collective modes of the pure bilayer \tj model. Instead, their dispersion and spectral characteristics are more naturally and consistently interpreted in terms of plasmons.







\bibliography{main} 

\begin{thebibliography}{66}%
\makeatletter
\providecommand \@ifxundefined [1]{%
 \@ifx{#1\undefined}
}%
\providecommand \@ifnum [1]{%
 \ifnum #1\expandafter \@firstoftwo
 \else \expandafter \@secondoftwo
 \fi
}%
\providecommand \@ifx [1]{%
 \ifx #1\expandafter \@firstoftwo
 \else \expandafter \@secondoftwo
 \fi
}%
\providecommand \natexlab [1]{#1}%
\providecommand \enquote  [1]{``#1''}%
\providecommand \bibnamefont  [1]{#1}%
\providecommand \bibfnamefont [1]{#1}%
\providecommand \citenamefont [1]{#1}%
\providecommand \href@noop [0]{\@secondoftwo}%
\providecommand \href [0]{\begingroup \@sanitize@url \@href}%
\providecommand \@href[1]{\@@startlink{#1}\@@href}%
\providecommand \@@href[1]{\endgroup#1\@@endlink}%
\providecommand \@sanitize@url [0]{\catcode `\\12\catcode `\$12\catcode
  `\&12\catcode `\#12\catcode `\^12\catcode `\_12\catcode `\%12\relax}%
\providecommand \@@startlink[1]{}%
\providecommand \@@endlink[0]{}%
\providecommand \url  [0]{\begingroup\@sanitize@url \@url }%
\providecommand \@url [1]{\endgroup\@href {#1}{\urlprefix }}%
\providecommand \urlprefix  [0]{URL }%
\providecommand \Eprint [0]{\href }%
\providecommand \doibase [0]{http://dx.doi.org/}%
\providecommand \selectlanguage [0]{\@gobble}%
\providecommand \bibinfo  [0]{\@secondoftwo}%
\providecommand \bibfield  [0]{\@secondoftwo}%
\providecommand \translation [1]{[#1]}%
\providecommand \BibitemOpen [0]{}%
\providecommand \bibitemStop [0]{}%
\providecommand \bibitemNoStop [0]{.\EOS\space}%
\providecommand \EOS [0]{\spacefactor3000\relax}%
\providecommand \BibitemShut  [1]{\csname bibitem#1\endcsname}%
\let\auto@bib@innerbib\@empty
\bibitem [{\citenamefont {Keimer}\ \emph {et~al.}(2015)\citenamefont {Keimer},
  \citenamefont {Kivelson}, \citenamefont {Norman}, \citenamefont {Uchida},\
  and\ \citenamefont {Zaanen}}]{keimer15}%
  \BibitemOpen
  \bibfield  {author} {\bibinfo {author} {\bibfnamefont {B.}~\bibnamefont
  {Keimer}}, \bibinfo {author} {\bibfnamefont {S.~A.}\ \bibnamefont
  {Kivelson}}, \bibinfo {author} {\bibfnamefont {M.~R.}\ \bibnamefont
  {Norman}}, \bibinfo {author} {\bibfnamefont {S.}~\bibnamefont {Uchida}}, \
  and\ \bibinfo {author} {\bibfnamefont {J.}~\bibnamefont {Zaanen}},\
  }\bibfield  {title} {\enquote {\bibinfo {title} {From quantum matter to
  high-temperature superconductivity in copper oxides},}\ }\href {\doibase
  10.1038/nature14165} {\bibfield  {journal} {\bibinfo  {journal} {Nature}\
  }\textbf {\bibinfo {volume} {518}},\ \bibinfo {pages} {179--186} (\bibinfo
  {year} {2015})}\BibitemShut {NoStop}%
\bibitem [{\citenamefont {Anderson}(1987)}]{anderson87}%
  \BibitemOpen
  \bibfield  {author} {\bibinfo {author} {\bibfnamefont {P.~W.}\ \bibnamefont
  {Anderson}},\ }\bibfield  {title} {\enquote {\bibinfo {title} {The resonating
  valence bond state in {La}$_{2}${CuO}$_{4}$ and superconductivity},}\ }\href
  {\doibase 10.1126/science.235.4793.1196} {\bibfield  {journal} {\bibinfo
  {journal} {Science}\ }\textbf {\bibinfo {volume} {235}},\ \bibinfo {pages}
  {1196--1198} (\bibinfo {year} {1987})}\BibitemShut {NoStop}%
\bibitem [{\citenamefont {Ament}\ \emph {et~al.}(2011)\citenamefont {Ament},
  \citenamefont {van Veenendaal}, \citenamefont {Devereaux}, \citenamefont
  {Hill},\ and\ \citenamefont {van~den Brink}}]{ament11}%
  \BibitemOpen
  \bibfield  {author} {\bibinfo {author} {\bibfnamefont {Luuk J.~P.}\
  \bibnamefont {Ament}}, \bibinfo {author} {\bibfnamefont {Michel}\
  \bibnamefont {van Veenendaal}}, \bibinfo {author} {\bibfnamefont {Thomas~P.}\
  \bibnamefont {Devereaux}}, \bibinfo {author} {\bibfnamefont {John~P.}\
  \bibnamefont {Hill}}, \ and\ \bibinfo {author} {\bibfnamefont {Jeroen}\
  \bibnamefont {van~den Brink}},\ }\bibfield  {title} {\enquote {\bibinfo
  {title} {Resonant inelastic x-ray scattering studies of elementary
  excitations},}\ }\href@noop {} {\bibfield  {journal} {\bibinfo  {journal}
  {Rev. Mod. Phys.}\ }\textbf {\bibinfo {volume} {83}},\ \bibinfo {pages}
  {705--767} (\bibinfo {year} {2011})}\BibitemShut {NoStop}%
\bibitem [{\citenamefont {de~Groot}\ \emph {et~al.}(2024)\citenamefont
  {de~Groot}, \citenamefont {Haverkort}, \citenamefont {Elnaggar},
  \citenamefont {Juhin}, \citenamefont {Zhou},\ and\ \citenamefont
  {Glatzel}}]{degroot24}%
  \BibitemOpen
  \bibfield  {author} {\bibinfo {author} {\bibfnamefont {Frank M.~F.}\
  \bibnamefont {de~Groot}}, \bibinfo {author} {\bibfnamefont {Maurits~W.}\
  \bibnamefont {Haverkort}}, \bibinfo {author} {\bibfnamefont {Hebatalla}\
  \bibnamefont {Elnaggar}}, \bibinfo {author} {\bibfnamefont {Am{\'e}lie}\
  \bibnamefont {Juhin}}, \bibinfo {author} {\bibfnamefont {Ke-Jin}\
  \bibnamefont {Zhou}}, \ and\ \bibinfo {author} {\bibfnamefont {Pieter}\
  \bibnamefont {Glatzel}},\ }\bibfield  {title} {\enquote {\bibinfo {title}
  {Resonant inelastic x-ray scattering},}\ }\href {\doibase
  10.1038/s43586-024-00322-6} {\bibfield  {journal} {\bibinfo  {journal}
  {Nature Reviews Methods Primers}\ }\textbf {\bibinfo {volume} {4}},\ \bibinfo
  {pages} {45} (\bibinfo {year} {2024})}\BibitemShut {NoStop}%
\bibitem [{\citenamefont {Ghiringhelli}\ \emph {et~al.}(2012)\citenamefont
  {Ghiringhelli}, \citenamefont {Le~Tacon}, \citenamefont {Minola},
  \citenamefont {Blanco-Canosa}, \citenamefont {Mazzoli}, \citenamefont
  {Brookes}, \citenamefont {De~Luca}, \citenamefont {Frano}, \citenamefont
  {Hawthorn}, \citenamefont {He}, \citenamefont {Loew}, \citenamefont {Sala},
  \citenamefont {Peets}, \citenamefont {Salluzzo}, \citenamefont {Schierle},
  \citenamefont {Sutarto}, \citenamefont {Sawatzky}, \citenamefont {Weschke},
  \citenamefont {Keimer},\ and\ \citenamefont {Braicovich}}]{ghiringhelli12}%
  \BibitemOpen
  \bibfield  {author} {\bibinfo {author} {\bibfnamefont {G.}~\bibnamefont
  {Ghiringhelli}}, \bibinfo {author} {\bibfnamefont {M.}~\bibnamefont
  {Le~Tacon}}, \bibinfo {author} {\bibfnamefont {M.}~\bibnamefont {Minola}},
  \bibinfo {author} {\bibfnamefont {S.}~\bibnamefont {Blanco-Canosa}}, \bibinfo
  {author} {\bibfnamefont {C.}~\bibnamefont {Mazzoli}}, \bibinfo {author}
  {\bibfnamefont {N.~B.}\ \bibnamefont {Brookes}}, \bibinfo {author}
  {\bibfnamefont {G.~M.}\ \bibnamefont {De~Luca}}, \bibinfo {author}
  {\bibfnamefont {A.}~\bibnamefont {Frano}}, \bibinfo {author} {\bibfnamefont
  {D.~G.}\ \bibnamefont {Hawthorn}}, \bibinfo {author} {\bibfnamefont
  {F.}~\bibnamefont {He}}, \bibinfo {author} {\bibfnamefont {T.}~\bibnamefont
  {Loew}}, \bibinfo {author} {\bibfnamefont {M.~Moretti}\ \bibnamefont {Sala}},
  \bibinfo {author} {\bibfnamefont {D.~C.}\ \bibnamefont {Peets}}, \bibinfo
  {author} {\bibfnamefont {M.}~\bibnamefont {Salluzzo}}, \bibinfo {author}
  {\bibfnamefont {E.}~\bibnamefont {Schierle}}, \bibinfo {author}
  {\bibfnamefont {R.}~\bibnamefont {Sutarto}}, \bibinfo {author} {\bibfnamefont
  {G.~A.}\ \bibnamefont {Sawatzky}}, \bibinfo {author} {\bibfnamefont
  {E.}~\bibnamefont {Weschke}}, \bibinfo {author} {\bibfnamefont
  {B.}~\bibnamefont {Keimer}}, \ and\ \bibinfo {author} {\bibfnamefont
  {L.}~\bibnamefont {Braicovich}},\ }\bibfield  {title} {\enquote {\bibinfo
  {title} {{Long-Range Incommensurate Charge Fluctuations in
  (Y,Nd)Ba$_2$Cu$_3$O$_{6+x}$}},}\ }\href {\doibase 10.1126/science.1223532}
  {\bibfield  {journal} {\bibinfo  {journal} {Science}\ }\textbf {\bibinfo
  {volume} {337}},\ \bibinfo {pages} {821--825} (\bibinfo {year}
  {2012})}\BibitemShut {NoStop}%
\bibitem [{\citenamefont {Chang}\ \emph {et~al.}(2012)\citenamefont {Chang},
  \citenamefont {Blackburn}, \citenamefont {Holmes}, \citenamefont
  {Christensen}, \citenamefont {Larsen}, \citenamefont {Mesot}, \citenamefont
  {Liang}, \citenamefont {Bonn}, \citenamefont {Hardy}, \citenamefont
  {Watenphul}, \citenamefont {Zimmermann}, \citenamefont {Forgan},\ and\
  \citenamefont {Hayden}}]{chang12}%
  \BibitemOpen
  \bibfield  {author} {\bibinfo {author} {\bibfnamefont {J.}~\bibnamefont
  {Chang}}, \bibinfo {author} {\bibfnamefont {E.}~\bibnamefont {Blackburn}},
  \bibinfo {author} {\bibfnamefont {A.~T.}\ \bibnamefont {Holmes}}, \bibinfo
  {author} {\bibfnamefont {N.~B.}\ \bibnamefont {Christensen}}, \bibinfo
  {author} {\bibfnamefont {J.}~\bibnamefont {Larsen}}, \bibinfo {author}
  {\bibfnamefont {J.}~\bibnamefont {Mesot}}, \bibinfo {author} {\bibfnamefont
  {Ruixing}\ \bibnamefont {Liang}}, \bibinfo {author} {\bibfnamefont {D.~A.}\
  \bibnamefont {Bonn}}, \bibinfo {author} {\bibfnamefont {W.~N.}\ \bibnamefont
  {Hardy}}, \bibinfo {author} {\bibfnamefont {A.}~\bibnamefont {Watenphul}},
  \bibinfo {author} {\bibfnamefont {M.~v.}\ \bibnamefont {Zimmermann}},
  \bibinfo {author} {\bibfnamefont {E.~M.}\ \bibnamefont {Forgan}}, \ and\
  \bibinfo {author} {\bibfnamefont {S.~M.}\ \bibnamefont {Hayden}},\ }\bibfield
   {title} {\enquote {\bibinfo {title} {{Direct observation of competition
  between superconductivity and charge density wave order in
  YBa$_2$Cu$_3$O$_{6.67}$}},}\ }\href@noop {} {\bibfield  {journal} {\bibinfo
  {journal} {Nat. Phys.}\ }\textbf {\bibinfo {volume} {8}},\ \bibinfo {pages}
  {871} (\bibinfo {year} {2012})}\BibitemShut {NoStop}%
\bibitem [{\citenamefont {Achkar}\ \emph {et~al.}(2012)\citenamefont {Achkar},
  \citenamefont {Sutarto}, \citenamefont {Mao}, \citenamefont {He},
  \citenamefont {Frano}, \citenamefont {Blanco-Canosa}, \citenamefont
  {Le~Tacon}, \citenamefont {Ghiringhelli}, \citenamefont {Braicovich},
  \citenamefont {Minola}, \citenamefont {Moretti~Sala}, \citenamefont
  {Mazzoli}, \citenamefont {Liang}, \citenamefont {Bonn}, \citenamefont
  {Hardy}, \citenamefont {Keimer}, \citenamefont {Sawatzky},\ and\
  \citenamefont {Hawthorn}}]{achkar12}%
  \BibitemOpen
  \bibfield  {author} {\bibinfo {author} {\bibfnamefont {A.~J.}\ \bibnamefont
  {Achkar}}, \bibinfo {author} {\bibfnamefont {R.}~\bibnamefont {Sutarto}},
  \bibinfo {author} {\bibfnamefont {X.}~\bibnamefont {Mao}}, \bibinfo {author}
  {\bibfnamefont {F.}~\bibnamefont {He}}, \bibinfo {author} {\bibfnamefont
  {A.}~\bibnamefont {Frano}}, \bibinfo {author} {\bibfnamefont
  {S.}~\bibnamefont {Blanco-Canosa}}, \bibinfo {author} {\bibfnamefont
  {M.}~\bibnamefont {Le~Tacon}}, \bibinfo {author} {\bibfnamefont
  {G.}~\bibnamefont {Ghiringhelli}}, \bibinfo {author} {\bibfnamefont
  {L.}~\bibnamefont {Braicovich}}, \bibinfo {author} {\bibfnamefont
  {M.}~\bibnamefont {Minola}}, \bibinfo {author} {\bibfnamefont
  {M.}~\bibnamefont {Moretti~Sala}}, \bibinfo {author} {\bibfnamefont
  {C.}~\bibnamefont {Mazzoli}}, \bibinfo {author} {\bibfnamefont {Ruixing}\
  \bibnamefont {Liang}}, \bibinfo {author} {\bibfnamefont {D.~A.}\ \bibnamefont
  {Bonn}}, \bibinfo {author} {\bibfnamefont {W.~N.}\ \bibnamefont {Hardy}},
  \bibinfo {author} {\bibfnamefont {B.}~\bibnamefont {Keimer}}, \bibinfo
  {author} {\bibfnamefont {G.~A.}\ \bibnamefont {Sawatzky}}, \ and\ \bibinfo
  {author} {\bibfnamefont {D.~G.}\ \bibnamefont {Hawthorn}},\ }\bibfield
  {title} {\enquote {\bibinfo {title} {{Distinct Charge Orders in the Planes
  and Chains of Ortho-III-Ordered
  ${\rm{YBa}}_{2}{\rm{Cu}}_{3}{\rm{O}}_{6+\ensuremath{\delta}}$ Superconductors
  Identified by Resonant Elastic X-ray Scattering}},}\ }\href@noop {}
  {\bibfield  {journal} {\bibinfo  {journal} {Phys. Rev. Lett.}\ }\textbf
  {\bibinfo {volume} {109}},\ \bibinfo {pages} {167001} (\bibinfo {year}
  {2012})}\BibitemShut {NoStop}%
\bibitem [{\citenamefont {Blackburn}\ \emph {et~al.}(2013)\citenamefont
  {Blackburn}, \citenamefont {Chang}, \citenamefont {H\"ucker}, \citenamefont
  {Holmes}, \citenamefont {Christensen}, \citenamefont {Liang}, \citenamefont
  {Bonn}, \citenamefont {Hardy}, \citenamefont {R\"utt}, \citenamefont
  {Gutowski}, \citenamefont {Zimmermann}, \citenamefont {Forgan},\ and\
  \citenamefont {Hayden}}]{blackburn13}%
  \BibitemOpen
  \bibfield  {author} {\bibinfo {author} {\bibfnamefont {E.}~\bibnamefont
  {Blackburn}}, \bibinfo {author} {\bibfnamefont {J.}~\bibnamefont {Chang}},
  \bibinfo {author} {\bibfnamefont {M.}~\bibnamefont {H\"ucker}}, \bibinfo
  {author} {\bibfnamefont {A.~T.}\ \bibnamefont {Holmes}}, \bibinfo {author}
  {\bibfnamefont {N.~B.}\ \bibnamefont {Christensen}}, \bibinfo {author}
  {\bibfnamefont {Ruixing}\ \bibnamefont {Liang}}, \bibinfo {author}
  {\bibfnamefont {D.~A.}\ \bibnamefont {Bonn}}, \bibinfo {author}
  {\bibfnamefont {W.~N.}\ \bibnamefont {Hardy}}, \bibinfo {author}
  {\bibfnamefont {U.}~\bibnamefont {R\"utt}}, \bibinfo {author} {\bibfnamefont
  {O.}~\bibnamefont {Gutowski}}, \bibinfo {author} {\bibfnamefont {M.~v.}\
  \bibnamefont {Zimmermann}}, \bibinfo {author} {\bibfnamefont {E.~M.}\
  \bibnamefont {Forgan}}, \ and\ \bibinfo {author} {\bibfnamefont {S.~M.}\
  \bibnamefont {Hayden}},\ }\bibfield  {title} {\enquote {\bibinfo {title}
  {{X-Ray Diffraction Observations of a Charge-Density-Wave Order in
  Superconducting Ortho-II ${\rm{YBa}}_{2}{\rm{Cu}}_{3}{\rm{O}}_{6.54}$ Single
  Crystals in Zero Magnetic Field}},}\ }\href@noop {} {\bibfield  {journal}
  {\bibinfo  {journal} {Phys. Rev. Lett.}\ }\textbf {\bibinfo {volume} {110}},\
  \bibinfo {pages} {137004} (\bibinfo {year} {2013})}\BibitemShut {NoStop}%
\bibitem [{\citenamefont {Blanco-Canosa}\ \emph {et~al.}(2014)\citenamefont
  {Blanco-Canosa}, \citenamefont {Frano}, \citenamefont {Schierle},
  \citenamefont {Porras}, \citenamefont {Loew}, \citenamefont {Minola},
  \citenamefont {Bluschke}, \citenamefont {Weschke}, \citenamefont {Keimer},\
  and\ \citenamefont {Le~Tacon}}]{blanco-canosa14}%
  \BibitemOpen
  \bibfield  {author} {\bibinfo {author} {\bibfnamefont {S.}~\bibnamefont
  {Blanco-Canosa}}, \bibinfo {author} {\bibfnamefont {A.}~\bibnamefont
  {Frano}}, \bibinfo {author} {\bibfnamefont {E.}~\bibnamefont {Schierle}},
  \bibinfo {author} {\bibfnamefont {J.}~\bibnamefont {Porras}}, \bibinfo
  {author} {\bibfnamefont {T.}~\bibnamefont {Loew}}, \bibinfo {author}
  {\bibfnamefont {M.}~\bibnamefont {Minola}}, \bibinfo {author} {\bibfnamefont
  {M.}~\bibnamefont {Bluschke}}, \bibinfo {author} {\bibfnamefont
  {E.}~\bibnamefont {Weschke}}, \bibinfo {author} {\bibfnamefont
  {B.}~\bibnamefont {Keimer}}, \ and\ \bibinfo {author} {\bibfnamefont
  {M.}~\bibnamefont {Le~Tacon}},\ }\bibfield  {title} {\enquote {\bibinfo
  {title} {{Resonant x-ray scattering study of charge-density wave correlations
  in ${\rm{YBa}}_{2}{\rm{Cu}}_{3}{\rm{O}}_{6+x}$}},}\ }\href {\doibase
  10.1103/PhysRevB.90.054513} {\bibfield  {journal} {\bibinfo  {journal} {Phys.
  Rev. B}\ }\textbf {\bibinfo {volume} {90}},\ \bibinfo {pages} {054513}
  (\bibinfo {year} {2014})}\BibitemShut {NoStop}%
\bibitem [{\citenamefont {Comin}\ \emph {et~al.}(2014)\citenamefont {Comin},
  \citenamefont {Frano}, \citenamefont {Yee}, \citenamefont {Yoshida},
  \citenamefont {Eisaki}, \citenamefont {Schierle}, \citenamefont {Weschke},
  \citenamefont {Sutarto}, \citenamefont {He}, \citenamefont {Soumyanarayanan},
  \citenamefont {He}, \citenamefont {Le~Tacon}, \citenamefont {Elfimov},
  \citenamefont {Hoffman}, \citenamefont {Sawatzky}, \citenamefont {Keimer},\
  and\ \citenamefont {Damascelli}}]{comin14}%
  \BibitemOpen
  \bibfield  {author} {\bibinfo {author} {\bibfnamefont {R.}~\bibnamefont
  {Comin}}, \bibinfo {author} {\bibfnamefont {A.}~\bibnamefont {Frano}},
  \bibinfo {author} {\bibfnamefont {M.~M.}\ \bibnamefont {Yee}}, \bibinfo
  {author} {\bibfnamefont {Y.}~\bibnamefont {Yoshida}}, \bibinfo {author}
  {\bibfnamefont {H.}~\bibnamefont {Eisaki}}, \bibinfo {author} {\bibfnamefont
  {E.}~\bibnamefont {Schierle}}, \bibinfo {author} {\bibfnamefont
  {E.}~\bibnamefont {Weschke}}, \bibinfo {author} {\bibfnamefont
  {R.}~\bibnamefont {Sutarto}}, \bibinfo {author} {\bibfnamefont
  {F.}~\bibnamefont {He}}, \bibinfo {author} {\bibfnamefont {A.}~\bibnamefont
  {Soumyanarayanan}}, \bibinfo {author} {\bibfnamefont {Yang}\ \bibnamefont
  {He}}, \bibinfo {author} {\bibfnamefont {M.}~\bibnamefont {Le~Tacon}},
  \bibinfo {author} {\bibfnamefont {I.~S.}\ \bibnamefont {Elfimov}}, \bibinfo
  {author} {\bibfnamefont {Jennifer~E.}\ \bibnamefont {Hoffman}}, \bibinfo
  {author} {\bibfnamefont {G.~A.}\ \bibnamefont {Sawatzky}}, \bibinfo {author}
  {\bibfnamefont {B.}~\bibnamefont {Keimer}}, \ and\ \bibinfo {author}
  {\bibfnamefont {A.}~\bibnamefont {Damascelli}},\ }\bibfield  {title}
  {\enquote {\bibinfo {title} {{Charge Order Driven by Fermi-Arc Instability in
  Bi$_2$Sr$_{2-x}$La$_x$CuO$_{6+\delta}$}},}\ }\href {\doibase
  10.1126/science.1242996} {\bibfield  {journal} {\bibinfo  {journal}
  {Science}\ }\textbf {\bibinfo {volume} {343}},\ \bibinfo {pages} {390--392}
  (\bibinfo {year} {2014})}\BibitemShut {NoStop}%
\bibitem [{\citenamefont {Tabis}\ \emph {et~al.}(2014)\citenamefont {Tabis},
  \citenamefont {Li}, \citenamefont {Tacon}, \citenamefont {Braicovich},
  \citenamefont {Kreyssig}, \citenamefont {Minola}, \citenamefont {Dellea},
  \citenamefont {Weschke}, \citenamefont {Veit}, \citenamefont {Ramazanoglu},
  \citenamefont {Goldman}, \citenamefont {Schmitt}, \citenamefont
  {Ghiringhelli}, \citenamefont {Barisic}, \citenamefont {Chan}, \citenamefont
  {Dorow}, \citenamefont {Yu}, \citenamefont {Zhao}, \citenamefont {Keimer},\
  and\ \citenamefont {Greven}}]{tabis14}%
  \BibitemOpen
  \bibfield  {author} {\bibinfo {author} {\bibfnamefont {W.}~\bibnamefont
  {Tabis}}, \bibinfo {author} {\bibfnamefont {Y.}~\bibnamefont {Li}}, \bibinfo
  {author} {\bibfnamefont {M.~Le}\ \bibnamefont {Tacon}}, \bibinfo {author}
  {\bibfnamefont {L.}~\bibnamefont {Braicovich}}, \bibinfo {author}
  {\bibfnamefont {A.}~\bibnamefont {Kreyssig}}, \bibinfo {author}
  {\bibfnamefont {M.}~\bibnamefont {Minola}}, \bibinfo {author} {\bibfnamefont
  {G.}~\bibnamefont {Dellea}}, \bibinfo {author} {\bibfnamefont
  {E.}~\bibnamefont {Weschke}}, \bibinfo {author} {\bibfnamefont {M.~J.}\
  \bibnamefont {Veit}}, \bibinfo {author} {\bibfnamefont {M.}~\bibnamefont
  {Ramazanoglu}}, \bibinfo {author} {\bibfnamefont {A.~I.}\ \bibnamefont
  {Goldman}}, \bibinfo {author} {\bibfnamefont {T.}~\bibnamefont {Schmitt}},
  \bibinfo {author} {\bibfnamefont {G.}~\bibnamefont {Ghiringhelli}}, \bibinfo
  {author} {\bibfnamefont {N.}~\bibnamefont {Barisic}}, \bibinfo {author}
  {\bibfnamefont {M.~K.}\ \bibnamefont {Chan}}, \bibinfo {author}
  {\bibfnamefont {C.~J.}\ \bibnamefont {Dorow}}, \bibinfo {author}
  {\bibfnamefont {G.}~\bibnamefont {Yu}}, \bibinfo {author} {\bibfnamefont
  {X.}~\bibnamefont {Zhao}}, \bibinfo {author} {\bibfnamefont {B.}~\bibnamefont
  {Keimer}}, \ and\ \bibinfo {author} {\bibfnamefont {M.}~\bibnamefont
  {Greven}},\ }\bibfield  {title} {\enquote {\bibinfo {title} {{Charge order
  and its connection with Fermi-liquid charge transport in a pristine
  high-$T_c$ cuprate}},}\ }\href {http://dx.doi.org/10.1038/ncomms6875}
  {\bibfield  {journal} {\bibinfo  {journal} {Nat. Commun.}\ }\textbf {\bibinfo
  {volume} {5}},\ \bibinfo {pages} {5875} (\bibinfo {year} {2014})}\BibitemShut
  {NoStop}%
\bibitem [{\citenamefont {da~Silva~Neto}\ \emph {et~al.}(2014)\citenamefont
  {da~Silva~Neto}, \citenamefont {Aynajian}, \citenamefont {Frano},
  \citenamefont {Comin}, \citenamefont {Schierle}, \citenamefont {Weschke},
  \citenamefont {Gyenis}, \citenamefont {Wen}, \citenamefont {Schneeloch},
  \citenamefont {Xu}, \citenamefont {Ono}, \citenamefont {Gu}, \citenamefont
  {Le~Tacon},\ and\ \citenamefont {Yazdani}}]{da-silva-neto14}%
  \BibitemOpen
  \bibfield  {author} {\bibinfo {author} {\bibfnamefont {Eduardo~H.}\
  \bibnamefont {da~Silva~Neto}}, \bibinfo {author} {\bibfnamefont {Pegor}\
  \bibnamefont {Aynajian}}, \bibinfo {author} {\bibfnamefont {Alex}\
  \bibnamefont {Frano}}, \bibinfo {author} {\bibfnamefont {Riccardo}\
  \bibnamefont {Comin}}, \bibinfo {author} {\bibfnamefont {Enrico}\
  \bibnamefont {Schierle}}, \bibinfo {author} {\bibfnamefont {Eugen}\
  \bibnamefont {Weschke}}, \bibinfo {author} {\bibfnamefont {Andr{\'a}s}\
  \bibnamefont {Gyenis}}, \bibinfo {author} {\bibfnamefont {Jinsheng}\
  \bibnamefont {Wen}}, \bibinfo {author} {\bibfnamefont {John}\ \bibnamefont
  {Schneeloch}}, \bibinfo {author} {\bibfnamefont {Zhijun}\ \bibnamefont {Xu}},
  \bibinfo {author} {\bibfnamefont {Shimpei}\ \bibnamefont {Ono}}, \bibinfo
  {author} {\bibfnamefont {Genda}\ \bibnamefont {Gu}}, \bibinfo {author}
  {\bibfnamefont {Mathieu}\ \bibnamefont {Le~Tacon}}, \ and\ \bibinfo {author}
  {\bibfnamefont {Ali}\ \bibnamefont {Yazdani}},\ }\bibfield  {title} {\enquote
  {\bibinfo {title} {Ubiquitous interplay between charge ordering and
  high-temperature superconductivity in cuprates},}\ }\href {\doibase
  10.1126/science.1243479} {\bibfield  {journal} {\bibinfo  {journal}
  {Science}\ }\textbf {\bibinfo {volume} {343}},\ \bibinfo {pages} {393--396}
  (\bibinfo {year} {2014})}\BibitemShut {NoStop}%
\bibitem [{\citenamefont {Hashimoto}\ \emph {et~al.}(2014)\citenamefont
  {Hashimoto}, \citenamefont {Ghiringhelli}, \citenamefont {Lee}, \citenamefont
  {Dellea}, \citenamefont {Amorese}, \citenamefont {Mazzoli}, \citenamefont
  {Kummer}, \citenamefont {Brookes}, \citenamefont {Moritz}, \citenamefont
  {Yoshida}, \citenamefont {Eisaki}, \citenamefont {Hussain}, \citenamefont
  {Devereaux}, \citenamefont {Shen},\ and\ \citenamefont
  {Braicovich}}]{hashimoto14}%
  \BibitemOpen
  \bibfield  {author} {\bibinfo {author} {\bibfnamefont {M.}~\bibnamefont
  {Hashimoto}}, \bibinfo {author} {\bibfnamefont {G.}~\bibnamefont
  {Ghiringhelli}}, \bibinfo {author} {\bibfnamefont {W.-S.}\ \bibnamefont
  {Lee}}, \bibinfo {author} {\bibfnamefont {G.}~\bibnamefont {Dellea}},
  \bibinfo {author} {\bibfnamefont {A.}~\bibnamefont {Amorese}}, \bibinfo
  {author} {\bibfnamefont {C.}~\bibnamefont {Mazzoli}}, \bibinfo {author}
  {\bibfnamefont {K.}~\bibnamefont {Kummer}}, \bibinfo {author} {\bibfnamefont
  {N.~B.}\ \bibnamefont {Brookes}}, \bibinfo {author} {\bibfnamefont
  {B.}~\bibnamefont {Moritz}}, \bibinfo {author} {\bibfnamefont
  {Y.}~\bibnamefont {Yoshida}}, \bibinfo {author} {\bibfnamefont
  {H.}~\bibnamefont {Eisaki}}, \bibinfo {author} {\bibfnamefont
  {Z.}~\bibnamefont {Hussain}}, \bibinfo {author} {\bibfnamefont {T.~P.}\
  \bibnamefont {Devereaux}}, \bibinfo {author} {\bibfnamefont {Z.-X.}\
  \bibnamefont {Shen}}, \ and\ \bibinfo {author} {\bibfnamefont
  {L.}~\bibnamefont {Braicovich}},\ }\bibfield  {title} {\enquote {\bibinfo
  {title} {{Direct observation of bulk charge modulations in optimally doped
  ${\mathrm{Bi}}_{1.5}{\mathrm{Pb}}_{0.6}{\mathrm{Sr}}_{1.54}{\mathrm{CaCu}}_{2}{\mathrm{O}}_{8+\ensuremath{\delta}}$}},}\
  }\href {\doibase 10.1103/PhysRevB.89.220511} {\bibfield  {journal} {\bibinfo
  {journal} {Phys. Rev. B}\ }\textbf {\bibinfo {volume} {89}},\ \bibinfo
  {pages} {220511} (\bibinfo {year} {2014})}\BibitemShut {NoStop}%
\bibitem [{\citenamefont {Peng}\ \emph {et~al.}(2016)\citenamefont {Peng},
  \citenamefont {Salluzzo}, \citenamefont {Sun}, \citenamefont {Ponti},
  \citenamefont {Betto}, \citenamefont {Ferretti}, \citenamefont {Fumagalli},
  \citenamefont {Kummer}, \citenamefont {Le~Tacon}, \citenamefont {Zhou},
  \citenamefont {Brookes}, \citenamefont {Braicovich},\ and\ \citenamefont
  {Ghiringhelli}}]{peng16}%
  \BibitemOpen
  \bibfield  {author} {\bibinfo {author} {\bibfnamefont {Y.~Y.}\ \bibnamefont
  {Peng}}, \bibinfo {author} {\bibfnamefont {M.}~\bibnamefont {Salluzzo}},
  \bibinfo {author} {\bibfnamefont {X.}~\bibnamefont {Sun}}, \bibinfo {author}
  {\bibfnamefont {A.}~\bibnamefont {Ponti}}, \bibinfo {author} {\bibfnamefont
  {D.}~\bibnamefont {Betto}}, \bibinfo {author} {\bibfnamefont {A.~M.}\
  \bibnamefont {Ferretti}}, \bibinfo {author} {\bibfnamefont {F.}~\bibnamefont
  {Fumagalli}}, \bibinfo {author} {\bibfnamefont {K.}~\bibnamefont {Kummer}},
  \bibinfo {author} {\bibfnamefont {M.}~\bibnamefont {Le~Tacon}}, \bibinfo
  {author} {\bibfnamefont {X.~J.}\ \bibnamefont {Zhou}}, \bibinfo {author}
  {\bibfnamefont {N.~B.}\ \bibnamefont {Brookes}}, \bibinfo {author}
  {\bibfnamefont {L.}~\bibnamefont {Braicovich}}, \ and\ \bibinfo {author}
  {\bibfnamefont {G.}~\bibnamefont {Ghiringhelli}},\ }\bibfield  {title}
  {\enquote {\bibinfo {title} {{Direct observation of charge order in
  underdoped and optimally doped
  ${\mathrm{Bi}}_{2}{(\mathrm{Sr},\mathrm{La})}_{2}{\mathrm{CuO}}_{6+\ensuremath{\delta}}$
  by resonant inelastic x-ray scattering}},}\ }\href {\doibase
  10.1103/PhysRevB.94.184511} {\bibfield  {journal} {\bibinfo  {journal} {Phys.
  Rev. B}\ }\textbf {\bibinfo {volume} {94}},\ \bibinfo {pages} {184511}
  (\bibinfo {year} {2016})}\BibitemShut {NoStop}%
\bibitem [{\citenamefont {Chaix}\ \emph {et~al.}(2017)\citenamefont {Chaix},
  \citenamefont {Ghiringhelli}, \citenamefont {Peng}, \citenamefont
  {Hashimoto}, \citenamefont {Moritz}, \citenamefont {Kummer}, \citenamefont
  {Brookes}, \citenamefont {He}, \citenamefont {Chen}, \citenamefont {Ishida},
  \citenamefont {Yoshida}, \citenamefont {Eisaki}, \citenamefont {Salluzzo},
  \citenamefont {Braicovich}, \citenamefont {Shen}, \citenamefont {Devereaux},\
  and\ \citenamefont {Lee}}]{chaix17}%
  \BibitemOpen
  \bibfield  {author} {\bibinfo {author} {\bibfnamefont {L.}~\bibnamefont
  {Chaix}}, \bibinfo {author} {\bibfnamefont {G.}~\bibnamefont {Ghiringhelli}},
  \bibinfo {author} {\bibfnamefont {Y.~Y.}\ \bibnamefont {Peng}}, \bibinfo
  {author} {\bibfnamefont {M.}~\bibnamefont {Hashimoto}}, \bibinfo {author}
  {\bibfnamefont {B.}~\bibnamefont {Moritz}}, \bibinfo {author} {\bibfnamefont
  {K.}~\bibnamefont {Kummer}}, \bibinfo {author} {\bibfnamefont {N.~B.}\
  \bibnamefont {Brookes}}, \bibinfo {author} {\bibfnamefont {Y.}~\bibnamefont
  {He}}, \bibinfo {author} {\bibfnamefont {S.}~\bibnamefont {Chen}}, \bibinfo
  {author} {\bibfnamefont {S.}~\bibnamefont {Ishida}}, \bibinfo {author}
  {\bibfnamefont {Y.}~\bibnamefont {Yoshida}}, \bibinfo {author} {\bibfnamefont
  {H.}~\bibnamefont {Eisaki}}, \bibinfo {author} {\bibfnamefont
  {M.}~\bibnamefont {Salluzzo}}, \bibinfo {author} {\bibfnamefont
  {L.}~\bibnamefont {Braicovich}}, \bibinfo {author} {\bibfnamefont {Z.~X.}\
  \bibnamefont {Shen}}, \bibinfo {author} {\bibfnamefont {T.~P.}\ \bibnamefont
  {Devereaux}}, \ and\ \bibinfo {author} {\bibfnamefont {W.~S.}\ \bibnamefont
  {Lee}},\ }\bibfield  {title} {\enquote {\bibinfo {title} {{Dispersive charge
  density wave excitations in Bi$_2$Sr$_2$CaCu$_2$O$_{8+\delta}$}},}\
  }\href@noop {} {\bibfield  {journal} {\bibinfo  {journal} {Nat. Phys.}\
  }\textbf {\bibinfo {volume} {13}},\ \bibinfo {pages} {952} (\bibinfo {year}
  {2017})}\BibitemShut {NoStop}%
\bibitem [{\citenamefont {Arpaia}\ \emph {et~al.}(2019)\citenamefont {Arpaia},
  \citenamefont {Caprara}, \citenamefont {Fumagalli}, \citenamefont {Vecchi},
  \citenamefont {Peng}, \citenamefont {Andersson}, \citenamefont {Betto},
  \citenamefont {Luca}, \citenamefont {Brookes}, \citenamefont {Lombardi},
  \citenamefont {Salluzzo}, \citenamefont {Braicovich}, \citenamefont {Castro},
  \citenamefont {Grilli},\ and\ \citenamefont {Ghiringhelli}}]{arpaia19}%
  \BibitemOpen
  \bibfield  {author} {\bibinfo {author} {\bibfnamefont {R.}~\bibnamefont
  {Arpaia}}, \bibinfo {author} {\bibfnamefont {S.}~\bibnamefont {Caprara}},
  \bibinfo {author} {\bibfnamefont {R.}~\bibnamefont {Fumagalli}}, \bibinfo
  {author} {\bibfnamefont {G.~De}\ \bibnamefont {Vecchi}}, \bibinfo {author}
  {\bibfnamefont {Y.~Y.}\ \bibnamefont {Peng}}, \bibinfo {author}
  {\bibfnamefont {E.}~\bibnamefont {Andersson}}, \bibinfo {author}
  {\bibfnamefont {D.}~\bibnamefont {Betto}}, \bibinfo {author} {\bibfnamefont
  {G.~M.~De}\ \bibnamefont {Luca}}, \bibinfo {author} {\bibfnamefont {N.~B.}\
  \bibnamefont {Brookes}}, \bibinfo {author} {\bibfnamefont {F.}~\bibnamefont
  {Lombardi}}, \bibinfo {author} {\bibfnamefont {M.}~\bibnamefont {Salluzzo}},
  \bibinfo {author} {\bibfnamefont {L.}~\bibnamefont {Braicovich}}, \bibinfo
  {author} {\bibfnamefont {C.~Di}\ \bibnamefont {Castro}}, \bibinfo {author}
  {\bibfnamefont {M.}~\bibnamefont {Grilli}}, \ and\ \bibinfo {author}
  {\bibfnamefont {G.}~\bibnamefont {Ghiringhelli}},\ }\bibfield  {title}
  {\enquote {\bibinfo {title} {Dynamical charge density fluctuations pervading
  the phase diagram of a {Cu-}based high-{$T_{c}$} superconductor},}\
  }\href@noop {} {\bibfield  {journal} {\bibinfo  {journal} {Science}\ }\textbf
  {\bibinfo {volume} {365}},\ \bibinfo {pages} {906--910} (\bibinfo {year}
  {2019})}\BibitemShut {NoStop}%
\bibitem [{\citenamefont {Yu}\ \emph {et~al.}(2020)\citenamefont {Yu},
  \citenamefont {Tabis}, \citenamefont {Bialo}, \citenamefont {Yakhou},
  \citenamefont {Brookes}, \citenamefont {Anderson}, \citenamefont {Tang},
  \citenamefont {Yu},\ and\ \citenamefont {Greven}}]{yu20}%
  \BibitemOpen
  \bibfield  {author} {\bibinfo {author} {\bibfnamefont {B.}~\bibnamefont
  {Yu}}, \bibinfo {author} {\bibfnamefont {W.}~\bibnamefont {Tabis}}, \bibinfo
  {author} {\bibfnamefont {I.}~\bibnamefont {Bialo}}, \bibinfo {author}
  {\bibfnamefont {F.}~\bibnamefont {Yakhou}}, \bibinfo {author} {\bibfnamefont
  {N.~B.}\ \bibnamefont {Brookes}}, \bibinfo {author} {\bibfnamefont
  {Z.}~\bibnamefont {Anderson}}, \bibinfo {author} {\bibfnamefont
  {Y.}~\bibnamefont {Tang}}, \bibinfo {author} {\bibfnamefont {G.}~\bibnamefont
  {Yu}}, \ and\ \bibinfo {author} {\bibfnamefont {M.}~\bibnamefont {Greven}},\
  }\bibfield  {title} {\enquote {\bibinfo {title} {Unusual dynamic charge
  correlations in simple-tetragonal
  {${\mathrm{HgBa}}_{2}{\mathrm{CuO}}_{4+\ensuremath{\delta}}$}},}\ }\href@noop
  {} {\bibfield  {journal} {\bibinfo  {journal} {Phys. Rev. X}\ }\textbf
  {\bibinfo {volume} {10}},\ \bibinfo {pages} {021059} (\bibinfo {year}
  {2020})}\BibitemShut {NoStop}%
\bibitem [{\citenamefont {Lee}\ \emph {et~al.}(2021)\citenamefont {Lee},
  \citenamefont {Zhou}, \citenamefont {Hepting}, \citenamefont {Li},
  \citenamefont {Nag}, \citenamefont {Walters}, \citenamefont
  {Garcia-Fernandez}, \citenamefont {Robarts}, \citenamefont {Hashimoto},
  \citenamefont {Lu}, \citenamefont {Nosarzewski}, \citenamefont {Song},
  \citenamefont {Eisaki}, \citenamefont {Shen}, \citenamefont {Moritz},
  \citenamefont {Zaanen},\ and\ \citenamefont {Devereaux}}]{wslee21}%
  \BibitemOpen
  \bibfield  {author} {\bibinfo {author} {\bibfnamefont {W.~S.}\ \bibnamefont
  {Lee}}, \bibinfo {author} {\bibfnamefont {Ke-Jin}\ \bibnamefont {Zhou}},
  \bibinfo {author} {\bibfnamefont {M.}~\bibnamefont {Hepting}}, \bibinfo
  {author} {\bibfnamefont {J.}~\bibnamefont {Li}}, \bibinfo {author}
  {\bibfnamefont {A.}~\bibnamefont {Nag}}, \bibinfo {author} {\bibfnamefont
  {A.~C.}\ \bibnamefont {Walters}}, \bibinfo {author} {\bibfnamefont
  {M.}~\bibnamefont {Garcia-Fernandez}}, \bibinfo {author} {\bibfnamefont
  {H.~C.}\ \bibnamefont {Robarts}}, \bibinfo {author} {\bibfnamefont
  {M.}~\bibnamefont {Hashimoto}}, \bibinfo {author} {\bibfnamefont
  {H.}~\bibnamefont {Lu}}, \bibinfo {author} {\bibfnamefont {B.}~\bibnamefont
  {Nosarzewski}}, \bibinfo {author} {\bibfnamefont {D.}~\bibnamefont {Song}},
  \bibinfo {author} {\bibfnamefont {H.}~\bibnamefont {Eisaki}}, \bibinfo
  {author} {\bibfnamefont {Z.~X.}\ \bibnamefont {Shen}}, \bibinfo {author}
  {\bibfnamefont {B.}~\bibnamefont {Moritz}}, \bibinfo {author} {\bibfnamefont
  {J.}~\bibnamefont {Zaanen}}, \ and\ \bibinfo {author} {\bibfnamefont {T.~P.}\
  \bibnamefont {Devereaux}},\ }\bibfield  {title} {\enquote {\bibinfo {title}
  {Spectroscopic fingerprint of charge order melting driven by quantum
  fluctuations in a cuprate},}\ }\href@noop {} {\bibfield  {journal} {\bibinfo
  {journal} {Nat. Phys.}\ }\textbf {\bibinfo {volume} {17}},\ \bibinfo {pages}
  {53--57} (\bibinfo {year} {2021})}\BibitemShut {NoStop}%
\bibitem [{\citenamefont {Lu}\ \emph {et~al.}(2022)\citenamefont {Lu},
  \citenamefont {Hashimoto}, \citenamefont {Chen}, \citenamefont {Ishida},
  \citenamefont {Song}, \citenamefont {Eisaki}, \citenamefont {Nag},
  \citenamefont {Garcia-Fernandez}, \citenamefont {Arpaia}, \citenamefont
  {Ghiringhelli}, \citenamefont {Braicovich}, \citenamefont {Zaanen},
  \citenamefont {Moritz}, \citenamefont {Kummer}, \citenamefont {Brookes},
  \citenamefont {Zhou}, \citenamefont {Shen}, \citenamefont {Devereaux},\ and\
  \citenamefont {Lee}}]{lu22}%
  \BibitemOpen
  \bibfield  {author} {\bibinfo {author} {\bibfnamefont {Haiyu}\ \bibnamefont
  {Lu}}, \bibinfo {author} {\bibfnamefont {Makoto}\ \bibnamefont {Hashimoto}},
  \bibinfo {author} {\bibfnamefont {Su-Di}\ \bibnamefont {Chen}}, \bibinfo
  {author} {\bibfnamefont {Shigeyuki}\ \bibnamefont {Ishida}}, \bibinfo
  {author} {\bibfnamefont {Dongjoon}\ \bibnamefont {Song}}, \bibinfo {author}
  {\bibfnamefont {Hiroshi}\ \bibnamefont {Eisaki}}, \bibinfo {author}
  {\bibfnamefont {Abhishek}\ \bibnamefont {Nag}}, \bibinfo {author}
  {\bibfnamefont {Mirian}\ \bibnamefont {Garcia-Fernandez}}, \bibinfo {author}
  {\bibfnamefont {Riccardo}\ \bibnamefont {Arpaia}}, \bibinfo {author}
  {\bibfnamefont {Giacomo}\ \bibnamefont {Ghiringhelli}}, \bibinfo {author}
  {\bibfnamefont {Lucio}\ \bibnamefont {Braicovich}}, \bibinfo {author}
  {\bibfnamefont {Jan}\ \bibnamefont {Zaanen}}, \bibinfo {author}
  {\bibfnamefont {Brian}\ \bibnamefont {Moritz}}, \bibinfo {author}
  {\bibfnamefont {Kurt}\ \bibnamefont {Kummer}}, \bibinfo {author}
  {\bibfnamefont {Nicholas~B.}\ \bibnamefont {Brookes}}, \bibinfo {author}
  {\bibfnamefont {Ke-Jin}\ \bibnamefont {Zhou}}, \bibinfo {author}
  {\bibfnamefont {Zhi-Xun}\ \bibnamefont {Shen}}, \bibinfo {author}
  {\bibfnamefont {Thomas~P.}\ \bibnamefont {Devereaux}}, \ and\ \bibinfo
  {author} {\bibfnamefont {Wei-Sheng}\ \bibnamefont {Lee}},\ }\bibfield
  {title} {\enquote {\bibinfo {title} {Identification of a characteristic
  doping for charge order phenomena in {Bi-2212} cuprates via {RIXS}},}\
  }\href@noop {} {\bibfield  {journal} {\bibinfo  {journal} {Phys. Rev. B}\
  }\textbf {\bibinfo {volume} {106}},\ \bibinfo {pages} {155109} (\bibinfo
  {year} {2022})}\BibitemShut {NoStop}%
\bibitem [{\citenamefont {Arpaia}\ \emph {et~al.}(2023)\citenamefont {Arpaia},
  \citenamefont {Martinelli}, \citenamefont {Sala}, \citenamefont {Caprara},
  \citenamefont {Nag}, \citenamefont {Brookes}, \citenamefont {Camisa},
  \citenamefont {Li}, \citenamefont {Gao}, \citenamefont {Zhou}, \citenamefont
  {Garcia-Fernandez}, \citenamefont {Zhou}, \citenamefont {Schierle},
  \citenamefont {Bauch}, \citenamefont {Peng}, \citenamefont {Di~Castro},
  \citenamefont {Grilli}, \citenamefont {Lombardi}, \citenamefont
  {Braicovich},\ and\ \citenamefont {Ghiringhelli}}]{arpaia23}%
  \BibitemOpen
  \bibfield  {author} {\bibinfo {author} {\bibfnamefont {Riccardo}\
  \bibnamefont {Arpaia}}, \bibinfo {author} {\bibfnamefont {Leonardo}\
  \bibnamefont {Martinelli}}, \bibinfo {author} {\bibfnamefont {Marco~Moretti}\
  \bibnamefont {Sala}}, \bibinfo {author} {\bibfnamefont {Sergio}\ \bibnamefont
  {Caprara}}, \bibinfo {author} {\bibfnamefont {Abhishek}\ \bibnamefont {Nag}},
  \bibinfo {author} {\bibfnamefont {Nicholas~B.}\ \bibnamefont {Brookes}},
  \bibinfo {author} {\bibfnamefont {Pietro}\ \bibnamefont {Camisa}}, \bibinfo
  {author} {\bibfnamefont {Qizhi}\ \bibnamefont {Li}}, \bibinfo {author}
  {\bibfnamefont {Qiang}\ \bibnamefont {Gao}}, \bibinfo {author} {\bibfnamefont
  {Xingjiang}\ \bibnamefont {Zhou}}, \bibinfo {author} {\bibfnamefont {Mirian}\
  \bibnamefont {Garcia-Fernandez}}, \bibinfo {author} {\bibfnamefont {Ke-Jin}\
  \bibnamefont {Zhou}}, \bibinfo {author} {\bibfnamefont {Enrico}\ \bibnamefont
  {Schierle}}, \bibinfo {author} {\bibfnamefont {Thilo}\ \bibnamefont {Bauch}},
  \bibinfo {author} {\bibfnamefont {Ying~Ying}\ \bibnamefont {Peng}}, \bibinfo
  {author} {\bibfnamefont {Carlo}\ \bibnamefont {Di~Castro}}, \bibinfo {author}
  {\bibfnamefont {Marco}\ \bibnamefont {Grilli}}, \bibinfo {author}
  {\bibfnamefont {Floriana}\ \bibnamefont {Lombardi}}, \bibinfo {author}
  {\bibfnamefont {Lucio}\ \bibnamefont {Braicovich}}, \ and\ \bibinfo {author}
  {\bibfnamefont {Giacomo}\ \bibnamefont {Ghiringhelli}},\ }\bibfield  {title}
  {\enquote {\bibinfo {title} {Signature of quantum criticality in cuprates by
  charge density fluctuations},}\ }\href@noop {} {\bibfield  {journal}
  {\bibinfo  {journal} {Nat. Commun.}\ }\textbf {\bibinfo {volume} {14}},\
  \bibinfo {pages} {7198} (\bibinfo {year} {2023})}\BibitemShut {NoStop}%
\bibitem [{mis({\natexlab{a}})}]{misc-LSCO}%
  \BibitemOpen
  \href@noop {} {} ({\natexlab{a}}),\ \bibinfo {note} {the charge order
  observed in La-based cuprates is often discussed in terms of the spin-charge
  stripe order \cite{tranquada95}. It is an interesting subject to investigate
  its possible connection to the newly observed charge ordering tendency around
  $q_{x} \sim 0.6\pi$ in other hole-doped cuprates.}\BibitemShut {Stop}%
\bibitem [{\citenamefont {da~Silva~Neto}\ \emph {et~al.}(2015)\citenamefont
  {da~Silva~Neto}, \citenamefont {Comin}, \citenamefont {He}, \citenamefont
  {Sutarto}, \citenamefont {Jiang}, \citenamefont {Greene}, \citenamefont
  {Sawatzky},\ and\ \citenamefont {Damascelli}}]{da-silva-neto15}%
  \BibitemOpen
  \bibfield  {author} {\bibinfo {author} {\bibfnamefont {Eduardo~H.}\
  \bibnamefont {da~Silva~Neto}}, \bibinfo {author} {\bibfnamefont {Riccardo}\
  \bibnamefont {Comin}}, \bibinfo {author} {\bibfnamefont {Feizhou}\
  \bibnamefont {He}}, \bibinfo {author} {\bibfnamefont {Ronny}\ \bibnamefont
  {Sutarto}}, \bibinfo {author} {\bibfnamefont {Yeping}\ \bibnamefont {Jiang}},
  \bibinfo {author} {\bibfnamefont {Richard~L.}\ \bibnamefont {Greene}},
  \bibinfo {author} {\bibfnamefont {George~A.}\ \bibnamefont {Sawatzky}}, \
  and\ \bibinfo {author} {\bibfnamefont {Andrea}\ \bibnamefont {Damascelli}},\
  }\bibfield  {title} {\enquote {\bibinfo {title} {{Charge ordering in the
  electron-doped superconductor Nd$_{2-x}$Ce$_x$CuO$_4$}},}\ }\href {\doibase
  10.1126/science.1256441} {\bibfield  {journal} {\bibinfo  {journal}
  {Science}\ }\textbf {\bibinfo {volume} {347}},\ \bibinfo {pages} {282--285}
  (\bibinfo {year} {2015})}\BibitemShut {NoStop}%
\bibitem [{\citenamefont {da~Silva~Neto}\ \emph {et~al.}(2016)\citenamefont
  {da~Silva~Neto}, \citenamefont {Yu}, \citenamefont {Minola}, \citenamefont
  {Sutarto}, \citenamefont {Schierle}, \citenamefont {Boschini}, \citenamefont
  {Zonno}, \citenamefont {Bluschke}, \citenamefont {Higgins}, \citenamefont
  {Li}, \citenamefont {Yu}, \citenamefont {Weschke}, \citenamefont {He},
  \citenamefont {Le~Tacon}, \citenamefont {Greene}, \citenamefont {Greven},
  \citenamefont {Sawatzky}, \citenamefont {Keimer},\ and\ \citenamefont
  {Damascelli}}]{da-silva-neto16}%
  \BibitemOpen
  \bibfield  {author} {\bibinfo {author} {\bibfnamefont {Eduardo~H.}\
  \bibnamefont {da~Silva~Neto}}, \bibinfo {author} {\bibfnamefont {Biqiong}\
  \bibnamefont {Yu}}, \bibinfo {author} {\bibfnamefont {Matteo}\ \bibnamefont
  {Minola}}, \bibinfo {author} {\bibfnamefont {Ronny}\ \bibnamefont {Sutarto}},
  \bibinfo {author} {\bibfnamefont {Enrico}\ \bibnamefont {Schierle}}, \bibinfo
  {author} {\bibfnamefont {Fabio}\ \bibnamefont {Boschini}}, \bibinfo {author}
  {\bibfnamefont {Marta}\ \bibnamefont {Zonno}}, \bibinfo {author}
  {\bibfnamefont {Martin}\ \bibnamefont {Bluschke}}, \bibinfo {author}
  {\bibfnamefont {Joshua}\ \bibnamefont {Higgins}}, \bibinfo {author}
  {\bibfnamefont {Yangmu}\ \bibnamefont {Li}}, \bibinfo {author} {\bibfnamefont
  {Guichuan}\ \bibnamefont {Yu}}, \bibinfo {author} {\bibfnamefont {Eugen}\
  \bibnamefont {Weschke}}, \bibinfo {author} {\bibfnamefont {Feizhou}\
  \bibnamefont {He}}, \bibinfo {author} {\bibfnamefont {Mathieu}\ \bibnamefont
  {Le~Tacon}}, \bibinfo {author} {\bibfnamefont {Richard~L.}\ \bibnamefont
  {Greene}}, \bibinfo {author} {\bibfnamefont {Martin}\ \bibnamefont {Greven}},
  \bibinfo {author} {\bibfnamefont {George~A.}\ \bibnamefont {Sawatzky}},
  \bibinfo {author} {\bibfnamefont {Bernhard}\ \bibnamefont {Keimer}}, \ and\
  \bibinfo {author} {\bibfnamefont {Andrea}\ \bibnamefont {Damascelli}},\
  }\bibfield  {title} {\enquote {\bibinfo {title} {Doping-dependent charge
  order correlations in electron-doped cuprates},}\ }\href {\doibase
  10.1126/sciadv.1600782} {\bibfield  {journal} {\bibinfo  {journal} {Science
  Advances}\ }\textbf {\bibinfo {volume} {2}},\ \bibinfo {pages} {e1600782}
  (\bibinfo {year} {2016})}\BibitemShut {NoStop}%
\bibitem [{\citenamefont {da~Silva~Neto}\ \emph {et~al.}(2018)\citenamefont
  {da~Silva~Neto}, \citenamefont {Minola}, \citenamefont {Yu}, \citenamefont
  {Tabis}, \citenamefont {Bluschke}, \citenamefont {Unruh}, \citenamefont
  {Suzuki}, \citenamefont {Li}, \citenamefont {Yu}, \citenamefont {Betto},
  \citenamefont {Kummer}, \citenamefont {Yakhou}, \citenamefont {Brookes},
  \citenamefont {Le~Tacon}, \citenamefont {Greven}, \citenamefont {Keimer},\
  and\ \citenamefont {Damascelli}}]{da-silva-neto18}%
  \BibitemOpen
  \bibfield  {author} {\bibinfo {author} {\bibfnamefont {E.~H.}\ \bibnamefont
  {da~Silva~Neto}}, \bibinfo {author} {\bibfnamefont {M.}~\bibnamefont
  {Minola}}, \bibinfo {author} {\bibfnamefont {B.}~\bibnamefont {Yu}}, \bibinfo
  {author} {\bibfnamefont {W.}~\bibnamefont {Tabis}}, \bibinfo {author}
  {\bibfnamefont {M.}~\bibnamefont {Bluschke}}, \bibinfo {author}
  {\bibfnamefont {D.}~\bibnamefont {Unruh}}, \bibinfo {author} {\bibfnamefont
  {H.}~\bibnamefont {Suzuki}}, \bibinfo {author} {\bibfnamefont
  {Y.}~\bibnamefont {Li}}, \bibinfo {author} {\bibfnamefont {G.}~\bibnamefont
  {Yu}}, \bibinfo {author} {\bibfnamefont {D.}~\bibnamefont {Betto}}, \bibinfo
  {author} {\bibfnamefont {K.}~\bibnamefont {Kummer}}, \bibinfo {author}
  {\bibfnamefont {F.}~\bibnamefont {Yakhou}}, \bibinfo {author} {\bibfnamefont
  {N.~B.}\ \bibnamefont {Brookes}}, \bibinfo {author} {\bibfnamefont
  {M.}~\bibnamefont {Le~Tacon}}, \bibinfo {author} {\bibfnamefont
  {M.}~\bibnamefont {Greven}}, \bibinfo {author} {\bibfnamefont
  {B.}~\bibnamefont {Keimer}}, \ and\ \bibinfo {author} {\bibfnamefont
  {A.}~\bibnamefont {Damascelli}},\ }\bibfield  {title} {\enquote {\bibinfo
  {title} {Coupling between dynamic magnetic and charge-order correlations in
  the cuprate superconductor
  {${\mathrm{Nd}}_{2\ensuremath{-}x}{\mathrm{Ce}}_{x}{\mathrm{CuO}}_{4}$}},}\
  }\href {\doibase 10.1103/PhysRevB.98.161114} {\bibfield  {journal} {\bibinfo
  {journal} {Phys. Rev. B}\ }\textbf {\bibinfo {volume} {98}},\ \bibinfo
  {pages} {161114} (\bibinfo {year} {2018})}\BibitemShut {NoStop}%
\bibitem [{\citenamefont {Ishii}\ \emph {et~al.}(2005)\citenamefont {Ishii},
  \citenamefont {Tsutsui}, \citenamefont {Endoh}, \citenamefont {Tohyama},
  \citenamefont {Maekawa}, \citenamefont {Hoesch}, \citenamefont {Kuzushita},
  \citenamefont {Tsubota}, \citenamefont {Inami}, \citenamefont {Mizuki},
  \citenamefont {Murakami},\ and\ \citenamefont {Yamada}}]{ishii05}%
  \BibitemOpen
  \bibfield  {author} {\bibinfo {author} {\bibfnamefont {K.}~\bibnamefont
  {Ishii}}, \bibinfo {author} {\bibfnamefont {K.}~\bibnamefont {Tsutsui}},
  \bibinfo {author} {\bibfnamefont {Y.}~\bibnamefont {Endoh}}, \bibinfo
  {author} {\bibfnamefont {T.}~\bibnamefont {Tohyama}}, \bibinfo {author}
  {\bibfnamefont {S.}~\bibnamefont {Maekawa}}, \bibinfo {author} {\bibfnamefont
  {M.}~\bibnamefont {Hoesch}}, \bibinfo {author} {\bibfnamefont
  {K.}~\bibnamefont {Kuzushita}}, \bibinfo {author} {\bibfnamefont
  {M.}~\bibnamefont {Tsubota}}, \bibinfo {author} {\bibfnamefont
  {T.}~\bibnamefont {Inami}}, \bibinfo {author} {\bibfnamefont
  {J.}~\bibnamefont {Mizuki}}, \bibinfo {author} {\bibfnamefont
  {Y.}~\bibnamefont {Murakami}}, \ and\ \bibinfo {author} {\bibfnamefont
  {K.}~\bibnamefont {Yamada}},\ }\bibfield  {title} {\enquote {\bibinfo {title}
  {{Momentum Dependence of Charge Excitations in the Electron-Doped
  Superconductor ${\mathrm{Nd}}_{1.85}{\mathrm{Ce}}_{0.15}{\mathrm{CuO}}_{4}$:
  A Resonant Inelastic X-Ray Scattering Study}},}\ }\href {\doibase
  10.1103/PhysRevLett.94.207003} {\bibfield  {journal} {\bibinfo  {journal}
  {Phys. Rev. Lett.}\ }\textbf {\bibinfo {volume} {94}},\ \bibinfo {pages}
  {207003} (\bibinfo {year} {2005})}\BibitemShut {NoStop}%
\bibitem [{\citenamefont {Ishii}\ \emph {et~al.}(2014)\citenamefont {Ishii},
  \citenamefont {Fujita}, \citenamefont {Sasaki}, \citenamefont {Minola},
  \citenamefont {Dellea}, \citenamefont {Mazzoli}, \citenamefont {Kummer},
  \citenamefont {Ghiringhelli}, \citenamefont {Braicovich}, \citenamefont
  {Tohyama}, \citenamefont {Tsutsumi}, \citenamefont {Sato}, \citenamefont
  {Kajimoto}, \citenamefont {Ikeuchi}, \citenamefont {Yamada}, \citenamefont
  {Yoshida}, \citenamefont {Kurooka},\ and\ \citenamefont {Mizuki}}]{ishii14}%
  \BibitemOpen
  \bibfield  {author} {\bibinfo {author} {\bibfnamefont {K.}~\bibnamefont
  {Ishii}}, \bibinfo {author} {\bibfnamefont {M.}~\bibnamefont {Fujita}},
  \bibinfo {author} {\bibfnamefont {T.}~\bibnamefont {Sasaki}}, \bibinfo
  {author} {\bibfnamefont {M.}~\bibnamefont {Minola}}, \bibinfo {author}
  {\bibfnamefont {G.}~\bibnamefont {Dellea}}, \bibinfo {author} {\bibfnamefont
  {C.}~\bibnamefont {Mazzoli}}, \bibinfo {author} {\bibfnamefont
  {K.}~\bibnamefont {Kummer}}, \bibinfo {author} {\bibfnamefont
  {G.}~\bibnamefont {Ghiringhelli}}, \bibinfo {author} {\bibfnamefont
  {L.}~\bibnamefont {Braicovich}}, \bibinfo {author} {\bibfnamefont
  {T.}~\bibnamefont {Tohyama}}, \bibinfo {author} {\bibfnamefont
  {K.}~\bibnamefont {Tsutsumi}}, \bibinfo {author} {\bibfnamefont
  {K.}~\bibnamefont {Sato}}, \bibinfo {author} {\bibfnamefont {R.}~\bibnamefont
  {Kajimoto}}, \bibinfo {author} {\bibfnamefont {K.}~\bibnamefont {Ikeuchi}},
  \bibinfo {author} {\bibfnamefont {K.}~\bibnamefont {Yamada}}, \bibinfo
  {author} {\bibfnamefont {M.}~\bibnamefont {Yoshida}}, \bibinfo {author}
  {\bibfnamefont {M.}~\bibnamefont {Kurooka}}, \ and\ \bibinfo {author}
  {\bibfnamefont {J.}~\bibnamefont {Mizuki}},\ }\bibfield  {title} {\enquote
  {\bibinfo {title} {{High-energy spin and charge excitations in electron-doped
  copper oxide superconductors}},}\ }\href {\doibase
  http://dx.doi.org/10.1038/ncomms4714 10.1038/ncomms4714} {\bibfield
  {journal} {\bibinfo  {journal} {Nat. Commun.}\ }\textbf {\bibinfo {volume}
  {5}},\ \bibinfo {pages} {3714} (\bibinfo {year} {2014})}\BibitemShut
  {NoStop}%
\bibitem [{\citenamefont {Lee}\ \emph {et~al.}(2014)\citenamefont {Lee},
  \citenamefont {Lee}, \citenamefont {Nowadnick}, \citenamefont {Gerber},
  \citenamefont {Tabis}, \citenamefont {Huang}, \citenamefont {Strocov},
  \citenamefont {Motoyama}, \citenamefont {Yu}, \citenamefont {Moritz},
  \citenamefont {Huang}, \citenamefont {Wang}, \citenamefont {Huang},
  \citenamefont {Wu}, \citenamefont {Chen}, \citenamefont {Huang},
  \citenamefont {Greven}, \citenamefont {Schmitt}, \citenamefont {Shen},\ and\
  \citenamefont {Devereaux}}]{wslee14}%
  \BibitemOpen
  \bibfield  {author} {\bibinfo {author} {\bibfnamefont {W.~S.}\ \bibnamefont
  {Lee}}, \bibinfo {author} {\bibfnamefont {J.~J.}\ \bibnamefont {Lee}},
  \bibinfo {author} {\bibfnamefont {E.~A.}\ \bibnamefont {Nowadnick}}, \bibinfo
  {author} {\bibfnamefont {S.}~\bibnamefont {Gerber}}, \bibinfo {author}
  {\bibfnamefont {W.}~\bibnamefont {Tabis}}, \bibinfo {author} {\bibfnamefont
  {S.~W.}\ \bibnamefont {Huang}}, \bibinfo {author} {\bibfnamefont {V.~N.}\
  \bibnamefont {Strocov}}, \bibinfo {author} {\bibfnamefont {E.~M.}\
  \bibnamefont {Motoyama}}, \bibinfo {author} {\bibfnamefont {G.}~\bibnamefont
  {Yu}}, \bibinfo {author} {\bibfnamefont {B.}~\bibnamefont {Moritz}}, \bibinfo
  {author} {\bibfnamefont {H.~Y.}\ \bibnamefont {Huang}}, \bibinfo {author}
  {\bibfnamefont {R.~P.}\ \bibnamefont {Wang}}, \bibinfo {author}
  {\bibfnamefont {Y.~B.}\ \bibnamefont {Huang}}, \bibinfo {author}
  {\bibfnamefont {W.~B.}\ \bibnamefont {Wu}}, \bibinfo {author} {\bibfnamefont
  {C.~T.}\ \bibnamefont {Chen}}, \bibinfo {author} {\bibfnamefont {D.~J.}\
  \bibnamefont {Huang}}, \bibinfo {author} {\bibfnamefont {M.}~\bibnamefont
  {Greven}}, \bibinfo {author} {\bibfnamefont {T.}~\bibnamefont {Schmitt}},
  \bibinfo {author} {\bibfnamefont {Z.~X.}\ \bibnamefont {Shen}}, \ and\
  \bibinfo {author} {\bibfnamefont {T.~P.}\ \bibnamefont {Devereaux}},\
  }\bibfield  {title} {\enquote {\bibinfo {title} {{Asymmetry of collective
  excitations in electron- and hole-doped cuprate superconductors}},}\ }\href
  {\doibase http://dx.doi.org/10.1038/nphys3117 10.1038/nphys3117} {\bibfield
  {journal} {\bibinfo  {journal} {Nat. Phys.}\ }\textbf {\bibinfo {volume}
  {10}},\ \bibinfo {pages} {883--889} (\bibinfo {year} {2014})}\BibitemShut
  {NoStop}%
\bibitem [{\citenamefont {Greco}\ \emph {et~al.}(2016)\citenamefont {Greco},
  \citenamefont {Yamase},\ and\ \citenamefont {Bejas}}]{greco16}%
  \BibitemOpen
  \bibfield  {author} {\bibinfo {author} {\bibfnamefont {Andr\'es}\
  \bibnamefont {Greco}}, \bibinfo {author} {\bibfnamefont {Hiroyuki}\
  \bibnamefont {Yamase}}, \ and\ \bibinfo {author} {\bibfnamefont
  {Mat\'{\i}as}\ \bibnamefont {Bejas}},\ }\bibfield  {title} {\enquote
  {\bibinfo {title} {{Plasmon excitations in layered high-${T}_{c}$
  cuprates}},}\ }\href {\doibase 10.1103/PhysRevB.94.075139} {\bibfield
  {journal} {\bibinfo  {journal} {Phys. Rev. B}\ }\textbf {\bibinfo {volume}
  {94}},\ \bibinfo {pages} {075139} (\bibinfo {year} {2016})}\BibitemShut
  {NoStop}%
\bibitem [{\citenamefont {Ishii}\ \emph {et~al.}(2017)\citenamefont {Ishii},
  \citenamefont {Tohyama}, \citenamefont {Asano}, \citenamefont {Sato},
  \citenamefont {Fujita}, \citenamefont {Wakimoto}, \citenamefont {Tustsui},
  \citenamefont {Sota}, \citenamefont {Miyawaki}, \citenamefont {Niwa},
  \citenamefont {Harada}, \citenamefont {Pelliciari}, \citenamefont {Huang},
  \citenamefont {Schmitt}, \citenamefont {Yamamoto},\ and\ \citenamefont
  {Mizuki}}]{ishii17}%
  \BibitemOpen
  \bibfield  {author} {\bibinfo {author} {\bibfnamefont {Kenji}\ \bibnamefont
  {Ishii}}, \bibinfo {author} {\bibfnamefont {Takami}\ \bibnamefont {Tohyama}},
  \bibinfo {author} {\bibfnamefont {Shun}\ \bibnamefont {Asano}}, \bibinfo
  {author} {\bibfnamefont {Kentaro}\ \bibnamefont {Sato}}, \bibinfo {author}
  {\bibfnamefont {Masaki}\ \bibnamefont {Fujita}}, \bibinfo {author}
  {\bibfnamefont {Shuichi}\ \bibnamefont {Wakimoto}}, \bibinfo {author}
  {\bibfnamefont {Kenji}\ \bibnamefont {Tustsui}}, \bibinfo {author}
  {\bibfnamefont {Shigetoshi}\ \bibnamefont {Sota}}, \bibinfo {author}
  {\bibfnamefont {Jun}\ \bibnamefont {Miyawaki}}, \bibinfo {author}
  {\bibfnamefont {Hideharu}\ \bibnamefont {Niwa}}, \bibinfo {author}
  {\bibfnamefont {Yoshihisa}\ \bibnamefont {Harada}}, \bibinfo {author}
  {\bibfnamefont {Jonathan}\ \bibnamefont {Pelliciari}}, \bibinfo {author}
  {\bibfnamefont {Yaobo}\ \bibnamefont {Huang}}, \bibinfo {author}
  {\bibfnamefont {Thorsten}\ \bibnamefont {Schmitt}}, \bibinfo {author}
  {\bibfnamefont {Yoshiya}\ \bibnamefont {Yamamoto}}, \ and\ \bibinfo {author}
  {\bibfnamefont {Jun'ichiro}\ \bibnamefont {Mizuki}},\ }\bibfield  {title}
  {\enquote {\bibinfo {title} {{Observation of momentum-dependent charge
  excitations in hole-doped cuprates using resonant inelastic x-ray scattering
  at the oxygen $K$ edge}},}\ }\href {\doibase 10.1103/PhysRevB.96.115148}
  {\bibfield  {journal} {\bibinfo  {journal} {Phys. Rev. B}\ }\textbf {\bibinfo
  {volume} {96}},\ \bibinfo {pages} {115148} (\bibinfo {year}
  {2017})}\BibitemShut {NoStop}%
\bibitem [{\citenamefont {Dellea}\ \emph {et~al.}(2017)\citenamefont {Dellea},
  \citenamefont {Minola}, \citenamefont {Galdi}, \citenamefont {Di~Castro},
  \citenamefont {Aruta}, \citenamefont {Brookes}, \citenamefont {Jia},
  \citenamefont {Mazzoli}, \citenamefont {Moretti~Sala}, \citenamefont
  {Moritz}, \citenamefont {Orgiani}, \citenamefont {Schlom}, \citenamefont
  {Tebano}, \citenamefont {Balestrino}, \citenamefont {Braicovich},
  \citenamefont {Devereaux}, \citenamefont {Maritato},\ and\ \citenamefont
  {Ghiringhelli}}]{dellea17}%
  \BibitemOpen
  \bibfield  {author} {\bibinfo {author} {\bibfnamefont {G.}~\bibnamefont
  {Dellea}}, \bibinfo {author} {\bibfnamefont {M.}~\bibnamefont {Minola}},
  \bibinfo {author} {\bibfnamefont {A.}~\bibnamefont {Galdi}}, \bibinfo
  {author} {\bibfnamefont {D.}~\bibnamefont {Di~Castro}}, \bibinfo {author}
  {\bibfnamefont {C.}~\bibnamefont {Aruta}}, \bibinfo {author} {\bibfnamefont
  {N.~B.}\ \bibnamefont {Brookes}}, \bibinfo {author} {\bibfnamefont {C.~J.}\
  \bibnamefont {Jia}}, \bibinfo {author} {\bibfnamefont {C.}~\bibnamefont
  {Mazzoli}}, \bibinfo {author} {\bibfnamefont {M.}~\bibnamefont
  {Moretti~Sala}}, \bibinfo {author} {\bibfnamefont {B.}~\bibnamefont
  {Moritz}}, \bibinfo {author} {\bibfnamefont {P.}~\bibnamefont {Orgiani}},
  \bibinfo {author} {\bibfnamefont {D.~G.}\ \bibnamefont {Schlom}}, \bibinfo
  {author} {\bibfnamefont {A.}~\bibnamefont {Tebano}}, \bibinfo {author}
  {\bibfnamefont {G.}~\bibnamefont {Balestrino}}, \bibinfo {author}
  {\bibfnamefont {L.}~\bibnamefont {Braicovich}}, \bibinfo {author}
  {\bibfnamefont {T.~P.}\ \bibnamefont {Devereaux}}, \bibinfo {author}
  {\bibfnamefont {L.}~\bibnamefont {Maritato}}, \ and\ \bibinfo {author}
  {\bibfnamefont {G.}~\bibnamefont {Ghiringhelli}},\ }\bibfield  {title}
  {\enquote {\bibinfo {title} {Spin and charge excitations in artificial hole-
  and electron-doped infinite layer cuprate superconductors},}\ }\href
  {\doibase 10.1103/PhysRevB.96.115117} {\bibfield  {journal} {\bibinfo
  {journal} {Phys. Rev. B}\ }\textbf {\bibinfo {volume} {96}},\ \bibinfo
  {pages} {115117} (\bibinfo {year} {2017})}\BibitemShut {NoStop}%
\bibitem [{\citenamefont {Hepting}\ \emph {et~al.}(2018)\citenamefont
  {Hepting}, \citenamefont {Chaix}, \citenamefont {Huang}, \citenamefont
  {Fumagalli}, \citenamefont {Peng}, \citenamefont {Moritz}, \citenamefont
  {Kummer}, \citenamefont {Brookes}, \citenamefont {Lee}, \citenamefont
  {Hashimoto}, \citenamefont {Sarkar}, \citenamefont {He}, \citenamefont
  {Rotundu}, \citenamefont {Lee}, \citenamefont {Greene}, \citenamefont
  {Braicovich}, \citenamefont {Ghiringhelli}, \citenamefont {Shen},
  \citenamefont {Devereaux},\ and\ \citenamefont {Lee}}]{hepting18}%
  \BibitemOpen
  \bibfield  {author} {\bibinfo {author} {\bibfnamefont {M.}~\bibnamefont
  {Hepting}}, \bibinfo {author} {\bibfnamefont {L.}~\bibnamefont {Chaix}},
  \bibinfo {author} {\bibfnamefont {E.~W.}\ \bibnamefont {Huang}}, \bibinfo
  {author} {\bibfnamefont {R.}~\bibnamefont {Fumagalli}}, \bibinfo {author}
  {\bibfnamefont {Y.~Y.}\ \bibnamefont {Peng}}, \bibinfo {author}
  {\bibfnamefont {B.}~\bibnamefont {Moritz}}, \bibinfo {author} {\bibfnamefont
  {K.}~\bibnamefont {Kummer}}, \bibinfo {author} {\bibfnamefont {N.~B.}\
  \bibnamefont {Brookes}}, \bibinfo {author} {\bibfnamefont {W.~C.}\
  \bibnamefont {Lee}}, \bibinfo {author} {\bibfnamefont {M.}~\bibnamefont
  {Hashimoto}}, \bibinfo {author} {\bibfnamefont {T.}~\bibnamefont {Sarkar}},
  \bibinfo {author} {\bibfnamefont {J.-F.}\ \bibnamefont {He}}, \bibinfo
  {author} {\bibfnamefont {C.~R.}\ \bibnamefont {Rotundu}}, \bibinfo {author}
  {\bibfnamefont {Y.~S.}\ \bibnamefont {Lee}}, \bibinfo {author} {\bibfnamefont
  {R.~L.}\ \bibnamefont {Greene}}, \bibinfo {author} {\bibfnamefont
  {L.}~\bibnamefont {Braicovich}}, \bibinfo {author} {\bibfnamefont
  {G.}~\bibnamefont {Ghiringhelli}}, \bibinfo {author} {\bibfnamefont {Z.~X.}\
  \bibnamefont {Shen}}, \bibinfo {author} {\bibfnamefont {T.~P.}\ \bibnamefont
  {Devereaux}}, \ and\ \bibinfo {author} {\bibfnamefont {W.~S.}\ \bibnamefont
  {Lee}},\ }\bibfield  {title} {\enquote {\bibinfo {title} {Three-dimensional
  collective charge excitations in electron-doped copper oxide
  superconductors},}\ }\href {\doibase 10.1038/s41586-018-0648-3} {\bibfield
  {journal} {\bibinfo  {journal} {Nature}\ }\textbf {\bibinfo {volume} {563}},\
  \bibinfo {pages} {374--378} (\bibinfo {year} {2018})}\BibitemShut {NoStop}%
\bibitem [{\citenamefont {Greco}\ \emph {et~al.}(2019)\citenamefont {Greco},
  \citenamefont {Yamase},\ and\ \citenamefont {Bejas}}]{greco19}%
  \BibitemOpen
  \bibfield  {author} {\bibinfo {author} {\bibfnamefont {Andr{\'e}s}\
  \bibnamefont {Greco}}, \bibinfo {author} {\bibfnamefont {Hiroyuki}\
  \bibnamefont {Yamase}}, \ and\ \bibinfo {author} {\bibfnamefont {Mat{\'i}as}\
  \bibnamefont {Bejas}},\ }\bibfield  {title} {\enquote {\bibinfo {title}
  {Origin of high-energy charge excitations observed by resonant inelastic
  x-ray scattering in cuprate superconductors},}\ }\href {\doibase
  10.1038/s42005-018-0099-z} {\bibfield  {journal} {\bibinfo  {journal}
  {Commun. Phys.}\ }\textbf {\bibinfo {volume} {2}},\ \bibinfo {pages} {3}
  (\bibinfo {year} {2019})}\BibitemShut {NoStop}%
\bibitem [{\citenamefont {Hepting}\ \emph {et~al.}(2022)\citenamefont
  {Hepting}, \citenamefont {Bejas}, \citenamefont {Nag}, \citenamefont
  {Yamase}, \citenamefont {Coppola}, \citenamefont {Betto}, \citenamefont
  {Falter}, \citenamefont {Garcia-Fernandez}, \citenamefont {Agrestini},
  \citenamefont {Zhou}, \citenamefont {Minola}, \citenamefont {Sacco},
  \citenamefont {Maritato}, \citenamefont {Orgiani}, \citenamefont {Wei},
  \citenamefont {Shen}, \citenamefont {Schlom}, \citenamefont {Galdi},
  \citenamefont {Greco},\ and\ \citenamefont {Keimer}}]{hepting22}%
  \BibitemOpen
  \bibfield  {author} {\bibinfo {author} {\bibfnamefont {M.}~\bibnamefont
  {Hepting}}, \bibinfo {author} {\bibfnamefont {M.}~\bibnamefont {Bejas}},
  \bibinfo {author} {\bibfnamefont {A.}~\bibnamefont {Nag}}, \bibinfo {author}
  {\bibfnamefont {H.}~\bibnamefont {Yamase}}, \bibinfo {author} {\bibfnamefont
  {N.}~\bibnamefont {Coppola}}, \bibinfo {author} {\bibfnamefont
  {D.}~\bibnamefont {Betto}}, \bibinfo {author} {\bibfnamefont
  {C.}~\bibnamefont {Falter}}, \bibinfo {author} {\bibfnamefont
  {M.}~\bibnamefont {Garcia-Fernandez}}, \bibinfo {author} {\bibfnamefont
  {S.}~\bibnamefont {Agrestini}}, \bibinfo {author} {\bibfnamefont {Ke-Jin}\
  \bibnamefont {Zhou}}, \bibinfo {author} {\bibfnamefont {M.}~\bibnamefont
  {Minola}}, \bibinfo {author} {\bibfnamefont {C.}~\bibnamefont {Sacco}},
  \bibinfo {author} {\bibfnamefont {L.}~\bibnamefont {Maritato}}, \bibinfo
  {author} {\bibfnamefont {P.}~\bibnamefont {Orgiani}}, \bibinfo {author}
  {\bibfnamefont {H.~I.}\ \bibnamefont {Wei}}, \bibinfo {author} {\bibfnamefont
  {K.~M.}\ \bibnamefont {Shen}}, \bibinfo {author} {\bibfnamefont {D.~G.}\
  \bibnamefont {Schlom}}, \bibinfo {author} {\bibfnamefont {A.}~\bibnamefont
  {Galdi}}, \bibinfo {author} {\bibfnamefont {A.}~\bibnamefont {Greco}}, \ and\
  \bibinfo {author} {\bibfnamefont {B.}~\bibnamefont {Keimer}},\ }\bibfield
  {title} {\enquote {\bibinfo {title} {Gapped collective charge excitations and
  interlayer hopping in cuprate superconductors},}\ }\href {\doibase
  10.1103/PhysRevLett.129.047001} {\bibfield  {journal} {\bibinfo  {journal}
  {Phys. Rev. Lett.}\ }\textbf {\bibinfo {volume} {129}},\ \bibinfo {pages}
  {047001} (\bibinfo {year} {2022})}\BibitemShut {NoStop}%
\bibitem [{\citenamefont {Greco}\ \emph {et~al.}(2020)\citenamefont {Greco},
  \citenamefont {Yamase},\ and\ \citenamefont {Bejas}}]{greco20}%
  \BibitemOpen
  \bibfield  {author} {\bibinfo {author} {\bibfnamefont {Andr\'es}\
  \bibnamefont {Greco}}, \bibinfo {author} {\bibfnamefont {Hiroyuki}\
  \bibnamefont {Yamase}}, \ and\ \bibinfo {author} {\bibfnamefont
  {Mat\'{\i}as}\ \bibnamefont {Bejas}},\ }\bibfield  {title} {\enquote
  {\bibinfo {title} {Close inspection of plasmon excitations in cuprate
  superconductors},}\ }\href {\doibase 10.1103/PhysRevB.102.024509} {\bibfield
  {journal} {\bibinfo  {journal} {Phys. Rev. B}\ }\textbf {\bibinfo {volume}
  {102}},\ \bibinfo {pages} {024509} (\bibinfo {year} {2020})}\BibitemShut
  {NoStop}%
\bibitem [{\citenamefont {Fetter}(1974)}]{fetter74}%
  \BibitemOpen
  \bibfield  {author} {\bibinfo {author} {\bibfnamefont {Alexander~L}\
  \bibnamefont {Fetter}},\ }\bibfield  {title} {\enquote {\bibinfo {title}
  {Electrodynamics of a layered electron gas. {II}. periodic array},}\ }\href
  {\doibase https://doi.org/10.1016/0003-4916(74)90397-2} {\bibfield  {journal}
  {\bibinfo  {journal} {Annals of Physics}\ }\textbf {\bibinfo {volume} {88}},\
  \bibinfo {pages} {1--25} (\bibinfo {year} {1974})}\BibitemShut {NoStop}%
\bibitem [{\citenamefont {Griffin}\ and\ \citenamefont
  {Pindor}(1989)}]{griffin89}%
  \BibitemOpen
  \bibfield  {author} {\bibinfo {author} {\bibfnamefont {A.}~\bibnamefont
  {Griffin}}\ and\ \bibinfo {author} {\bibfnamefont {A.~J.}\ \bibnamefont
  {Pindor}},\ }\bibfield  {title} {\enquote {\bibinfo {title} {Plasmon
  dispersion relations and the induced electron interaction in oxide
  superconductors: Numerical results},}\ }\href {\doibase
  10.1103/PhysRevB.39.11503} {\bibfield  {journal} {\bibinfo  {journal} {Phys.
  Rev. B}\ }\textbf {\bibinfo {volume} {39}},\ \bibinfo {pages} {11503--11514}
  (\bibinfo {year} {1989})}\BibitemShut {NoStop}%
\bibitem [{\citenamefont {Iyo}\ \emph {et~al.}(2007)\citenamefont {Iyo},
  \citenamefont {Tanaka}, \citenamefont {Kito}, \citenamefont {Kodama},
  \citenamefont {M.~Shirage}, \citenamefont {D.~Shivagan}, \citenamefont
  {Matsuhata}, \citenamefont {Tokiwa},\ and\ \citenamefont {Watanabe}}]{iyo07}%
  \BibitemOpen
  \bibfield  {author} {\bibinfo {author} {\bibfnamefont {Akira}\ \bibnamefont
  {Iyo}}, \bibinfo {author} {\bibfnamefont {Yasumoto}\ \bibnamefont {Tanaka}},
  \bibinfo {author} {\bibfnamefont {Hijiri}\ \bibnamefont {Kito}}, \bibinfo
  {author} {\bibfnamefont {Yasuharu}\ \bibnamefont {Kodama}}, \bibinfo {author}
  {\bibfnamefont {Parasharam}\ \bibnamefont {M.~Shirage}}, \bibinfo {author}
  {\bibfnamefont {Dilip}\ \bibnamefont {D.~Shivagan}}, \bibinfo {author}
  {\bibfnamefont {Hirofumi}\ \bibnamefont {Matsuhata}}, \bibinfo {author}
  {\bibfnamefont {Kazuyasu}\ \bibnamefont {Tokiwa}}, \ and\ \bibinfo {author}
  {\bibfnamefont {Tsuneo}\ \bibnamefont {Watanabe}},\ }\bibfield  {title}
  {\enquote {\bibinfo {title} {{Tc vs n Relationship for Multilayered High-Tc
  Superconductors}},}\ }\href {\doibase 10.1143/JPSJ.76.094711} {\bibfield
  {journal} {\bibinfo  {journal} {Journal of the Physical Society of Japan}\
  }\textbf {\bibinfo {volume} {76}},\ \bibinfo {pages} {094711} (\bibinfo
  {year} {2007})}\BibitemShut {NoStop}%
\bibitem [{\citenamefont {Bejas}\ \emph {et~al.}(2024)\citenamefont {Bejas},
  \citenamefont {Zimmermann}, \citenamefont {Betto}, \citenamefont {Boyko},
  \citenamefont {Green}, \citenamefont {Loew}, \citenamefont {Brookes},
  \citenamefont {Cristiani}, \citenamefont {Logvenov}, \citenamefont {Minola},
  \citenamefont {Keimer}, \citenamefont {Yamase}, \citenamefont {Greco},\ and\
  \citenamefont {Hepting}}]{bejas24}%
  \BibitemOpen
  \bibfield  {author} {\bibinfo {author} {\bibfnamefont {M.}~\bibnamefont
  {Bejas}}, \bibinfo {author} {\bibfnamefont {V.}~\bibnamefont {Zimmermann}},
  \bibinfo {author} {\bibfnamefont {D.}~\bibnamefont {Betto}}, \bibinfo
  {author} {\bibfnamefont {T.~D.}\ \bibnamefont {Boyko}}, \bibinfo {author}
  {\bibfnamefont {R.~J.}\ \bibnamefont {Green}}, \bibinfo {author}
  {\bibfnamefont {T.}~\bibnamefont {Loew}}, \bibinfo {author} {\bibfnamefont
  {N.~B.}\ \bibnamefont {Brookes}}, \bibinfo {author} {\bibfnamefont
  {G.}~\bibnamefont {Cristiani}}, \bibinfo {author} {\bibfnamefont
  {G.}~\bibnamefont {Logvenov}}, \bibinfo {author} {\bibfnamefont
  {M.}~\bibnamefont {Minola}}, \bibinfo {author} {\bibfnamefont
  {B.}~\bibnamefont {Keimer}}, \bibinfo {author} {\bibfnamefont
  {H.}~\bibnamefont {Yamase}}, \bibinfo {author} {\bibfnamefont
  {A.}~\bibnamefont {Greco}}, \ and\ \bibinfo {author} {\bibfnamefont
  {M.}~\bibnamefont {Hepting}},\ }\bibfield  {title} {\enquote {\bibinfo
  {title} {Plasmon dispersion in bilayer cuprate superconductors},}\ }\href
  {\doibase 10.1103/PhysRevB.109.144516} {\bibfield  {journal} {\bibinfo
  {journal} {Phys. Rev. B}\ }\textbf {\bibinfo {volume} {109}},\ \bibinfo
  {pages} {144516} (\bibinfo {year} {2024})}\BibitemShut {NoStop}%
\bibitem [{\citenamefont {Yamase}(2025)}]{yamase25}%
  \BibitemOpen
  \bibfield  {author} {\bibinfo {author} {\bibfnamefont {Hiroyuki}\
  \bibnamefont {Yamase}},\ }\bibfield  {title} {\enquote {\bibinfo {title}
  {Theory of charge dynamics in bilayer electron system with long-range coulomb
  interaction},}\ }\href {\doibase 10.1103/PhysRevB.111.085138} {\bibfield
  {journal} {\bibinfo  {journal} {Phys. Rev. B}\ }\textbf {\bibinfo {volume}
  {111}},\ \bibinfo {pages} {085138} (\bibinfo {year} {2025})}\BibitemShut
  {NoStop}%
\bibitem [{\citenamefont {Sellati}\ and\ \citenamefont
  {Benfatto}(2025)}]{sellati25}%
  \BibitemOpen
  \bibfield  {author} {\bibinfo {author} {\bibfnamefont {Niccol\`o}\
  \bibnamefont {Sellati}}\ and\ \bibinfo {author} {\bibfnamefont {Lara}\
  \bibnamefont {Benfatto}},\ }\bibfield  {title} {\enquote {\bibinfo {title}
  {Ghost josephson plasmon in bilayer superconductors},}\ }\href {\doibase
  10.1103/PhysRevB.111.104509} {\bibfield  {journal} {\bibinfo  {journal}
  {Phys. Rev. B}\ }\textbf {\bibinfo {volume} {111}},\ \bibinfo {pages}
  {104509} (\bibinfo {year} {2025})}\BibitemShut {NoStop}%
\bibitem [{\citenamefont {Nakata}\ \emph {et~al.}(2025)\citenamefont {Nakata},
  \citenamefont {Bejas}, \citenamefont {Okamoto}, \citenamefont {Yamamoto},
  \citenamefont {Shiga}, \citenamefont {Takahashi}, \citenamefont {Huang},
  \citenamefont {Kumigashira}, \citenamefont {Wadati}, \citenamefont
  {Miyawaki}, \citenamefont {Ishida}, \citenamefont {Eisaki}, \citenamefont
  {Fujimori}, \citenamefont {Greco}, \citenamefont {Yamase}, \citenamefont
  {Huang},\ and\ \citenamefont {Suzuki}}]{nakata25}%
  \BibitemOpen
  \bibfield  {author} {\bibinfo {author} {\bibfnamefont {S.}~\bibnamefont
  {Nakata}}, \bibinfo {author} {\bibfnamefont {M.}~\bibnamefont {Bejas}},
  \bibinfo {author} {\bibfnamefont {J.}~\bibnamefont {Okamoto}}, \bibinfo
  {author} {\bibfnamefont {K.}~\bibnamefont {Yamamoto}}, \bibinfo {author}
  {\bibfnamefont {D.}~\bibnamefont {Shiga}}, \bibinfo {author} {\bibfnamefont
  {R.}~\bibnamefont {Takahashi}}, \bibinfo {author} {\bibfnamefont {H.~Y.}\
  \bibnamefont {Huang}}, \bibinfo {author} {\bibfnamefont {H.}~\bibnamefont
  {Kumigashira}}, \bibinfo {author} {\bibfnamefont {H.}~\bibnamefont {Wadati}},
  \bibinfo {author} {\bibfnamefont {J.}~\bibnamefont {Miyawaki}}, \bibinfo
  {author} {\bibfnamefont {S.}~\bibnamefont {Ishida}}, \bibinfo {author}
  {\bibfnamefont {H.}~\bibnamefont {Eisaki}}, \bibinfo {author} {\bibfnamefont
  {A.}~\bibnamefont {Fujimori}}, \bibinfo {author} {\bibfnamefont
  {A.}~\bibnamefont {Greco}}, \bibinfo {author} {\bibfnamefont
  {H.}~\bibnamefont {Yamase}}, \bibinfo {author} {\bibfnamefont {D.~J.}\
  \bibnamefont {Huang}}, \ and\ \bibinfo {author} {\bibfnamefont
  {H.}~\bibnamefont {Suzuki}},\ }\bibfield  {title} {\enquote {\bibinfo {title}
  {{Out-of-phase plasmon excitations in the trilayer cuprate
  ${\mathrm{Bi}}_{2}{\mathrm{Sr}}_{2}{\mathrm{Ca}}_{2}{\mathrm{Cu}}_{3}{\mathrm{O}}_{10+\ensuremath{\delta}}$}},}\
  }\href {\doibase 10.1103/PhysRevB.111.165141} {\bibfield  {journal} {\bibinfo
   {journal} {Phys. Rev. B}\ }\textbf {\bibinfo {volume} {111}},\ \bibinfo
  {pages} {165141} (\bibinfo {year} {2025})}\BibitemShut {NoStop}%
\bibitem [{\citenamefont {Mitrano}\ \emph {et~al.}(2018)\citenamefont
  {Mitrano}, \citenamefont {Husain}, \citenamefont {Vig}, \citenamefont
  {Kogar}, \citenamefont {Rak}, \citenamefont {Rubeck}, \citenamefont
  {Schmalian}, \citenamefont {Uchoa}, \citenamefont {Schneeloch}, \citenamefont
  {Zhong}, \citenamefont {Gu},\ and\ \citenamefont {Abbamonte}}]{mitrano18}%
  \BibitemOpen
  \bibfield  {author} {\bibinfo {author} {\bibfnamefont {M.}~\bibnamefont
  {Mitrano}}, \bibinfo {author} {\bibfnamefont {A.~A.}\ \bibnamefont {Husain}},
  \bibinfo {author} {\bibfnamefont {S.}~\bibnamefont {Vig}}, \bibinfo {author}
  {\bibfnamefont {A.}~\bibnamefont {Kogar}}, \bibinfo {author} {\bibfnamefont
  {M.~S.}\ \bibnamefont {Rak}}, \bibinfo {author} {\bibfnamefont {S.~I.}\
  \bibnamefont {Rubeck}}, \bibinfo {author} {\bibfnamefont {J.}~\bibnamefont
  {Schmalian}}, \bibinfo {author} {\bibfnamefont {B.}~\bibnamefont {Uchoa}},
  \bibinfo {author} {\bibfnamefont {J.}~\bibnamefont {Schneeloch}}, \bibinfo
  {author} {\bibfnamefont {R.}~\bibnamefont {Zhong}}, \bibinfo {author}
  {\bibfnamefont {G.~D.}\ \bibnamefont {Gu}}, \ and\ \bibinfo {author}
  {\bibfnamefont {P.}~\bibnamefont {Abbamonte}},\ }\bibfield  {title} {\enquote
  {\bibinfo {title} {Anomalous density fluctuations in a strange metal},}\
  }\href {\doibase 10.1073/pnas.1721495115} {\bibfield  {journal} {\bibinfo
  {journal} {Proc. Natl. Acad. Sci. U. S. A.}\ }\textbf {\bibinfo {volume}
  {115}},\ \bibinfo {pages} {5392--5396} (\bibinfo {year} {2018})}\BibitemShut
  {NoStop}%
\bibitem [{\citenamefont {Romero-Berm\'{u}dez}\ \emph
  {et~al.}(2019)\citenamefont {Romero-Berm\'{u}dez}, \citenamefont {Krikun},
  \citenamefont {Schalm},\ and\ \citenamefont {Zaanen}}]{romero-bermudez19}%
  \BibitemOpen
  \bibfield  {author} {\bibinfo {author} {\bibfnamefont {Aurelio}\ \bibnamefont
  {Romero-Berm\'{u}dez}}, \bibinfo {author} {\bibfnamefont {Alexander}\
  \bibnamefont {Krikun}}, \bibinfo {author} {\bibfnamefont {Koenraad}\
  \bibnamefont {Schalm}}, \ and\ \bibinfo {author} {\bibfnamefont {Jan}\
  \bibnamefont {Zaanen}},\ }\bibfield  {title} {\enquote {\bibinfo {title}
  {Anomalous attenuation of plasmons in strange metals and holography},}\
  }\href {\doibase 10.1103/PhysRevB.99.235149} {\bibfield  {journal} {\bibinfo
  {journal} {Phys. Rev. B}\ }\textbf {\bibinfo {volume} {99}},\ \bibinfo
  {pages} {235149} (\bibinfo {year} {2019})}\BibitemShut {NoStop}%
\bibitem [{\citenamefont {den Eede}\ \emph {et~al.}(2024)\citenamefont {den
  Eede}, \citenamefont {van Stralen}, \citenamefont {Flipse},\ and\
  \citenamefont {Stoof}}]{vandeneede24}%
  \BibitemOpen
  \bibfield  {author} {\bibinfo {author} {\bibfnamefont {S.~T.~Van}\
  \bibnamefont {den Eede}}, \bibinfo {author} {\bibfnamefont {T.~J.~N.}\
  \bibnamefont {van Stralen}}, \bibinfo {author} {\bibfnamefont {C.~F.~J.}\
  \bibnamefont {Flipse}}, \ and\ \bibinfo {author} {\bibfnamefont {H.~T.~C.}\
  \bibnamefont {Stoof}},\ }\bibfield  {title} {\enquote {\bibinfo {title}
  {Plasmons in a layered strange metal using the gauge-gravity duality},}\
  }\href {\doibase 10.1103/PhysRevB.109.085119} {\bibfield  {journal} {\bibinfo
   {journal} {Phys. Rev. B}\ }\textbf {\bibinfo {volume} {109}},\ \bibinfo
  {pages} {085119} (\bibinfo {year} {2024})}\BibitemShut {NoStop}%
\bibitem [{\citenamefont {Sun}\ \emph {et~al.}(2023)\citenamefont {Sun},
  \citenamefont {Huo}, \citenamefont {Hu}, \citenamefont {Li}, \citenamefont
  {Liu}, \citenamefont {Han}, \citenamefont {Tang}, \citenamefont {Mao},
  \citenamefont {Yang}, \citenamefont {Wang}, \citenamefont {Cheng},
  \citenamefont {Yao}, \citenamefont {Zhang},\ and\ \citenamefont
  {Wang}}]{sun23}%
  \BibitemOpen
  \bibfield  {author} {\bibinfo {author} {\bibfnamefont {H.}~\bibnamefont
  {Sun}}, \bibinfo {author} {\bibfnamefont {M.}~\bibnamefont {Huo}}, \bibinfo
  {author} {\bibfnamefont {X.}~\bibnamefont {Hu}}, \bibinfo {author}
  {\bibfnamefont {J.}~\bibnamefont {Li}}, \bibinfo {author} {\bibfnamefont
  {Z.}~\bibnamefont {Liu}}, \bibinfo {author} {\bibfnamefont {Y.}~\bibnamefont
  {Han}}, \bibinfo {author} {\bibfnamefont {L.}~\bibnamefont {Tang}}, \bibinfo
  {author} {\bibfnamefont {Z.}~\bibnamefont {Mao}}, \bibinfo {author}
  {\bibfnamefont {P.}~\bibnamefont {Yang}}, \bibinfo {author} {\bibfnamefont
  {B.}~\bibnamefont {Wang}}, \bibinfo {author} {\bibfnamefont {J.}~\bibnamefont
  {Cheng}}, \bibinfo {author} {\bibfnamefont {D.-X.}\ \bibnamefont {Yao}},
  \bibinfo {author} {\bibfnamefont {G.-M.}\ \bibnamefont {Zhang}}, \ and\
  \bibinfo {author} {\bibfnamefont {M.}~\bibnamefont {Wang}},\ }\bibfield
  {title} {\enquote {\bibinfo {title} {{Signatures of superconductivity near 80
  K in a nickelate under high pressure}},}\ }\href {\doibase
  10.1038/s41586-023-06408-7} {\bibfield  {journal} {\bibinfo  {journal}
  {Nature}\ }\textbf {\bibinfo {volume} {621}},\ \bibinfo {pages} {493--498}
  (\bibinfo {year} {2023})}\BibitemShut {NoStop}%
\bibitem [{\citenamefont {Hou}\ \emph {et~al.}(2023)\citenamefont {Hou},
  \citenamefont {Yang}, \citenamefont {Liu}, \citenamefont {Li}, \citenamefont
  {Shan}, \citenamefont {Ma}, \citenamefont {Wang}, \citenamefont {Wang},
  \citenamefont {Guo}, \citenamefont {Sun}, \citenamefont {Uwatoko},
  \citenamefont {Wang}, \citenamefont {Zhang}, \citenamefont {Wang},\ and\
  \citenamefont {Cheng}}]{hou23}%
  \BibitemOpen
  \bibfield  {author} {\bibinfo {author} {\bibfnamefont {J.}~\bibnamefont
  {Hou}}, \bibinfo {author} {\bibfnamefont {P.-T.}\ \bibnamefont {Yang}},
  \bibinfo {author} {\bibfnamefont {Z.-Y.}\ \bibnamefont {Liu}}, \bibinfo
  {author} {\bibfnamefont {J.-Y.}\ \bibnamefont {Li}}, \bibinfo {author}
  {\bibfnamefont {P.-F.}\ \bibnamefont {Shan}}, \bibinfo {author}
  {\bibfnamefont {L.}~\bibnamefont {Ma}}, \bibinfo {author} {\bibfnamefont
  {G.}~\bibnamefont {Wang}}, \bibinfo {author} {\bibfnamefont {N.-N.}\
  \bibnamefont {Wang}}, \bibinfo {author} {\bibfnamefont {H.-Z.}\ \bibnamefont
  {Guo}}, \bibinfo {author} {\bibfnamefont {J.-P.}\ \bibnamefont {Sun}},
  \bibinfo {author} {\bibfnamefont {Y.}~\bibnamefont {Uwatoko}}, \bibinfo
  {author} {\bibfnamefont {M.}~\bibnamefont {Wang}}, \bibinfo {author}
  {\bibfnamefont {G.-M.}\ \bibnamefont {Zhang}}, \bibinfo {author}
  {\bibfnamefont {B.-S.}\ \bibnamefont {Wang}}, \ and\ \bibinfo {author}
  {\bibfnamefont {J.-G.}\ \bibnamefont {Cheng}},\ }\bibfield  {title} {\enquote
  {\bibinfo {title} {{Emergence of High-Temperature Superconducting Phase in
  Pressurized La3Ni2O7 Crystals}},}\ }\href {\doibase
  10.1088/0256-307X/40/11/117302} {\bibfield  {journal} {\bibinfo  {journal}
  {Chinese Phys. Lett.}\ }\textbf {\bibinfo {volume} {40}},\ \bibinfo {pages}
  {117302} (\bibinfo {year} {2023})}\BibitemShut {NoStop}%
\bibitem [{\citenamefont {Foussats}\ and\ \citenamefont
  {Greco}(2002)}]{foussats02}%
  \BibitemOpen
  \bibfield  {author} {\bibinfo {author} {\bibfnamefont {Adriana}\ \bibnamefont
  {Foussats}}\ and\ \bibinfo {author} {\bibfnamefont {Andr\'es}\ \bibnamefont
  {Greco}},\ }\bibfield  {title} {\enquote {\bibinfo {title} {{Large-$N$
  expansion based on the Hubbard operator path integral representation and its
  application to the {$t$-$J$} model}},}\ }\href@noop {} {\bibfield  {journal}
  {\bibinfo  {journal} {Phys. Rev. B}\ }\textbf {\bibinfo {volume} {65}},\
  \bibinfo {pages} {195107} (\bibinfo {year} {2002})}\BibitemShut {NoStop}%
\bibitem [{\citenamefont {Nag}\ \emph {et~al.}(2020)\citenamefont {Nag},
  \citenamefont {Zhu}, \citenamefont {Bejas}, \citenamefont {Li}, \citenamefont
  {Robarts}, \citenamefont {Yamase}, \citenamefont {Petsch}, \citenamefont
  {Song}, \citenamefont {Eisaki}, \citenamefont {Walters}, \citenamefont
  {Garc\'{\i}a-Fern\'andez}, \citenamefont {Greco}, \citenamefont {Hayden},\
  and\ \citenamefont {Zhou}}]{nag20}%
  \BibitemOpen
  \bibfield  {author} {\bibinfo {author} {\bibfnamefont {Abhishek}\
  \bibnamefont {Nag}}, \bibinfo {author} {\bibfnamefont {M.}~\bibnamefont
  {Zhu}}, \bibinfo {author} {\bibfnamefont {Mat\'{\i}as}\ \bibnamefont
  {Bejas}}, \bibinfo {author} {\bibfnamefont {J.}~\bibnamefont {Li}}, \bibinfo
  {author} {\bibfnamefont {H.~C.}\ \bibnamefont {Robarts}}, \bibinfo {author}
  {\bibfnamefont {Hiroyuki}\ \bibnamefont {Yamase}}, \bibinfo {author}
  {\bibfnamefont {A.~N.}\ \bibnamefont {Petsch}}, \bibinfo {author}
  {\bibfnamefont {D.}~\bibnamefont {Song}}, \bibinfo {author} {\bibfnamefont
  {H.}~\bibnamefont {Eisaki}}, \bibinfo {author} {\bibfnamefont {A.~C.}\
  \bibnamefont {Walters}}, \bibinfo {author} {\bibfnamefont {M.}~\bibnamefont
  {Garc\'{\i}a-Fern\'andez}}, \bibinfo {author} {\bibfnamefont {Andr\'es}\
  \bibnamefont {Greco}}, \bibinfo {author} {\bibfnamefont {S.~M.}\ \bibnamefont
  {Hayden}}, \ and\ \bibinfo {author} {\bibfnamefont {Ke-Jin}\ \bibnamefont
  {Zhou}},\ }\bibfield  {title} {\enquote {\bibinfo {title} {{Detection of
  Acoustic Plasmons in Hole-Doped Lanthanum and Bismuth Cuprate Superconductors
  Using Resonant Inelastic X-Ray Scattering}},}\ }\href {\doibase
  10.1103/PhysRevLett.125.257002} {\bibfield  {journal} {\bibinfo  {journal}
  {Phys. Rev. Lett.}\ }\textbf {\bibinfo {volume} {125}},\ \bibinfo {pages}
  {257002} (\bibinfo {year} {2020})}\BibitemShut {NoStop}%
\bibitem [{\citenamefont {Yamase}(2021)}]{yamase21c}%
  \BibitemOpen
  \bibfield  {author} {\bibinfo {author} {\bibfnamefont {Hiroyuki}\
  \bibnamefont {Yamase}},\ }\bibfield  {title} {\enquote {\bibinfo {title}
  {Theoretical insights into electronic nematic order, bond-charge orders, and
  plasmons in cuprate superconductors},}\ }\href@noop {} {\bibfield  {journal}
  {\bibinfo  {journal} {J. Phys. Soc. Jpn.}\ }\textbf {\bibinfo {volume}
  {90}},\ \bibinfo {pages} {111011} (\bibinfo {year} {2021})}\BibitemShut
  {NoStop}%
\bibitem [{\citenamefont {Hepting}\ \emph {et~al.}(2023)\citenamefont
  {Hepting}, \citenamefont {Boyko}, \citenamefont {Zimmermann}, \citenamefont
  {Bejas}, \citenamefont {Suyolcu}, \citenamefont {Puphal}, \citenamefont
  {Green}, \citenamefont {Zinni}, \citenamefont {Kim}, \citenamefont {Casa},
  \citenamefont {Upton}, \citenamefont {Wong}, \citenamefont {Schulz},
  \citenamefont {Bartkowiak}, \citenamefont {Habicht}, \citenamefont
  {Pomjakushina}, \citenamefont {Cristiani}, \citenamefont {Logvenov},
  \citenamefont {Minola}, \citenamefont {Yamase}, \citenamefont {Greco},\ and\
  \citenamefont {Keimer}}]{hepting23}%
  \BibitemOpen
  \bibfield  {author} {\bibinfo {author} {\bibfnamefont {M.}~\bibnamefont
  {Hepting}}, \bibinfo {author} {\bibfnamefont {T.~D.}\ \bibnamefont {Boyko}},
  \bibinfo {author} {\bibfnamefont {V.}~\bibnamefont {Zimmermann}}, \bibinfo
  {author} {\bibfnamefont {M.}~\bibnamefont {Bejas}}, \bibinfo {author}
  {\bibfnamefont {Y.~E.}\ \bibnamefont {Suyolcu}}, \bibinfo {author}
  {\bibfnamefont {P.}~\bibnamefont {Puphal}}, \bibinfo {author} {\bibfnamefont
  {R.~J.}\ \bibnamefont {Green}}, \bibinfo {author} {\bibfnamefont
  {L.}~\bibnamefont {Zinni}}, \bibinfo {author} {\bibfnamefont
  {J.}~\bibnamefont {Kim}}, \bibinfo {author} {\bibfnamefont {D.}~\bibnamefont
  {Casa}}, \bibinfo {author} {\bibfnamefont {M.~H.}\ \bibnamefont {Upton}},
  \bibinfo {author} {\bibfnamefont {D.}~\bibnamefont {Wong}}, \bibinfo {author}
  {\bibfnamefont {C.}~\bibnamefont {Schulz}}, \bibinfo {author} {\bibfnamefont
  {M.}~\bibnamefont {Bartkowiak}}, \bibinfo {author} {\bibfnamefont
  {K.}~\bibnamefont {Habicht}}, \bibinfo {author} {\bibfnamefont
  {E.}~\bibnamefont {Pomjakushina}}, \bibinfo {author} {\bibfnamefont
  {G.}~\bibnamefont {Cristiani}}, \bibinfo {author} {\bibfnamefont
  {G.}~\bibnamefont {Logvenov}}, \bibinfo {author} {\bibfnamefont
  {M.}~\bibnamefont {Minola}}, \bibinfo {author} {\bibfnamefont
  {H.}~\bibnamefont {Yamase}}, \bibinfo {author} {\bibfnamefont
  {A.}~\bibnamefont {Greco}}, \ and\ \bibinfo {author} {\bibfnamefont
  {B.}~\bibnamefont {Keimer}},\ }\bibfield  {title} {\enquote {\bibinfo {title}
  {Evolution of plasmon excitations across the phase diagram of the cuprate
  superconductor
  {${\mathrm{La}}_{2\ensuremath{-}x}{\mathrm{Sr}}_{x}{\mathrm{CuO}}_{4}$}},}\
  }\href {\doibase 10.1103/PhysRevB.107.214516} {\bibfield  {journal} {\bibinfo
   {journal} {Phys. Rev. B}\ }\textbf {\bibinfo {volume} {107}},\ \bibinfo
  {pages} {214516} (\bibinfo {year} {2023})}\BibitemShut {NoStop}%
\bibitem [{\citenamefont {Nag}\ \emph {et~al.}(2024)\citenamefont {Nag},
  \citenamefont {Zinni}, \citenamefont {Choi}, \citenamefont {Li},
  \citenamefont {Tu}, \citenamefont {Walters}, \citenamefont {Agrestini},
  \citenamefont {Hayden}, \citenamefont {Bejas}, \citenamefont {Lin},
  \citenamefont {Yamase}, \citenamefont {Jin}, \citenamefont
  {Garc\'{\i}a-Fern\'andez}, \citenamefont {Fink}, \citenamefont {Greco},\ and\
  \citenamefont {Zhou}}]{nag24}%
  \BibitemOpen
  \bibfield  {author} {\bibinfo {author} {\bibfnamefont {Abhishek}\
  \bibnamefont {Nag}}, \bibinfo {author} {\bibfnamefont {Luciano}\ \bibnamefont
  {Zinni}}, \bibinfo {author} {\bibfnamefont {Jaewon}\ \bibnamefont {Choi}},
  \bibinfo {author} {\bibfnamefont {J.}~\bibnamefont {Li}}, \bibinfo {author}
  {\bibfnamefont {Sijia}\ \bibnamefont {Tu}}, \bibinfo {author} {\bibfnamefont
  {A.~C.}\ \bibnamefont {Walters}}, \bibinfo {author} {\bibfnamefont
  {S.}~\bibnamefont {Agrestini}}, \bibinfo {author} {\bibfnamefont {S.~M.}\
  \bibnamefont {Hayden}}, \bibinfo {author} {\bibfnamefont {Mat\'{\i}as}\
  \bibnamefont {Bejas}}, \bibinfo {author} {\bibfnamefont {Zefeng}\
  \bibnamefont {Lin}}, \bibinfo {author} {\bibfnamefont {H.}~\bibnamefont
  {Yamase}}, \bibinfo {author} {\bibfnamefont {Kui}\ \bibnamefont {Jin}},
  \bibinfo {author} {\bibfnamefont {M.}~\bibnamefont
  {Garc\'{\i}a-Fern\'andez}}, \bibinfo {author} {\bibfnamefont
  {J.}~\bibnamefont {Fink}}, \bibinfo {author} {\bibfnamefont {Andr\'es}\
  \bibnamefont {Greco}}, \ and\ \bibinfo {author} {\bibfnamefont {Ke-Jin}\
  \bibnamefont {Zhou}},\ }\bibfield  {title} {\enquote {\bibinfo {title}
  {Impact of electron correlations on two-particle charge response in electron-
  and hole-doped cuprates},}\ }\href {\doibase
  10.1103/PhysRevResearch.6.043184} {\bibfield  {journal} {\bibinfo  {journal}
  {Phys. Rev. Res.}\ }\textbf {\bibinfo {volume} {6}},\ \bibinfo {pages}
  {043184} (\bibinfo {year} {2024})}\BibitemShut {NoStop}%
\bibitem [{\citenamefont {Bejas}\ \emph {et~al.}(2025)\citenamefont {Bejas},
  \citenamefont {Wu}, \citenamefont {Chakraborty}, \citenamefont {Schnyder},\
  and\ \citenamefont {Greco}}]{bejas25}%
  \BibitemOpen
  \bibfield  {author} {\bibinfo {author} {\bibfnamefont {Mat\'{\i}as}\
  \bibnamefont {Bejas}}, \bibinfo {author} {\bibfnamefont {Xianxin}\
  \bibnamefont {Wu}}, \bibinfo {author} {\bibfnamefont {Debmalya}\ \bibnamefont
  {Chakraborty}}, \bibinfo {author} {\bibfnamefont {Andreas~P.}\ \bibnamefont
  {Schnyder}}, \ and\ \bibinfo {author} {\bibfnamefont {Andr\'es}\ \bibnamefont
  {Greco}},\ }\bibfield  {title} {\enquote {\bibinfo {title} {Out-of-plane
  bond-order phase, superconductivity, and their competition in the $t$-
  ${J}_{\ensuremath{\parallel}}\text{\ensuremath{-}}{J}_{\ensuremath{\perp}}$
  model: Possible implications for bilayer nickelates},}\ }\href {\doibase
  10.1103/PhysRevB.111.144514} {\bibfield  {journal} {\bibinfo  {journal}
  {Phys. Rev. B}\ }\textbf {\bibinfo {volume} {111}},\ \bibinfo {pages}
  {144514} (\bibinfo {year} {2025})}\BibitemShut {NoStop}%
\bibitem [{\citenamefont {Prelov\ifmmode~\check{s}\else \v{s}\fi{}ek}\ and\
  \citenamefont {Horsch}(1999)}]{prelovsek99}%
  \BibitemOpen
  \bibfield  {author} {\bibinfo {author} {\bibfnamefont {P.}~\bibnamefont
  {Prelov\ifmmode~\check{s}\else \v{s}\fi{}ek}}\ and\ \bibinfo {author}
  {\bibfnamefont {P.}~\bibnamefont {Horsch}},\ }\bibfield  {title} {\enquote
  {\bibinfo {title} {Electron-energy loss spectra and plasmon resonance in
  cuprates},}\ }\href {\doibase 10.1103/PhysRevB.60.R3735} {\bibfield
  {journal} {\bibinfo  {journal} {Phys. Rev. B}\ }\textbf {\bibinfo {volume}
  {60}},\ \bibinfo {pages} {R3735--R3738} (\bibinfo {year} {1999})}\BibitemShut
  {NoStop}%
\bibitem [{mis({\natexlab{b}})}]{misc-perp}%
  \BibitemOpen
  \href@noop {} {} ({\natexlab{b}}),\ \bibinfo {note} {in reality, one would
  expect $J_{\perp}/J \sim 0.1$ \cite{hayden96,reznik96}, yielding $J_{\perp}/t
  \sim 0.03$, which, however, may be small enough to be neglected.}\BibitemShut
  {Stop}%
\bibitem [{\citenamefont {Bejas}\ \emph {et~al.}(2012)\citenamefont {Bejas},
  \citenamefont {Greco},\ and\ \citenamefont {Yamase}}]{bejas12}%
  \BibitemOpen
  \bibfield  {author} {\bibinfo {author} {\bibfnamefont {Mat\'{\i}as}\
  \bibnamefont {Bejas}}, \bibinfo {author} {\bibfnamefont {Andr\'es}\
  \bibnamefont {Greco}}, \ and\ \bibinfo {author} {\bibfnamefont {Hiroyuki}\
  \bibnamefont {Yamase}},\ }\bibfield  {title} {\enquote {\bibinfo {title}
  {Possible charge instabilities in two-dimensional doped {Mott} insulators},}\
  }\href@noop {} {\bibfield  {journal} {\bibinfo  {journal} {Phys. Rev. B}\
  }\textbf {\bibinfo {volume} {86}},\ \bibinfo {pages} {224509} (\bibinfo
  {year} {2012})}\BibitemShut {NoStop}%
\bibitem [{\citenamefont {Bejas}\ \emph {et~al.}(2014)\citenamefont {Bejas},
  \citenamefont {Greco},\ and\ \citenamefont {Yamase}}]{bejas14}%
  \BibitemOpen
  \bibfield  {author} {\bibinfo {author} {\bibfnamefont {Mat{\'i}as}\
  \bibnamefont {Bejas}}, \bibinfo {author} {\bibfnamefont {Andr{\'e}s}\
  \bibnamefont {Greco}}, \ and\ \bibinfo {author} {\bibfnamefont {Hiroyuki}\
  \bibnamefont {Yamase}},\ }\bibfield  {title} {\enquote {\bibinfo {title}
  {{Strong particle-hole asymmetry of charge instabilities in doped Mott
  insulators}},}\ }\href@noop {} {\bibfield  {journal} {\bibinfo  {journal}
  {New J. Phys.}\ }\textbf {\bibinfo {volume} {16}},\ \bibinfo {pages} {123002}
  (\bibinfo {year} {2014})}\BibitemShut {NoStop}%
\bibitem [{\citenamefont {Hubbard}(1963)}]{hubbard63}%
  \BibitemOpen
  \bibfield  {author} {\bibinfo {author} {\bibfnamefont {J.}~\bibnamefont
  {Hubbard}},\ }\bibfield  {title} {\enquote {\bibinfo {title} {Electron
  correlations in narrow energy bands},}\ }\href {\doibase
  10.1098/rspa.1963.0204} {\bibfield  {journal} {\bibinfo  {journal} {Proc. R.
  Soc. Lond. A}\ }\textbf {\bibinfo {volume} {276}},\ \bibinfo {pages}
  {238--257} (\bibinfo {year} {1963})}\BibitemShut {NoStop}%
\bibitem [{\citenamefont {Faddeev}\ and\ \citenamefont
  {Jackiw}(1988)}]{faddeev88}%
  \BibitemOpen
  \bibfield  {author} {\bibinfo {author} {\bibfnamefont {L.}~\bibnamefont
  {Faddeev}}\ and\ \bibinfo {author} {\bibfnamefont {R.}~\bibnamefont
  {Jackiw}},\ }\bibfield  {title} {\enquote {\bibinfo {title} {Hamiltonian
  reduction of unconstrained and constrained systems},}\ }\href@noop {}
  {\bibfield  {journal} {\bibinfo  {journal} {Phys. Rev. Lett.}\ }\textbf
  {\bibinfo {volume} {60}},\ \bibinfo {pages} {1692--1694} (\bibinfo {year}
  {1988})}\BibitemShut {NoStop}%
\bibitem [{\citenamefont {Sundermeyer}(1982)}]{sundermeyer82}%
  \BibitemOpen
  \bibfield  {author} {\bibinfo {author} {\bibfnamefont {K.}~\bibnamefont
  {Sundermeyer}},\ }\href@noop {} {\emph {\bibinfo {title} {Constrained
  Dynamics}}}\ (\bibinfo  {publisher} {Berlin: Springer},\ \bibinfo {year}
  {1982})\BibitemShut {NoStop}%
\bibitem [{\citenamefont {Govaerts}(1990)}]{govaerts90}%
  \BibitemOpen
  \bibfield  {author} {\bibinfo {author} {\bibfnamefont {Jan}\ \bibnamefont
  {Govaerts}},\ }\bibfield  {title} {\enquote {\bibinfo {title} {Hamiltonian
  reduction of first-order actions},}\ }\href {\doibase
  10.1142/S0217751X90001574} {\bibfield  {journal} {\bibinfo  {journal} {Int.
  J. Mod. Phys. A}\ }\textbf {\bibinfo {volume} {05}},\ \bibinfo {pages}
  {3625--3640} (\bibinfo {year} {1990})}\BibitemShut {NoStop}%
\bibitem [{\citenamefont {Foussats}\ and\ \citenamefont
  {Greco}(2004)}]{foussats04}%
  \BibitemOpen
  \bibfield  {author} {\bibinfo {author} {\bibfnamefont {Adriana}\ \bibnamefont
  {Foussats}}\ and\ \bibinfo {author} {\bibfnamefont {Andr\'es}\ \bibnamefont
  {Greco}},\ }\bibfield  {title} {\enquote {\bibinfo {title} {{Large-$N$
  expansion based on the Hubbard operator path integral representation and its
  application to the $t\text{\ensuremath{-}}J$ model. II. The case for finite
  $J$}},}\ }\href {\doibase 10.1103/PhysRevB.70.205123} {\bibfield  {journal}
  {\bibinfo  {journal} {Phys. Rev. B}\ }\textbf {\bibinfo {volume} {70}},\
  \bibinfo {pages} {205123} (\bibinfo {year} {2004})}\BibitemShut {NoStop}%
\bibitem [{\citenamefont {Yamase}\ \emph {et~al.}(2021)\citenamefont {Yamase},
  \citenamefont {Bejas},\ and\ \citenamefont {Greco}}]{yamase21a}%
  \BibitemOpen
  \bibfield  {author} {\bibinfo {author} {\bibfnamefont {Hiroyuki}\
  \bibnamefont {Yamase}}, \bibinfo {author} {\bibfnamefont {Mat\'{\i}as}\
  \bibnamefont {Bejas}}, \ and\ \bibinfo {author} {\bibfnamefont {Andr\'es}\
  \bibnamefont {Greco}},\ }\bibfield  {title} {\enquote {\bibinfo {title}
  {{Electron self-energy from quantum charge fluctuations in the layered
  $t\ensuremath{-}J$ model with long-range Coulomb interaction}},}\ }\href
  {\doibase 10.1103/PhysRevB.104.045141} {\bibfield  {journal} {\bibinfo
  {journal} {Phys. Rev. B}\ }\textbf {\bibinfo {volume} {104}},\ \bibinfo
  {pages} {045141} (\bibinfo {year} {2021})}\BibitemShut {NoStop}%
\bibitem [{\citenamefont {Bejas}\ \emph {et~al.}(2017)\citenamefont {Bejas},
  \citenamefont {Yamase},\ and\ \citenamefont {Greco}}]{bejas17}%
  \BibitemOpen
  \bibfield  {author} {\bibinfo {author} {\bibfnamefont {Mat\'{\i}as}\
  \bibnamefont {Bejas}}, \bibinfo {author} {\bibfnamefont {Hiroyuki}\
  \bibnamefont {Yamase}}, \ and\ \bibinfo {author} {\bibfnamefont {Andr\'es}\
  \bibnamefont {Greco}},\ }\bibfield  {title} {\enquote {\bibinfo {title} {Dual
  structure in the charge excitation spectrum of electron-doped cuprates},}\
  }\href {\doibase 10.1103/PhysRevB.96.214513} {\bibfield  {journal} {\bibinfo
  {journal} {Phys. Rev. B}\ }\textbf {\bibinfo {volume} {96}},\ \bibinfo
  {pages} {214513} (\bibinfo {year} {2017})}\BibitemShut {NoStop}%
\bibitem [{\citenamefont {Tranquada}\ \emph {et~al.}(1995)\citenamefont
  {Tranquada}, \citenamefont {Sternlieb}, \citenamefont {Axe}, \citenamefont
  {Nakamura},\ and\ \citenamefont {Uchida}}]{tranquada95}%
  \BibitemOpen
  \bibfield  {author} {\bibinfo {author} {\bibfnamefont {J.~M.}\ \bibnamefont
  {Tranquada}}, \bibinfo {author} {\bibfnamefont {B.~J.}\ \bibnamefont
  {Sternlieb}}, \bibinfo {author} {\bibfnamefont {J.~D.}\ \bibnamefont {Axe}},
  \bibinfo {author} {\bibfnamefont {Y.}~\bibnamefont {Nakamura}}, \ and\
  \bibinfo {author} {\bibfnamefont {S.}~\bibnamefont {Uchida}},\ }\bibfield
  {title} {\enquote {\bibinfo {title} {Evidence for stripe correlations of
  spins and holes in copper oxide superconductors},}\ }\href {\doibase
  10.1038/375561a0} {\bibfield  {journal} {\bibinfo  {journal} {Nature}\
  }\textbf {\bibinfo {volume} {375}},\ \bibinfo {pages} {561--563} (\bibinfo
  {year} {1995})}\BibitemShut {NoStop}%
\bibitem [{\citenamefont {Hayden}\ \emph {et~al.}(1996)\citenamefont {Hayden},
  \citenamefont {Aeppli}, \citenamefont {Perring}, \citenamefont {Mook},\ and\
  \citenamefont {Do\ifmmode~\breve{g}\else \u{g}\fi{}an}}]{hayden96}%
  \BibitemOpen
  \bibfield  {author} {\bibinfo {author} {\bibfnamefont {S.~M.}\ \bibnamefont
  {Hayden}}, \bibinfo {author} {\bibfnamefont {G.}~\bibnamefont {Aeppli}},
  \bibinfo {author} {\bibfnamefont {T.~G.}\ \bibnamefont {Perring}}, \bibinfo
  {author} {\bibfnamefont {H.~A.}\ \bibnamefont {Mook}}, \ and\ \bibinfo
  {author} {\bibfnamefont {F.}~\bibnamefont {Do\ifmmode~\breve{g}\else
  \u{g}\fi{}an}},\ }\bibfield  {title} {\enquote {\bibinfo {title}
  {{High-frequency spin waves in
  Y${\mathrm{Ba}}_{2}$${\mathrm{Cu}}_{3}$${\mathrm{O}}_{6.15}$}},}\ }\href
  {\doibase 10.1103/PhysRevB.54.R6905} {\bibfield  {journal} {\bibinfo
  {journal} {Phys. Rev. B}\ }\textbf {\bibinfo {volume} {54}},\ \bibinfo
  {pages} {R6905--R6908} (\bibinfo {year} {1996})}\BibitemShut {NoStop}%
\bibitem [{\citenamefont {Reznik}\ \emph {et~al.}(1996)\citenamefont {Reznik},
  \citenamefont {Bourges}, \citenamefont {Fong}, \citenamefont {Regnault},
  \citenamefont {Bossy}, \citenamefont {Vettier}, \citenamefont {Milius},
  \citenamefont {Aksay},\ and\ \citenamefont {Keimer}}]{reznik96}%
  \BibitemOpen
  \bibfield  {author} {\bibinfo {author} {\bibfnamefont {D.}~\bibnamefont
  {Reznik}}, \bibinfo {author} {\bibfnamefont {P.}~\bibnamefont {Bourges}},
  \bibinfo {author} {\bibfnamefont {H.~F.}\ \bibnamefont {Fong}}, \bibinfo
  {author} {\bibfnamefont {L.~P.}\ \bibnamefont {Regnault}}, \bibinfo {author}
  {\bibfnamefont {J.}~\bibnamefont {Bossy}}, \bibinfo {author} {\bibfnamefont
  {C.}~\bibnamefont {Vettier}}, \bibinfo {author} {\bibfnamefont {D.~L.}\
  \bibnamefont {Milius}}, \bibinfo {author} {\bibfnamefont {I.~A.}\
  \bibnamefont {Aksay}}, \ and\ \bibinfo {author} {\bibfnamefont
  {B.}~\bibnamefont {Keimer}},\ }\bibfield  {title} {\enquote {\bibinfo {title}
  {{Direct observation of optical magnons in
  Y${\mathrm{Ba}}_{2}$${\mathrm{Cu}}_{3}$${\mathrm{O}}_{6.2}$}},}\ }\href
  {\doibase 10.1103/PhysRevB.53.R14741} {\bibfield  {journal} {\bibinfo
  {journal} {Phys. Rev. B}\ }\textbf {\bibinfo {volume} {53}},\ \bibinfo
  {pages} {R14741--R14744} (\bibinfo {year} {1996})}\BibitemShut {NoStop}%
\end{thebibliography}%

\end{document}